\documentclass[12pt,a4paper]{iopart}

\usepackage[english]{babel}

\usepackage{iopams}
\usepackage{verbatim}
\usepackage{color}

\usepackage{color}

\usepackage[colorlinks,citecolor=blue,linkcolor=black,urlcolor=blue]{hyperref}

\def\iotabar{\lower3pt\hbox{$\mathchar'26$}\mkern-7mu\iota}
\newcommand {\aplt}{\ {\raise-.5ex\hbox{$\buildrel<\over\sim$}}\ }
\newcommand{\dd}{\mbox{d}}

\newcommand{\spe}{{\sigma}}
\newcommand{\lw}{{\rm lw}}
\newcommand{\sw}{{\rm sw}}
\newcommand{\eq}[1]{(\ref{#1})}
\newcommand{\bun}{\hat{\mathbf{b}}}
\newcommand{\eun}{\hat{\mathbf{e}}}

\newcommand{\phiwig}{\widetilde{\phi}}
\newcommand{\Phiwig}{\widetilde{\Phi}}
\newcommand{\boldr}{\mathbf{r}}
\newcommand{\bv}{\mathbf{v}}
\newcommand{\bA}{\mathbf{A}}
\newcommand{\bR}{\mathbf{R}}

\newcommand{\bB}{\mathbf{B}}
\newcommand{\bE}{\mathbf{E}}
\newcommand{\bZ}{\mathbf{Z}}
\newcommand{\bK}{\mathbf{K}}

\newcommand{\bX}{\mathbf{X}}
\newcommand{\bV}{\mathbf{V}}

\newcommand{\matrixtop}[1]{\buildrel\leftrightarrow\over{#1}}
\newcommand{\matI}{\matrixtop{\mathbf{I}}}
\newcommand{\matW}{\matrixtop{\mathbf{W}}}
\newcommand{\matM}{\matrixtop{\mathbf{M}}}

\newcommand{\dotcross}{ \raise 0.65ex\hbox{${\scriptstyle {{_{\displaystyle \cdot}}\atop\times}}$} }
\newcommand{\crossdot}{ \raise 0.5ex\hbox{${\scriptstyle {{_\times}\atop{\displaystyle \cdot}}}$} }
\newcommand{\rhobf}{\mbox{\boldmath$\rho$}}

\newcommand{\kappabf}{\mbox{\boldmath$\kappa$}}
\newcommand{\zetabf}{\mbox{\boldmath$\zeta$}}

\newcommand{\zun}{\hat{\zetabf}}

\newcommand{\cT}{{\cal T}}
\newcommand{\modTinv}{{\mathbb T_{\spe,0}}}
\newcommand{\modTinvprime}{{\mathbb T_{\spe',0}}}

\newcommand{\sumsig}{ \raise -1.3ex\hbox{${{\displaystyle \sum}\atop{\scriptstyle \sigma}}$} }
\newcounter{appnumb}

\begin{document}

\title[Long-wavelength limit of gyrokinetics in a turbulent tokamak]
      {Long-wavelength limit of gyrokinetics in a turbulent tokamak and
        its intrinsic ambipolarity}

\author{Iv\'an Calvo$^{1}$}
\eads{\mailto{ivan.calvo@ciemat.es}}

\author{Felix I Parra$^{2}$}
\eads{\mailto{fparra@mit.edu}}

\vspace{1cm}

\address{$^1$Laboratorio Nacional de Fusi\'on, Asociaci\'on
EURATOM-CIEMAT, 28040 Madrid, Spain}
\address{$^2$Plasma Science and Fusion Center, MIT, Cambridge, MA 02139, USA}

\begin{abstract}
Recently, the electrostatic gyrokinetic Hamiltonian
    and change of coordinates have been computed to order $\epsilon^2$
    in general magnetic geometry. Here $\epsilon$ is the gyrokinetic
    expansion parameter, the gyroradius over the macroscopic scale
    length. Starting from these results, the long-wavelength limit of
    the gyrokinetic Fokker-Planck and quasineutrality equations is
    taken for tokamak geometry. Employing the set of equations derived
    in the present article, it is possible to calculate the
    long-wavelength components of the distribution functions and of
    the poloidal electric field to order $\epsilon^2$. These
    higher-order pieces contain both neoclassical and turbulent
    contributions, and constitute one of the necessary ingredients
    (the other is given by the short-wavelength components up to
    second order) that will eventually enter a complete model for
    the radial transport of toroidal angular momentum in a
    tokamak in the low flow ordering. Finally, we provide an explicit
    and detailed proof that the system consisting of second-order
    gyrokinetic Fokker-Planck and quasineutrality equations leaves the
    long-wavelength radial electric field undetermined; that is, the
    turbulent tokamak is intrinsically ambipolar.
\end{abstract}


\vskip 3cm


\maketitle

\section{Introduction}
\label{sec:Introduction}

Gyrokinetic theory~\cite{catto78} and gyrokinetic codes
\cite{dimits96, dorland00, dannert05, candy03, chen03, peeters09} are
recognized as the fundamental tools for the description of
microturbulence in fusion and astrophysical plasmas. Gyrokinetic
theory consists of the elimination of the degree of freedom associated
to the gyration of the charged particle around the magnetic field
order by order in an asymptotic expansion in $\epsilon = \rho/L \ll
1$, where $\rho$ is the gyroradius and $L$ is the macroscopic scale
length of the problem. This procedure reduces the phase-space
dimension and, more importantly, the degree of freedom averaged out is
precisely the one with the shortest time scale.  The savings in
computational time that gyrokinetics has provided have made it
possible to simulate kinetic plasma turbulence. Derivations of the
gyrokinetic equations by iterative methods can be found in
references~\cite{frieman82, lee83, wwlee83, bernstein85, parra08}, and
via Hamiltonian and Lagrangian methods in references~\cite{dubin83,
  hahm88, brizard07,ParraCalvo2011}. A recent review of gyrokinetic
theory is given in \cite{krommes2012}.

The gyrokinetic equations have typically been solved only for the
turbulent components of the distribution function and the
electrostatic potential (we restrict our discussion to electrostatic
gyrokinetics), but in recent years growing supercomputer capabilities
have motivated an increasing interest in the extension of gyrokinetic
calculations to longer wavelengths and transport time scales. However,
at least for a tokamak~\footnote{Throughout this paper ``tokamak''
  means ``axisymmetric tokamak''.}, this is a subtle issue, as
F.~I. Parra and P.~J.  Catto have discussed in a series of papers
\cite{parra08,parra09b,parra09c,parra10b,parra10c,parra10a}. The main
lines of the argument can be stated in a succinct way. The
perpendicular component of the long-wavelength piece of the plasma
velocity depends on the long-wavelength radial electric field through
the $\bE \times \bB$ drift. The momentum conservation equation can be
used to obtain the three components of the velocity, and from it,
derive the radial electric field. The plasma velocity is to lowest
order parallel to the flux surfaces because the radial particle drift
is small.  Then, the poloidal and toroidal components of the momentum
conservation equation are sufficient to calculate the velocity to the
order of interest, and by decomposing it in parallel and perpendicular
components, the radial electric field can be obtained by making the
perpendicular component equal to the $\bE \times \bB$ drift plus the
diamagnetic velocity. The poloidal component of the velocity is
strongly damped by collisions because the poloidal direction is not a
direction of symmetry. The poloidal velocity is determined by setting
the collisional viscosity in the poloidal direction equal to zero,
giving a poloidal velocity proportional to the ion temperature
gradient unless collisionality is really small and turbulence can
compete with the collisional damping~\cite{parra09b, parra10a}.
Unfortunately, the toroidal component of the momentum equation that
would give the toroidal component of the velocity and completely
determine the radial electric field is identically satisfied to order
$\epsilon^2$ by any toroidal velocity \cite{parra09b, parra10b}. Since
gyrokinetic equations are customarily derived and solved to order
$\epsilon$, the tokamak long-wavelength radial electric field cannot
be correctly obtained from the standard set of gyrokinetic equations
available in the literature.

In the limit in which the velocity is of the order of the diamagnetic
velocity, known as low flow limit, the calculation of the radial flux
of toroidal angular momentum, which we need to compute the radial
electric field, is especially demanding because
this flux is smaller than the radial flux of particles and
energy in the expansion in $\epsilon$. The low flow limit is relevant
in the study of intrinsic rotation \cite{parra11c,
  parra11d,ParraBarnesCalvoCatto2012}. In references~\cite{parra10a,
  parra11d}, a method to calculate the toroidal angular momentum
conservation equation in the low flow limit to the order in which it
is not identically zero is proposed. With the toroidal angular
momentum equation to this order, it is possible to obtain the toroidal
rotation and hence calculate the radial electric field. The formula
for the radial flux of toroidal angular momentum in \cite{parra10a,
  parra11d} is given as a sum of several integrals over the first- and
second-order pieces of the distribution functions and the
electrostatic potential. To avoid calculating these second-order
pieces in complete detail, a subsidiary expansion in $B_p/B \ll 1$ was
employed, where $B_p$ is the poloidal magnetic field and $B$ is the
total magnetic field.  With the derivation for the first time of the
gyrokinetic equations and change of coordinates in general magnetic
geometry up to second order~\cite{ParraCalvo2011}, it has become
possible to calculate the second-order pieces without resorting to a
subsidiary expansion. In this article, we present the equations that
need to be solved to obtain the long-wavelength second order
pieces. These equations have not been explicitly written before.  They
contain neoclassical \cite{hinton76, helander02bk} and turbulent
contributions. The turbulent contributions have never been considered
to our knowledge, and the complete neoclassical equations have only
been used in the Pfirsch-Schl\"{u}ter limit in
\cite{catto05}. Calculations of the neoclassical radial flux of
toroidal angular momentum in other collisionality regimes have relied
on the $B_p/B \ll 1$ expansion \cite{wong09}.

We emphasize that the equations derived here are the first step
towards a complete model for the computation of radial transport of
toroidal angular momentum in a tokamak. The second step, that will be
taken in a future publication, includes the derivation of the
equations determining the short-wavelength components of the
distribution functions and electrostatic potential to second
order. To ease the reading of the paper, we advance
in this introduction which are the equations that we derive, and that
will eventually enter the aforementioned complete model for toroidal
angular momentum transport in a tokamak. They are the long-wavelength
Fokker-Planck equations to second order, \eq{eq:Vlasovorder1gyroav4}
and \eq{eq:eqH2sigma}, that give the long-wavelength
component of the distribution functions; the quasineutrality equation
up to second-order \eq{eq:gyroPoissonlw2order0}, \eq
{eq:quasineutralityWithCorrectionsOrder1}, and
\eq{eq:quasineutralityWithCorrectionsOrder2}, that determines the
first and second-order pieces of the long-wavelength poloidal electric
field; and the transport equations for density
\eq{eq:transportEq_density} and energy
\eq{eq:transportEq_totalEnergy}. The first-order pieces of the
short-wavelength components of the distribution functions and
electrostatic potential appear in \eq{eq:eqH2sigma}, and we give the
equations for them in \eq {eq:sworder1distfunction} and
\eq{eq:quasinautralitySWorder1appendix}.

Carrying the expansion to second order in $\epsilon$ at long
wavelengths also clarifies the issues with the radial electric field
raised in references
\cite{parra08,parra09b,parra09c,parra10b,parra10c,parra10a}, pointed
out at the beginning of this introduction. Along with the derivation
of the equations we give an explicit proof of the indeterminacy of the
radial electric field, showing that it cannot be found from the
long-wavelength gyrokinetic Fokker-Planck and quasineutrality
equations correct to second order. This property, known as intrinsic
ambipolarity, was first proven for neoclassical transport in
\cite{kovrizhnykh69, rutherford70} and it was shown to hold for
turbulent tokamaks in \cite{parra09b} using the identical cancellation
of the toroidal angular momentum conservation equation to the order of
interest. The intrinsic ambipolarity of purely turbulent particle
fluxes was shown to hold in \cite{Sugama96}, even electromagnetically
and in general magnetic geometry (that is why the long-wavelength
radial electric field in non-quasisymmetric stellarators is determined
from neoclassical theory). This is, however, the first direct,
explicit, and general proof for turbulent tokamaks. Instead of
resorting to the toroidal angular momentum equation, we write the
long-wavelength equations order by order and show that they can be
solved for any radial electric field, leaving it undetermined. Those
readers who are familiar with the Chapman-Enskog results on the
derivation of fluid equations from kinetic theory (see the classical
monograph \cite{ChapmanCowling}) will find that the approach that we
adopt at some stages of the proof is very similar. The analogy becomes
especially clear in Section \ref{sec:transportEqs}. In previous
sections the long-wavelength Fokker-Planck and quasineutrality
equations have been derived up to second order. In Section
\ref{sec:transportEqs} we inspect the second-order piece of the
long-wavelength Fokker-Planck equation and learn that it possesses
solvability conditions, i.e. the existence of solutions of this
equation imposes constraints on lowest-order quantities. These
constraints are transport equations for particle and energy
density. The way of obtaining them and of showing that we have
actually found all the solvability conditions are the aspects
particularly reminiscent of the Chapman-Enskog
techniques. Nevertheless, we have written the paper in a
self-contained fashion and no prior knowledge of the Chapman-Enskog
theory is assumed.

The rest of the paper is organized as follows. In Section
\ref{sec:secondordergyrokinetics} we introduce the gyrokinetic
formulation and the essential results and notation from
\cite{ParraCalvo2011} that will be needed here. An important element
of our derivation is the scale separation between the turbulent
short-wavelength fluctuations and the equilibrium long-wavelength
profiles. In Section 2 we also discuss the implications of this scale
separation and formalize the notion of ``taking the long-wavelength
limit of gyrokinetics''. The most laborious part of this work
corresponds to explicitly taking the long-wavelength limit of the
gyrokinetic system of equations in tokamak geometry by employing the
results of \cite{ParraCalvo2011}. In Section
\ref{sec:FokkerPlancklong} we do it for the Fokker-Planck equation and
in Section \ref{sec:longwavePoisson} for the quasineutrality equation.
Reaching the final expressions for the long-wavelength limit of the
gyrokinetic system to second order involves enormous amounts of
algebra, and in order to ease a first reading of the paper the most
cumbersome parts of the calculation have been collected in the
appendices. Using the results of Sections \ref{sec:FokkerPlancklong}
and \ref{sec:longwavePoisson} we prove in Section
\ref{sec:indeterminacyRadialElectricField} that the long-wavelength
tokamak radial electric field is not determined by second-order
Fokker-Planck and quasineutrality equations. A complete proof requires
computing the solvability conditions imposed by the second-order
long-wavelength Fokker-Planck equation, contained in Subsection
\ref{sec:transportEqs}. These conditions are transport equations for
particle and energy densities, as mentioned above.  With these
transport equations, we show in Section
\ref{sec:timeevolutionquasineutrality} that the well-known
neoclassical intrinsic ambipolarity property of the tokamak is not
broken by the turbulent terms that are specific to gyrokinetics, that
is, the radial electric field is left undetermined by a gyrokinetic
system of equations correct to second order in $\epsilon$. Section
\ref{sec:conclusions} is devoted to a discussion of the results and
the conclusions.

\section{Second-order electrostatic gyrokinetics}
\label{sec:secondordergyrokinetics}

In this section we state and justify the
assumptions of the theory, and we summarize the results
from reference \cite{ParraCalvo2011} that will be needed.

\subsection{Kinetic description of a plasma in a static magnetic field}
\label{sec:kineticdescriptionplasma}

The kinetic description of a plasma in the electrostatic approximation
involves the Fokker-Planck equation for each species $\spe$,
\begin{eqnarray}\label{eq:FPinitial}
\fl\partial_t f_\spe + \bv\cdot\nabla_\boldr f_\spe
+\frac{Z_\spe e}{m_\spe}\left(-\nabla_\boldr\varphi +
c^{-1}\, \bv\times\bB\right)\cdot\nabla_\bv f_\spe =\nonumber\\[5pt]
\fl\hspace{1cm}
\sum_{\spe'}C_{\spe \spe'}[f_\spe,f_{\spe'}](\boldr,\bv),
\end{eqnarray}
and Poisson's equation,
\begin{eqnarray}\label{eq:Poisson}
\nabla^2_\boldr \varphi (\boldr,t)= -4\pi e\sum_\spe Z_\spe\int \,
f_\spe(\boldr,\bv,t)\dd ^3 v.
\end{eqnarray}
Here $c$ is the speed of light, $e$ the charge of the proton,
$\varphi(\boldr, t)$ the electrostatic potential,
$\bB(\boldr)=\nabla_\boldr\times\bA(\boldr)$ a time-independent
magnetic field, $f_\spe(\boldr,\bv,t)$ the phase-space probability
distribution, and $Z_\spe e$ and $m_\spe$ are the
charge and the mass of species $\spe$. We recall that the Landau
collision operator between species $\spe$ and $\spe'$ reads
\begin{eqnarray}\label{eq:collisionoperator}
C_{\spe \spe'}[f_\spe,f_{\spe'}](\boldr,\bv) =
\nonumber\\[5pt]
\hspace{1cm}
\frac{\gamma_{\spe\spe'}}{m_\sigma}
\nabla_\bv\cdot
\int
\matW(\bv-\bv')
\cdot
\Bigg(
\frac{1}{m_\spe}f_{\spe'}(\boldr,\bv',t)\nabla_\bv f_\spe(\boldr,\bv,t)
\nonumber\\[5pt]
\hspace{1cm}
-
\frac{1}{m_{\spe'}}
f_{\spe}(\boldr,\bv,t)\nabla_{\bv'} f_{\spe'}(\boldr,\bv',t)
\Bigg)
\dd^3v',
\end{eqnarray}
where
\begin{equation}
\gamma_{\spe\spe'}:= 2\pi Z_\spe^2 Z_{\spe'}^2 e^4 \ln\Lambda,
\end{equation}
\begin{equation}
{\bf \matW}({\bold w})
 := \frac{|{\bold w}|^2{\matI}-{\bold w}{\bold w}}{|{\bold w}|^3},
\end{equation}
$\ln\Lambda$ is the Coulomb logarithm, and $\matI$ is the identity
matrix. A direct check shows that the Fokker-Planck equation can also
be written as
\begin{eqnarray}
\partial_t f_\spe + \{f_\spe,H_\spe\}_\bX =
\sum_{\spe'}C_{\spe \spe'}[f_\spe,f_{\spe'}](\bX),
\end{eqnarray}
where we designate by $\bX\equiv (\boldr,\bv)$ a set of euclidean
coordinates in phase-space,
\begin{eqnarray}
H_\spe(\boldr,\bv,t) = \frac{1}{2}m_\spe \bv^2 + Z_\spe
e\varphi(\boldr,t)
\end{eqnarray}
is the Hamiltonian of species $\spe$, and the Poisson bracket of
two functions on phase space, $g_1(\boldr,\bv)$,
$g_2(\boldr,\bv)$, is
\begin{eqnarray}
\{g_1,g_2\}_\bX &= \frac{1}{m_\spe}\left(\nabla_\boldr g_1\cdot\nabla_\bv g_2 -
\nabla_\bv g_1\cdot\nabla_\boldr g_2\right)\nonumber\\[5pt]
 &+ \frac{Z_\spe e}{m_\spe^2c} \bB\cdot(\nabla_\bv g_1\times\nabla_\bv g_2).
\end{eqnarray}

\subsection{Dimensionless variables}
\label{sec:dimensionlessvariables}

In most of what follows we find it convenient to work with
non-dimensionalized variables~\cite{ParraCalvo2011}. The
species-independent normalization
\begin{eqnarray} \label{norm_spindep}
\underline{t} = \frac{c_s t}{L}, \ \underline{\boldr} =
\frac{\boldr}{L}, \ \underline{\bA} = \frac{\bA}{B_0 L}, \
\underline{\varphi} = \frac{e \varphi}{\epsilon_s T_{e0}}, \
\nonumber\\[5pt]
\underline{H_\spe} = \frac{H_\spe}{T_{e0}}, \
\underline{n_{\sigma}}=\frac{n_\sigma}{n_{e0}}, \
\underline{T_{\sigma}}=\frac{T_\sigma}{T_{e0}},
\end{eqnarray}
\noindent is employed for time, space, vector potential, electrostatic
potential, Hamiltonian, particle density, and temperature; and the
species-dependent normalization
\begin{equation} \label{norm_spdep}
\underline{\bv_\spe} = \frac{\bv_\spe}{v_{t\spe}}, \
\underline{f_\spe} = \frac{v_{t\spe}^3}{n_{e0}} f_\spe,
\end{equation}
\noindent for velocities and distribution functions. In the previous
expressions $L\sim|\nabla_\boldr\ln|\bB||^{-1}$ is the typical length
of variation of the magnetic field, $B_0$ a typical value of the
magnetic field strength, $c_s=\sqrt{T_{e0}/m_i}$ the sound speed,
$T_{e0}$ a typical electron temperature, $n_{e0}$ a typical electron
density, and $m_i$ the mass of the dominant ion species, that we
assume singly charged. Finally, $v_{t\spe}$ is the thermal speed of
species $\spe$, $\epsilon_s=\rho_s/L$, where $\rho_s=c_s/\Omega_i$ is
a characteristic sound gyroradius, and $\Omega_i={eB_0/(m_ic)}$ is a
characteristic ion gyrofrequency. We take
$v_{t\spe}=\sqrt{T_{e0}/m_\spe}$ as the expression for the typical
thermal speed, i.e. we assume that $T_{e 0}$, the characteristic
temperature of electrons, is also the characteristic temperature for
all species. This assumption is justified when the time between
collisions is shorter than the transport time scale, leading to
thermal equilibration between species. The normalization of the
electrostatic potential might seem strange at this point but it
will be explained in the next subsection.

The natural, species-independent expansion parameter in gyrokinetic
theory is $\epsilon_s$. Many expressions, however, are more
conveniently written in terms of the species-dependent parameter
$\epsilon_\spe =\rho_\spe/L$, where $\rho_\spe=v_{t\spe}/\Omega_\spe$
is a characteristic gyroradius of species $\spe$ and
$\Omega_\spe=Z_\spe e B_0/(m_\spe c)$ a characteristic
gyrofrequency. Observe that the relation between $\epsilon_\spe$ and
$\epsilon_s$ is $\epsilon_s=\lambda_\spe\epsilon_\spe$, with
\begin{equation}
\lambda_\spe = \frac{\rho_s}{\rho_\spe} = Z_\spe
\sqrt{\frac{m_i}{m_\spe}}.
\end{equation}

In dimensionless variables, the Fokker-Planck
equation~\eq{eq:FPinitial} becomes
\begin{eqnarray}\label{eq:FPnon-dim}
\partial_{\underline{t}}\, \underline{f_\spe} +
\tau_\spe
\left\{\underline{f_\spe},\underline{H_\spe}
\right\}_{\underline{\bX}} = \tau_\spe
\sum_{\spe'}\underline{C_{\spe \spe'}}
[\underline{f_\spe},\underline{f_{\spe'}}]
(\underline{\boldr},\underline{\bv}),
\end{eqnarray}
where
\begin{equation}
\tau_\spe = \frac{v_{t\spe}}{c_s} =
\sqrt{\frac{m_i}{m_\spe }}\, ,
\end{equation}
and the Poisson bracket of two functions
$g_1(\underline{\boldr},\underline{\bv})$,
$g_2(\underline{\boldr},\underline{\bv})$ (we no longer write the
subindex $\spe$ in $\bv_\spe$) is defined by
\begin{eqnarray}
\{g_1,g_2\}_{\underline{\bX}} &=
\left(\nabla_{\underline{\boldr}}
        g_1\cdot\nabla_{\underline{\bv}} g_2 - \nabla_{\underline{\bv}}
        g_1\cdot\nabla_{\underline{\boldr}} g_2\right) 
\nonumber\\[5pt]
&+
        \frac{1}{\epsilon_\spe}
        \underline{\bB}\cdot(\nabla_{\underline{\bv}}
        g_1\times\nabla_{\underline{\bv}} g_2).
\end{eqnarray}
Here $\underline{\bX}\equiv (\underline{\boldr},\underline{\bv})$ are
the dimensionless cartesian coordinates. The normalized collision
operator is
\begin{eqnarray}\label{eq:collisionoperatornondim}
\underline{C_{\spe \spe'}}
[\underline{f_\spe},\underline{f_{\spe'}}]
(\underline{\boldr},\underline{\bv}) =\nonumber\\[5pt]
\hspace{1cm}\underline{\gamma_{\spe\spe'}} \nabla_{\underline{\bv}}\cdot
\int \matW\left(\tau_\spe\underline{\bv} -
\tau_{\spe^\prime}\underline{\bv'}\right) \cdot
\Bigg( \tau_\spe\underline{f_{\spe'}}(\underline{\boldr},\underline{\bv'},
\underline{t})
\nabla_{\underline{\bv}}
\underline{f_\spe}(\underline{\boldr},\underline{\bv},\underline{t})
\nonumber\\[5pt]
\hspace{1cm}
-
\tau_{\spe'}
\underline{f_{\spe}}(\underline{\boldr},\underline{\bv},\underline{t})
\nabla_{\underline{\bv'}} \underline{f_{\spe'}}
(\underline{\boldr},\underline{\bv'},\underline{t}) \Bigg)
\dd^3\underline{v'},\nonumber\\
\end{eqnarray}
with
\begin{equation}
\underline{\gamma_{\spe\spe'}}:= \frac{2 \pi
Z_\spe^2 Z_{\spe'}^2 n_{e0} e^4 L}{T_{e0}^2}\ln\Lambda.
\end{equation}
Note in passing that $\underline{\gamma_{\spe\spe'}}$ is the usual
collisionality parameter $\nu_{*\spe \spe^\prime}$ up to a factor of
order unity. We use the following definition of $\nu_{*\spe
  \spe^\prime}$:
\begin{equation}
\nu_{*\spe \spe^\prime}:=L\nu_{\spe \spe^\prime}/v_{t\spe},
\end{equation}
where the collision frequency is
\begin{equation}
\nu_{\spe \spe^\prime}:= \frac{4\sqrt{2\pi}}{3}
\frac{Z_\spe^2 Z_{\spe'}^2 n_{e0} e^4 }{m_\spe^{1/2} T_{\spe}^{3/2}}\ln\Lambda,
\end{equation}
which coincides with Braginskii's definition~\cite{Braginskii65} for
$\spe=e$ and $\spe'=i$.

As for equation~\eq{eq:Poisson},
\begin{eqnarray}\label{eq:Poissonnondim}
 \frac{\epsilon_s \lambda_{De}^2}{L^2} \nabla_{\underline{r}}^2\, 
\underline{\varphi} (\underline{\boldr}, \underline{t}) = -\sum_\spe Z_\spe
\int
\underline{f_\spe}(\underline{\boldr},
\underline{\bv}, \underline{t}) \dd^3 \underline{v},
\end{eqnarray}
where
\begin{equation}
\lambda_{De} = \sqrt{\frac{T_{e0}}{4\pi e^2 n_{e0}}}
\end{equation}
is the electron Debye length. We assume that the Debye
length is sufficiently small that we can neglect
the left-hand side of \eq{eq:Poissonnondim}, so quasineutrality
\begin{eqnarray}\label{eq:Quasineutralitynondim}
 \sum_\spe Z_\spe
\int \underline{f_\spe}(\underline{\boldr}, \underline{\bv},
\underline{t}) \dd^3 \underline{v}=0
\end{eqnarray}
 holds. 

\subsection{Gyrokinetic ordering and separation of scales}
\label{sec:orderingANDseparationOFscales}

In strongly magnetized plasmas a small quantity,
$\epsilon_\spe=\rho_\spe/L\ll 1$, naturally arises for each
species. The smallness of $\epsilon_\spe$ implies that two very
different length scales exist: the gyroradius scale and the
macroscopic scale. Also, strong magnetization makes the time scale
associated to the gyromotion around a field line, $\Omega_\spe^{-1}$,
very small compared to microturbulence time scales. It is therefore
justified to try to average over the irrelevant gyromotion without
losing non-zero gyroradius effects. Gyrokinetics is
  the theory that results from averaging over the gyromotion when the
  parameter $\epsilon_\spe$ (or more precisely $\epsilon_s$) is
  small. We assume that
  $\underline{\gamma_{\spe\spe'}}\sim\lambda_\spe\sim\tau_\spe\sim 1$
  for all $\spe,\spe'$. That is, the only formal expansion parameter
  is $\epsilon_s$. This is a maximal expansion in the sense that the
  different physically reasonable and customary subsidiary expansions
  (such as expansions in mass ratios) are contained in our results and
  could be eventually performed in order to simplify the equations.

As most gyrokinetic derivations, this article relies
  on a set of {\it ans\" atze} about scale separation and ordering
  that we proceed to explain.

First we define a transport or coarse-grain average, that for a given
function extracts the axisymmetric component (recall that our aim is
to fully work out the axisymmetric case) corresponding to long
wavelengths and small frequencies.  Let $\{\psi,\Theta,\zeta\}$ be a
set of flux coordinates, where $\psi$ is the poloidal magnetic flux,
$\Theta$ is the poloidal angle, and $\zeta$ is the toroidal angle. A
working definition of this averaging operation can be given by
\begin{equation}
\fl\langle \dots \rangle_{\rm{T}} 
= \frac{1}{2\pi\Delta t \Delta \psi \Delta\Theta}
\int_{\Delta t}\dd t
\int_{\Delta \psi}\dd \psi
\int_{\Delta \Theta}\dd \Theta\int_0^{2\pi}\dd\zeta(\dots),
\end{equation}
where $\epsilon_s\ll \Delta\psi/\psi \ll 1$, $\epsilon_s\ll \Delta\Theta \ll
1$, and $L/c_s\ll\Delta t\ll \tau_E$. Here $\tau_E \sim \epsilon_s^{-2} L/c_s$ is the transport time scale. For any function
$g(\boldr,t)$, we define
\begin{eqnarray}
g^\lw &:= \left\langle g
\right\rangle_{\rm{T}}\nonumber\\[5pt]
g^\sw &:= g - g^\lw.
\end{eqnarray}
The following obvious properties will be repeatedly employed:
\begin{eqnarray}
&&\left[g^\lw\right]^\lw = g^\lw,\nonumber\\[5pt]
&&\left[g^\sw\right]^\lw = 0,\nonumber\\[5pt]
&&\left[g h\right]^\lw = g^\lw h^\lw + \left[g^\sw h^\sw\right]^\lw, 
\end{eqnarray}
for any two functions $g(\boldr,t)$ and $h(\boldr,t)$. We decompose
the fields of our theory using the coarse-grain average:
\begin{eqnarray}
f_\spe& = f_\spe^\lw + f_\spe^\sw,\nonumber\\
\varphi& = \varphi^\lw
+ \varphi^\sw.
\end{eqnarray}

An {\it ansatz} is made about the relative size of
  the long-wavelength and the
  short-wavelength components. The
  long-wavelength component of the distribution function
  is assumed to be larger than the short-wavelength
  piece by a factor of $\epsilon_s^{-1} \gg 1$; the
  long-wavelength piece of the potential is itself
  comparable to the kinetic energy of the particles and its
  short-wavelength component is also small in
  $\epsilon_s$. Summarizing,
\begin{eqnarray} \label{orderings}
\frac{v_{t\spe}^3 f_\spe^\sw}{n_{e0}}\sim \frac{Z_\spe e
\varphi^\sw}{m_\spe v_{t\spe}^2} \sim \epsilon_s,
\nonumber\\[5pt]
\frac{v_{t\spe}^3 f_\spe^\lw}{n_{e0}}\sim
\frac{Z_\spe e \varphi^\lw}{m_\spe v_{t\spe}^2}
\sim 1.
\end{eqnarray}

We need also an {\it ansatz} about the size of the space and time
derivatives of the long- and short-wavelength components of our
fields. The long-wavelength components $f_\spe^\lw$ and $\varphi^\lw$ are
characterized by large spatial scales, of the order of the macroscopic
scale $L$, and long time scales, of the order of the transport time
scale, $\tau_E:=L/(c_s\epsilon_s^2)$, i.e.
\begin{eqnarray}
\nabla_\boldr \ln f_\spe^\lw, \ \nabla_\boldr
\ln \varphi^\lw
\sim 1/L, \nonumber\\[5pt]
\partial_t \ln f_\spe^\lw ,\
\partial_t \ln \varphi^\lw \sim\epsilon_s^2 c_s/L.
\end{eqnarray}
The short-wavelength components $f_\spe^\sw$ and $\varphi^\sw$ have
perpendicular wavelengths of the order of the sound gyroradius, and
short time scales, of the order of the turbulent correlation time. The
parallel correlation length of the short-wavelength component is much
longer than its characteristic perpendicular
wavelength, and it is comparable to the size of the machine. In short,
$f_\spe^\sw$ and $\varphi^\sw$ are characterized by
\begin{eqnarray}
\bun\cdot\nabla_\boldr \ln f_\spe^\sw,\ \bun\cdot\nabla_\boldr
\ln \varphi^\sw \sim 1/L,
\nonumber\\[5pt]
\nabla_{\boldr_\perp} \ln f_\spe^\sw,\ \nabla_{\boldr_\perp}
\ln \varphi^\sw \sim 1/\rho_s,
\nonumber\\[5pt]
\partial_t \ln f_\spe^\sw, \ \partial_t 
\ln \varphi^\sw \sim
c_s/L.
\end{eqnarray}
The magnetic field only contains long-wavelength
components,
\begin{eqnarray}
\nabla_\boldr \ln |\bB| \sim 1/L.
\end{eqnarray}

The above assumptions make the elimination of the gyrophase order by
order in $\epsilon_s$ possible and the resulting equations
consistent. These assumptions are based on experimental and
theoretical evidence. In experiments it has been possible to confirm
that the characteristic correlation length of the turbulence is of the
order of and scales with the ion gyroradius \cite{mckee01}. The same
measurements showed that the size of the turbulent fluctuations scales
with the ion gyroradius. The characteristic length of the turbulent
eddies and the size of the fluctuations are related to each other by
the background gradient. An eddy of length $\ell_\bot \sim \rho_s$
mixes the plasma contained within it. In the presence of a gradient
this eddy will lead to fluctuations on top of the background density
of order $\delta n_e \sim \ell_\bot |\nabla n_e| \sim \epsilon_s n_e
\ll n_e$.

In addition to the experimental measurements, there exist strong
theoretical arguments in favor of the assumptions above. The equations
obtained using these assumptions lead to a nonlinear system of
gyrokinetic equations for the fluctuations. These equations can be
implemented in numerical simulations that encompass several ion
gyroradii, as is done in \cite{dorland00, dannert05, candy03,
  peeters09}. These simulations converge for numerical domains that
are sufficiently large to contain the largest turbulent eddies. The
model is consistent if the domain size is only several gyroradii
across, proving that for sufficiently small gyroradius the turbulence
eddies will scale with the gyroradius. The flux tube simulations
converge, and the characteristic size of the turbulent eddies is
indeed of the order of the ion gyroradius. In \cite{barnes11} the
fluctuation spectrum of the turbulence is studied by varying different
parameters. The final result is that the spectrum peaks around
wavelengths proportional to the ion gyroradius, and although the
constant of proportionality depends on the density and temperature
gradients and the magnetic field characteristics, it is of order
unity. The size of the fluctuations is also of order $\epsilon_s$.

Observe that in dimensionless variables the short-wavelength
electrostatic potential and distribution functions satisfy
\begin{eqnarray}\label{eq:ordering_sw_dimensionless}
\underline{\varphi}^\sw(\underline{\boldr}, \underline{t})\sim 1,
\nonumber\\[5pt]
\underline{f_\spe^\sw} (\underline{\boldr},\underline{\bv},
\underline{t})
\sim \epsilon_s,\nonumber\\[5pt]
{\bun}(\underline{\boldr}) \cdot \nabla_{\underline{\boldr}}\,
\underline{\varphi}^\sw (\underline{\boldr}, \underline{t}) \sim 1,
\nonumber\\[5pt]
{\bun}(\underline{\boldr}) \cdot \nabla_{\underline{\boldr}}\,
\underline{f_\spe^\sw} (\underline{\boldr},\underline{\bv},
\underline{t}) \sim \epsilon_s,
\nonumber\\[5pt]
\nabla_{\underline{\boldr}_{\perp}}\, \underline{\varphi}^\sw
 (\underline{\boldr},
\underline{t})  \sim 1/\epsilon_s,\nonumber\\[5pt]
\nabla_{\underline{\boldr}_{\perp}}\, \underline{f_\spe^\sw}
 (\underline{\boldr},\underline{\bv},
\underline{t})  \sim 1,
\nonumber\\[5pt] \partial_{\underline{t}} \underline{\varphi}^\sw (\underline{\boldr}, \underline{t}) \sim 1,
\nonumber\\[5pt] \partial_{\underline{t}} \underline{f_\spe^\sw} (\underline{\boldr},\underline{\bv},
\underline{t}) \sim \epsilon_s.
\end{eqnarray}
The normalized functions $\underline{\varphi}^\sw$ and
$\underline{f_\spe}^\sw$ are of different size due to our choice of
dimensionless variables, consistent with \cite{ParraCalvo2011} (see
\eq{norm_spindep} and \eq{norm_spdep}).

As for the long-wavelength
components,
\begin{eqnarray}\label{eq:ordering_lw_dimensionless}
  \underline{\varphi}^\lw(\underline{\boldr},
  \underline{t})\sim 1/\epsilon_s,\nonumber\\[5pt]
  \underline{f_\spe^\lw}(\underline{\boldr},\underline{\bv},
  \underline{t})\sim 1,\nonumber\\[5pt]
  \nabla_{\underline{\boldr}}\, \underline{\varphi}^\lw
  (\underline{\boldr},
  \underline{t})  \sim 1/\epsilon_s,\nonumber\\[5pt]
  \nabla_{\underline{\boldr}}\, \underline{f_\spe^\lw}
  (\underline{\boldr},
  \underline{t})  \sim 1,
  \nonumber\\[5pt] \partial_{\underline{t}} \underline{\varphi}^\lw (\underline{\boldr}, \underline{t}) \sim \epsilon_s,
  \nonumber\\[5pt] \partial_{\underline{t}} \underline{f_\spe^\lw}(\underline{\boldr},\underline{\bv},
  \underline{t}) \sim \epsilon_s^2.
\end{eqnarray}

The following convention is adopted when we expand
$\underline{\varphi}^\lw(\underline{\boldr},
\underline{t})$ in powers of $\epsilon_s$:
\begin{equation}\label{eq:potentiallw1}
\underline{\varphi}^\lw(\underline{\boldr},
\underline{t}) :=
\frac{1}{\epsilon_s}\underline{\varphi_0}(\underline{\boldr},
\underline{t}) +
\underline{\varphi_1}^\lw(\underline{\boldr},
\underline{t}) + \epsilon_s
\underline{\varphi_2}^\lw(\underline{\boldr},
\underline{t}) +
O(\epsilon_s^2).
\end{equation}
Similarly,
\begin{equation}
\underline{\varphi}^\sw(\underline{\boldr},
\underline{t}) :=
\underline{\varphi_1}^\sw(\underline{\boldr},
\underline{t}) + \epsilon_s
\underline{\varphi_2}^\sw(\underline{\boldr},
\underline{t}) +
O(\epsilon_s^2).
\end{equation}

From now on we do not underline variables but assume that we are
working with the dimensionless ones unless otherwise stated.

\subsection{Gyrokinetic expansion to second order}
\label{sec:gyrokineticstosecondorder}

The complete calculation of the gyrokinetic system of equations to
second order is given for the first time in reference
\cite{ParraCalvo2011} in the phase-space Lagrangian formalism. The
latter was applied to the problem of guiding-center motion by
Littlejohn~\cite{littlejohn83} and has been used extensively in modern
formulations of gyrokinetics~\cite{brizard07}. In reference
\cite{ParraCalvo2011} we perform a change of variables in
\eq{eq:FPnon-dim} and \eq{eq:Quasineutralitynondim} that decouples the
fast degree of freedom (the gyrophase) from the slow ones in the
absence of collisions. This decoupling is achieved by eliminating the
dependence on the gyration order by order in $\epsilon_\spe$. Let us
denote the transformation~\footnote{We warn the reader that we call
  $\cT_\spe$ to the transformation that is often called
  $\cT_\spe^{-1}$ in the literature.} from the new phase-space
coordinates $\bZ\equiv\{\bR,u,\mu,\theta\}$ to the euclidean ones
$\bX\equiv\{\boldr,\bv\}$ by $\cT_\spe$,
\begin{equation}
(\boldr,\bv) = \cT_\spe(\bR,u,\mu,\theta,t).
\end{equation}
The transformation is, in general, explicitly time-dependent and
is expressed as a power series in $\epsilon_\spe$.
Here $\bR$ is the position of the gyrocenter, and
$u$, $\mu$, and $\theta$ are deformations of the parallel velocity,
magnetic moment, and gyrophase. We recall that in
\cite{ParraCalvo2011} the gyrokinetic transformation is written as
the composition of two transformations. First, the {\it
non-perturbative
  transformation}, $(\boldr, \bv) = \cT_{NP, \spe} (\bZ_g) \equiv
\cT_{NP,\spe} (\bR_g, v_{||g}, \mu_g, \theta_g)$,
\begin{eqnarray}\label{eq:changeNonPert}
\boldr &=& \bR_g  + \epsilon_\spe \rhobf ( \bR_g, \mu_g, \theta_g),
\nonumber\\[5pt]
\bv &=& v_{||g} \bun(\bR_g) + \rhobf ( \bR_g, \mu_g, \theta_g)
\times \bB (\bR_g),
\end{eqnarray}
with the gyroradius vector defined as
\begin{equation}\label{eq:defrho}
\rhobf ( \bR_g, \mu_g, \theta_g ) = - \sqrt{\frac{2
\mu_g}{B(\bR_g)}} \left [ \sin \theta_g \eun_1 (\bR_g) - \cos
\theta_g \eun_2 (\bR_g) \right ].
\end{equation}
The unit vectors $\eun_1 (\boldr)$ and $\eun_2 (\boldr)$ are
orthogonal to each other and to $\bun = \bB/B$, and satisfy $\eun_1
\times \eun_2 = \bun$ at every location $\boldr$. Second, the
 {\it perturbative transformation}
\begin{equation}
(\bR_g, v_{||g}, \mu_g, \theta_g) = \cT_{P, \spe} (\bR, u, \mu, \theta, t),
\end{equation}
that we express as
\begin{eqnarray}\label{newvar}
\bR_g &= \bR + \sum_{i=1}^n \epsilon_\spe^{i+1} {\bR}_{i+1},
\;
\nonumber\\
v_{||g} &= u + \sum_{i=1}^n \epsilon_\spe^i {u}_i, \;
\nonumber\\
\mu_g &=
\mu + \sum_{i=1}^n \epsilon_\spe^i {\mu}_i, \;
\nonumber\\
\theta_g &= \theta
+ \sum_{i=1}^n \epsilon_\spe^i {\theta}_i.
\end{eqnarray}
The gyrokinetic transformation is
\begin{equation}\label{eq:gyrotransform_composition}
\cT_\spe = \cT_{NP,\spe} \cT_{P,\spe}.
\end{equation}

At this point we need to mention that the derivation of $\cT_\spe$ in
\cite{ParraCalvo2011} assumed that the electrostatic potential had
only a short-wavelength component, i.e. we assumed $\varphi =
\varphi^\sw$ and $\varphi^\lw = 0$. Since $\varphi^\sw$ is small in
$\epsilon_s$, this assumption lead to normalizing the electrostatic
potential with $\epsilon_s T_{e0}/e$. It is easy to relax the
assumption in \cite{ParraCalvo2011} that $\varphi = \varphi^\sw$ and
include $\varphi^\lw$. In equations (68) and (69) of
\cite{ParraCalvo2011} we display the phase-space Lagrangian after the
non-perturbative transformation. The Hamiltonian is given by
\begin{equation}
H = H^{(0)} + \epsilon_\sigma H^{(1)},
\end{equation}
with
\begin{equation}
H^{(0)} = \frac{1}{2}v_{||g}^2 + \mu_g B (\bR_g)
\end{equation}
and 
\begin{equation}
  H^{(1)} = Z_\spe \lambda_\spe \varphi^\sw
 ( \bR_g + \epsilon_\spe \rhobf (\bR_g, \mu_g, \theta_g), t).
\end{equation} 
Using this expression, it is possible to obtain the perturbative
change of variables $\cT_{P,\spe}$ by expanding in $\epsilon_\spe$. To
do so, $H^{(1)}$ must be of order unity, and if instead of
$\varphi = \varphi^\sw$ we have a long wavelength piece $\varphi^\lw
\sim 1/\epsilon_s \gg 1$, it would seem that the condition
$H^{(1)}\sim 1$ is not satisfied. Fortunately, it is possible to
redefine $H^{(0)}$ and $H^{(1)}$ so that the expansion can be
performed. The new Hamiltonian is given by
\begin{equation}
H^{(0)} = \frac{1}{2}v_{||g}^2 + \mu_g B (\bR_g)
 + Z_\spe \lambda_\spe \epsilon_\spe \langle \phi_\sigma \rangle (\bR_g, \mu_g, t)
\end{equation}
and 
\begin{equation}
H^{(1)} = Z_\spe \lambda_\spe \tilde\phi_\spe (\bR_g, \mu_g, \theta_g, t),
\end{equation} 
where the function $\phi_\spe$ is defined
as
\begin{equation}
\phi_\spe(\bR_g,\mu_g,\theta_g,t) :=
\varphi(\bR_g+\epsilon_\spe\rhobf(\bR_g,\mu_g,\theta_g),t).
\end{equation}
From it we can calculate
\begin{equation}
\tilde\phi_\spe(\bR_g,\mu_g,\theta_g,t) :=
\phi_\spe(\bR_g,\mu_g,\theta_g,t)
 - \langle\phi_\spe\rangle(\bR_g,\mu_g,t)
\end{equation}
and
\begin{equation}
\langle\phi_\spe\rangle(\bR_g,\mu_g,t) :=
\frac{1}{2\pi}\int_0^{2\pi}\phi_\spe(\bR_g,\mu_g,\theta_g,t)
\dd\theta_g.
\end{equation}
Here $\langle\cdot\rangle$ stands for the average over the
gyrophase. We now prove that $H^{(1)}$ is indeed of order unity. From
the ordering and scale separation assumptions on $\varphi$,
equations~\eq{eq:ordering_sw_dimensionless} and
\eq{eq:ordering_lw_dimensionless}, we obtain that the turbulent
component of $\phi_\spe$ is $O(1)$, i.e.
\begin{eqnarray}
\phi^\sw_\spe = \phi^\sw_{\spe 1} + O(\epsilon_s),\nonumber\\
\tilde\phi^\sw_\spe = \tilde\phi^\sw_{\spe 1} + O(\epsilon_s).
\end{eqnarray}
For the long wavelength
piece $\phi^\lw_\spe$ we use that it is possible to Taylor expand
around $\boldr = \bR$ to find
\begin{eqnarray}\label{eq:potentiallw2}
\fl\langle\phi_\spe^\lw\rangle(\bR_g,\mu_g,t) =
\frac{1}{\epsilon_s}\varphi_0(\bR_g,t)
+\varphi_1^\lw(\bR_g,t)
\nonumber\\[5pt]
\fl
\hspace{0.5cm}
+\epsilon_s \left( \frac{\mu_g}{2\lambda_\spe^2 B(\bR_g)}
 (\matI-\bun(\bR_g)\bun(\bR_g)):\nabla_{\bR_g}\nabla_{\bR_g}
\varphi_0(\bR_g,t)
+\varphi_2^\lw(\bR_g,t) \right)
\nonumber\\[5pt]
\fl
\hspace{0.5cm}
 +O(\epsilon_s^2)
\end{eqnarray}
and
\begin{equation}\label{eq:potentiallw3}
\fl\tilde\phi_\spe^\lw(\bR_g,\mu_g,\theta_g,t) =
\frac{1}{\lambda_\spe} \rhobf(\bR_g,\mu_g,\theta_g)
\cdot\nabla_{\bR_g}\varphi_0(\bR_g,t) + O(\epsilon_s),
\end{equation}
giving $\tilde\phi^\lw_\spe = O(1)$ as expected. We have expanded up
to first order in $\epsilon_s$ in \eq{eq:potentiallw2} because it will
be needed later in this paper. Our double-dot convention for arbitrary
matrix $\matM$ is $\mathbf u\mathbf v :\matM = \mathbf v \cdot\matM
\cdot \mathbf u$. The authors of references \cite{Dimits92} and
\cite{Dimits12} already pointed out the usefulness of the separation
of the electrostatic potential into a large gyrophase-independent
piece and a small gyrophase-dependent one, and exploited it in their
derivations.

  We want to write the Fokker-Planck equation in gyrokinetic
  coordinates. Denote by $\cT^*_\spe$ the pull-back transformation
  induced by $\cT_\spe$. Acting on a function $g(\bX,
    t)$, $\cT_\spe^* g (\bZ,t)$ is simply the function $g$ written in
  the coordinates $\bZ$, i.e.
\begin{equation}
\cT_\spe^* g (\bZ,t) = g(\cT_\spe(\bZ,t), t).
\end{equation}
Now, defining $ F_\spe:=\cT_\spe^* f_\spe $, we transform
\eq{eq:FPnon-dim} and get:
\begin{eqnarray}\label{eq:FokkerPlancknondimgyro}
\fl\partial_{t} F_\spe + \tau_\spe
\left\{F_\spe,\overline{H}_\spe \right\}_\bZ =
\tau_\spe \sum_{\spe'}\cT^{*}_{\spe} C_{\spe
\spe'} [\cT^{-1*}_{\spe} F_\spe,\cT^{-1*}_{\spe'}F_{\spe'}]
(\bZ, t),
\end{eqnarray}
where $\cT^{-1*}_{\spe}$ is the pull-back transformation that
corresponds to $\cT_\spe^{-1}$, i.e. $\cT_\spe^{-1*} F_\spe (\bX,t) =
F_\spe(\cT_\spe^{-1}(\bX,t),t)$, and the Poisson bracket in the new
coordinates is expressed as
\begin{eqnarray} \label{eq:poissonbracket}
\fl
\{G_1,G_2\}_\bZ &=
\frac{1}{\epsilon_\spe}\left(\partial_\mu G_1
\partial_\theta G_2
 - \partial_\theta
G_1\partial_\mu G_2
\right) 
\nonumber
\\[5pt]
\fl&
+
\frac{1}{B_{||,\spe}^*}{\bf B}_\spe^* \cdot
\left(\nabla^*_{\bR}G_1\partial_u G_2
-
\partial_u G_1
\nabla_{\bR}^*G_2\right) \nonumber
\\[5pt]
\fl&
+ \frac{\epsilon_\spe }{B_{||}^*} \nabla_{\bR}^*G_1 \cdot (
\bun \times \nabla_{\bR}^*G_2 ),
\end{eqnarray}
with
\begin{equation} \label{Bstar}
\bB_\spe^* (\bR, u, \mu) := \bB(\bR) + \epsilon_\spe u \nabla_\bR \times
\bun (\bR)
 - \epsilon^2_\spe \mu \nabla_\bR \times \bK (\bR),
\end{equation}
\begin{eqnarray} \label{Bstar_par}
\fl B^*_{||,\spe} (\bR, u, \mu) := \bB_\spe^* (\bR, u, \mu) \cdot \bun (\bR)
\nonumber \\[5pt] = B(\bR) + \epsilon_\spe u \bun(\bR) \cdot \nabla_\bR
\times \bun (\bR)
\nonumber \\[5pt]
 - \epsilon^2_\spe \mu \bun(\bR) \cdot \nabla_\bR
\times \bK (\bR),
\end{eqnarray}
\begin{equation}
\nabla_{\bR}^* := \nabla_{\bR} - {\bf
K}(\bR)\partial_\theta,
\end{equation}
and
\begin{equation}\label{eq:defvectorK}
\bK (\bR) = \frac{1}{2} \bun(\bR) \bun(\bR) \cdot \nabla_\bR
\times \bun(\bR) - \nabla_\bR \eun_2 (\bR) \cdot \eun_1 (\bR).
\end{equation}
The above expression \eq{eq:poissonbracket} for the Poisson bracket is
customary in the literature~\cite{brizard07}. The gyrokinetic change
of coordinates is not unique, in the sense that there are infinitely
many transformations such that the gyromotion is decoupled from the
rest of degrees of freedom and such that the coordinate $\mu$ is an
adiabatic invariant. To make comparisons with standard references in
the literature easy, we have made use of this flexibility by choosing
our change of variables so that the Poisson bracket takes the form
\eq{eq:poissonbracket}.

The main achievement of \cite{ParraCalvo2011} was the computation
of the gyrokinetic Hamiltonian $\overline{H}_\spe
=\sum_{n=0}^\infty \epsilon_\spe^n \overline{H}_\spe^{(n)}$ to order
$\epsilon_\spe^2$. The result is:
\begin{eqnarray}\label{eq:Hgyro0}
\fl \overline{H}^{(0)}_\spe = \frac{1}{2}u^2+\mu
B,
\end{eqnarray}
\begin{eqnarray}\label{eq:Hgyro1}
\fl \overline{H}^{(1)}_\spe = Z_\spe\lambda_\spe\langle\phi_\spe\rangle,
\end{eqnarray}
\begin{eqnarray}\label{eq:Hgyro2}
\fl \overline{H}^{(2)}_\spe = 
Z_\spe^2 \lambda_\spe^2 \Psi_{\phi,\spe} +Z_\spe\lambda_\spe
\Psi_{\phi B,\spe} + \Psi_{B,\spe},
\end{eqnarray}
with
\begin{eqnarray}
\fl \Psi_{\phi,\spe} &= \frac{1}{2 B^2} \left
\langle \nabla_{(\bR_\bot/\epsilon_\spe)} \Phiwig_\spe \cdot \left
( \bun \times \nabla_{(\bR_\bot/\epsilon_\spe)} \phiwig_\spe
\right ) \right\rangle
\nonumber\\ \fl &
 - \frac{1}{2 B} \partial_\mu \langle
\phiwig^2_\spe \rangle, \label{Psi2_phi}
\end{eqnarray}
\begin{eqnarray}
\fl \Psi_{\phi B,\spe}& =  - \frac{u}{B} \left
\langle \left ( \nabla_{(\bR_\bot/\epsilon_\spe)} \phiwig_\spe
\times \bun \right ) \cdot \nabla_\bR \bun \cdot \rhobf \right
\rangle
\nonumber\\ \fl &
 - \frac{\mu}{2 B^2} \nabla_\bR B \cdot
\nabla_{(\bR_\bot/\epsilon_\spe)} \langle\phi_\spe\rangle
- \frac{1}{B} \nabla_\bR B
\cdot \langle \phiwig_\spe\, \rhobf \rangle
\nonumber\\ \fl & - \frac{1}{4 B} \left \langle
\nabla_{(\bR_\bot/\epsilon_\spe)} \phiwig_\spe \cdot \left [
\rhobf \rhobf - ( \rhobf \times \bun ) (\rhobf \times \bun) \right
] \cdot \nabla_\bR B \right \rangle  \nonumber\\ \fl &
- \frac{u^2}{B} \bun \cdot \nabla_\bR \bun \cdot
\left \langle \partial_\mu \phiwig_\spe\, \rhobf \right \rangle -
\frac{u^2}{2 \mu B} \bun \cdot \nabla_\bR \bun \cdot \langle
\phiwig_\spe\, \rhobf \rangle \nonumber\\\fl &
+
\frac{u}{4} \nabla_\bR \bun : \left \langle \partial_\mu
\phiwig_\spe\, \left [ \rhobf ( \rhobf \times \bun ) + (\rhobf
\times \bun) \rhobf \right ] \right \rangle \nonumber\\\fl & +
\frac{u}{4 \mu} \nabla_\bR \bun : \left \langle \phiwig_\spe\,
\left [ \rhobf ( \rhobf \times \bun ) + (\rhobf \times \bun)
\rhobf \right ] \right \rangle \label{Psi2_phiB}
\end{eqnarray}
and
\begin{eqnarray}
\fl \Psi_{B,\spe} &=  - \frac{3u^2 \mu}{2B^2} \bun \cdot \nabla_\bR
\bun \cdot \nabla_\bR B
\nonumber \\
\fl & 
 + \frac{\mu^2}{4B} (\matI - \bun \bun) :
\nabla_\bR \nabla_\bR \bB \cdot \bun
\nonumber \\
\fl &
 - \frac{3\mu^2}{4B^2}
|\nabla_{\bR_\bot} B|^2 
+ \frac{u^2 \mu}{2B}
\nabla_\bR \bun : \nabla_\bR \bun
\nonumber \\
\fl &  + \left(\frac{\mu^2}{8} -
\frac{u^2 \mu}{4B}\right) \nabla_{\bR_\perp} \bun : (\nabla_{\bR_\perp}
\bun)^\mathrm{T}
\nonumber \\
\fl & 
 - \left(\frac{3 u^2 \mu}{8B} +
\frac{\mu^2}{16}\right) (\nabla_\bR \cdot \bun)^2 \nonumber \\ \fl
& + \left(\frac{3 u^2 \mu}{2B}-\frac{u^4}{2B^2}\right) |\bun \cdot
\nabla_\bR \bun|^2 
\nonumber \\
\fl & 
+ \left(\frac{ u^2 \mu}{8B} -
\frac{\mu^2}{16}\right) (\bun \cdot \nabla_\bR \times \bun)^2.
\label{Psi2_B}
\end{eqnarray}
Here $\matrixtop{\mathbf{M}}^\mathrm{T}$ is the transpose of an
arbitrary matrix matrix $\matrixtop{\mathbf{M}}$, the magnetic field
quantities $\bB (\bR)$, $\bun (\bR)$ and $B(\bR)$ are evaluated at
$\bR$ instead of $\boldr$, the functions $\phi_{\spe} (\bR, \mu,
\theta, t)$, $\langle \phi_\spe \rangle (\bR, \mu, t)$ and
$\tilde\phi_\spe (\bR, \mu, \theta, t)$ are evaluated at $\bR$, $\mu$
and $\theta$ instead of $\bR_g$, $\mu_g$ and $\theta_g$, and
\begin{equation}
\Phiwig_\spe(\bR,\mu,\theta,t) :=
\int^\theta\phiwig_\spe(\bR,\mu,\theta',t)\dd\theta'\ ,
\end{equation}
where the lower limit of the integral is chosen such that
$\langle\Phiwig_\spe\rangle = 0$. The second-order Hamiltonian is
sufficient to obtain the long-wavelength component of the distribution
function to order $\epsilon_\spe^2$. To check this, the reader
can follow the calculation in this article assuming that
$\overline{H}^{(n)}_\spe$ for $n>2$ are known, and finding that these
higher-order terms do not enter the final equations.

In gyrokinetic variables the quasineutrality equation reads
\begin{eqnarray}\label{eq:gyroQuasineutrality}
\fl\sum_\spe Z_\spe  \int |\det\left(J_{\spe}\right)|
F_\spe
\delta\Big(\pi^{\boldr}\Big(\cT_{\spe}(\bZ,
t)\Big)-\boldr\Big)\dd ^6Z = 0,
\end{eqnarray}
with $\pi^\boldr(\boldr,\bv):=\boldr$, and the Jacobian of the
transformation to order $\epsilon_\spe ^2$ is
\begin{eqnarray}\label{eq:Jacobian}
|\det(J_{\spe} )| \equiv B_{||,\spe}^*.
\end{eqnarray}

The expressions for the corrections $\bR_2$, $u_1$, $\mu_1$,
and $\theta_1$ found in \cite{ParraCalvo2011} are
\begin{eqnarray}
\fl\bR_{\spe,2} &=& - \frac{2u}{B} \bun \bun \cdot \nabla_\bR \bun \cdot
(\rhobf \times \bun)- \frac{u}{B} \bun \times \nabla_\bR \bun \cdot \rhobf
\nonumber\\
\fl&&
 - \frac{1}{8} \bun \left [ \rhobf \rhobf -
(\rhobf \times \bun) (\rhobf \times \bun) \right ]: \nabla_\bR
\bun 
\nonumber\\
\fl&&- \frac{1}{2B} \rhobf \rhobf \cdot \nabla_\bR B
- \frac{Z_\spe \lambda_\spe}{B^2}
 \bun \times \nabla_{(\bR_\bot/\epsilon_\spe)} \Phiwig_\spe,
\label{R2}
\\[5pt]
\fl u_{\spe,1} &=& u \bun \cdot \nabla_\bR \bun \cdot \rhobf
\nonumber\\
\fl&&
 - \frac{B}{4}
\left [ \rhobf (\rhobf \times \bun) + (\rhobf \times \bun) \rhobf
\right ] : \nabla_\bR \bun, \label{u1}
\\[5pt]
\fl\mu_{\spe,1} &=& 
- \frac{Z_\spe\lambda_\spe\phiwig_\spe}{B}
- \frac{u^2}{B} \bun \cdot \nabla_\bR \bun
\cdot \rhobf
\nonumber\\
\fl&&
 + \frac{u}{4} \left [ \rhobf (\rhobf \times \bun) +
(\rhobf \times \bun) \rhobf \right ]: \nabla_\bR \bun,
\label{mu1}
\\[5pt]
\fl\theta_{\spe,1} &=& \frac{Z_\spe \lambda_\spe}{B}
\partial_\mu \Phiwig_\spe
+\frac{u^2}{2\mu B} \bun \cdot \nabla_\bR \bun \cdot
(\rhobf \times \bun)
\nonumber\\
\fl&&
 + \frac{u}{8\mu} \left [ \rhobf \rhobf -
(\rhobf \times \bun) (\rhobf \times \bun) \right ]: \nabla_\bR
\bun
\nonumber\\
\fl&&+ \frac{1}{B} (\rhobf \times \bun) \cdot \nabla_\bR B
 .
\label{theta1}
\end{eqnarray}
These corrections are needed to find the gyrokinetic quasineutrality
equation to the order or interest. Although it might
seem that we also need the next order corrections $\bR_{\spe,3}$,
$u_{\spe,2}$, $\mu_{\spe,2}$, and $\theta_{\spe,2}$, it will be shown that
they do not contribute in the long-wavelength limit.

In the following sections we take the
long-wavelength limit of \eq{eq:FokkerPlancknondimgyro} and
\eq{eq:gyroQuasineutrality} up to second-order in the expansion
parameter.

\section{Fokker-Planck equation at long wavelengths}
\label{sec:FokkerPlancklong}

The objective in this section is to take the long-wavelength limit of
the gyrokinetic Fokker-Planck equation \eq{eq:FokkerPlancknondimgyro}
up to second order in tokamak geometry. As a preliminary step we must
write \eq{eq:FokkerPlancknondimgyro} order by order; for this we will
expand $F_\spe$ as
\begin{eqnarray}\label{eq:orderingF}
F_\spe&=\sum_{n=0}^\infty
\epsilon_\spe^n F_{\spe n} = 
\sum_{n=0}^\infty
\epsilon_\spe^n F^\lw_{\spe n}
+
\sum_{n=1}^\infty
\epsilon_\spe^n F^\sw_{\spe n}.
\end{eqnarray}
From the scale separation and
ordering assumptions enumerated in Section
\ref{sec:secondordergyrokinetics} it follows that
\begin{eqnarray}
F_{\spe n}\sim 1,\  n\ge 0, \nonumber\\[5pt]
\bun (\bR) \cdot \nabla_{\bR} F_{\spe n}\sim 1,\  n\ge 0.
\end{eqnarray}
Also, the long-wavelength component
of every $F_{\spe n}$ must have perpendicular derivatives of order
unity in normalized variables, i.e.
\begin{eqnarray}
\nabla_{\bR_\perp}F_{\spe n}^\lw\sim 1,\  n\ge 0.
\end{eqnarray}
Finally, the zeroth-order distribution function must have an identically
vanishing short-wavelength component, and the perpendicular gradient of
the rest of the short-wavelength components is of order $\epsilon_\spe^{-1}$,
\begin{eqnarray}
F_{\spe 0}^\sw\equiv 0, \nonumber\\[5pt]
 \nabla_{\bR_\perp}F_{\spe n}^\sw\sim\epsilon_\spe^{-1},\  n\ge 1.
\end{eqnarray}
Then, one must use expression \eq{eq:poissonbracket} for the Poisson
bracket in gyrokinetic coordinates and the form of $\overline{H}_\spe$
given in equations~\eq{eq:Hgyro0}, \eq{eq:Hgyro1}, and
\eq{eq:Hgyro2}. With the help of \ref{sec:eqsofmotion}, writing
\eq{eq:FokkerPlancknondimgyro} order by order is relatively
straightforward. In addition to writing the equations order by order,
we manipulate them to make them as close as possible to the results
obtained in neoclassical theory \cite{hinton76, helander02bk}. This
form of the equations will be useful when we calculate the transport
equations for particles and energy in subsection
\ref{sec:transportEqs}.

Recall that along this paper we assume
$\gamma_{\spe\spe'}\sim\lambda_\spe\sim\tau_\spe\sim 1$ for all
$\spe,\spe'$, so that the only formal expansion parameter is
$\epsilon_s$.

\subsection{Long-wavelength Fokker-Planck equation to 
$O(\epsilon_\spe^{-1})$}
\label{sec:FokkerPlancklongminus1}

The coefficient of $\epsilon_\spe^{-1}$ in
\eq{eq:FokkerPlancknondimgyro} simply gives
\begin{eqnarray}
-\tau_\spe B\partial_\theta F_{\spe 0} = 0,
\end{eqnarray}
implying that $F_{\spe 0}$ is independent of $\theta$.

\subsection{Long-wavelength Fokker-Planck equation to 
$O(\epsilon_\spe^0)$}
\label{sec:FokkerPlancklong0}

Equation~\eq{eq:FokkerPlancknondimgyro} to order $\epsilon_\spe^0$
involves the collision operator, which is written in coordinates
$\bX\equiv (\boldr,\bv)$. Therefore, either we transform the collision
operator to gyrokinetic coordinates $\bZ\equiv (\bR,u,\mu,\theta)$ or
transform the gyrokinetic distribution function to coordinates
$\bX$. We choose the second option. To write order by order the
collision operator we need to obtain certain coefficients of the
Taylor expansion of the gyrokinetic change of coordinates $\cT_\spe$
and its inverse, $\cT_\spe^{-1}$,
\begin{eqnarray}
\fl\bX = \cT_{\spe}(\bZ,t) 
= \cT_{\spe,0}(\bZ,t)  + \epsilon_\spe \cT_{\spe,1}(\bZ,t)  +
O(\epsilon_\spe^2),\\[5pt]
\fl\bZ = \cT_{\spe}^{-1}(\bX,t) = \cT_{\spe,0}^{-1}(\bX,t)
+ \epsilon_\spe \cT_{\spe,1}^{-1}(\bX,t)
\nonumber\\ 
 +
\epsilon_\spe^2 \cT_{\spe,2}^{-1}(\bX,t) 
+O(\epsilon_\spe^3).
\end{eqnarray}
In the present subsection we need $\cT_{\spe,0}$,
the transformation $\cT_\spe$,
equation~\eq{eq:gyrotransform_composition}, for $\epsilon_\spe = 0$:
\begin{eqnarray}\label{eq:transformation_oderzero}
\cT_{\spe,0}(\bR,u,\mu,\theta)=(\bR,u\bun(\bR)+\rhobf(\bR,\mu,\theta)\times
\bB(\bR)).
\end{eqnarray}
The remaining terms in the Taylor expansions will be employed in
subsequent subsections and are computed in the appendices.

We write the collision operator in the zeroth-order long-wavelength
Fokker-Planck equation by employing the pull-back of
\eq{eq:transformation_oderzero} and its inverse, so the equation
reads:
\begin{eqnarray}\label{eq:Vlasovorder0}
\fl\hspace{1.5cm} u\bun\cdot\nabla_\bR F_{\spe 0}
-\bun\cdot\left(\mu\nabla_\bR B + Z_\spe
\nabla_\bR\varphi_0\right)
 \partial_u F_{\spe 0} - B\partial_\theta F_{\spe 1}^\lw
\nonumber\\[5pt]
\fl\hspace{3cm}=\sum_{\spe'}\cT^{*}_{\spe,0}
C_{\spe \spe'} [\cT^{-1*}_{\spe,0} F_{\spe
0},\cT^{-1*}_{\spe',0}F_{\spe' 0}] (\bR,u,\mu,\theta).
\end{eqnarray}
Using that $F_{\spe 0}$ is gyrophase independent and the isotropy
property of the collision operator (by which it gives a
gyrophase-independent function when acting on a gyrophase-independent
function) we immediately deduce that
\begin{equation}
\partial_\theta F_{\spe 1}^\lw = 0,
\end{equation}
i.e. $F_{\spe 1}^\lw$ is gyrophase-independent. Actually, it is trivial to
prove from the zeroth-order short-wavelength component of equation
\eq{eq:FokkerPlancknondimgyro} that also $\partial_\theta F_{\spe
  1}^\sw = 0$, so
\begin{equation}
\partial_\theta F_{\spe 1} = 0.
\end{equation}

We proceed to prove that the solution to \eq{eq:Vlasovorder0} is a
stationary Maxwellian. Multiplying \eq{eq:Vlasovorder0} by $-B\ln
F_{\spe 0}$ and integrating over $u,\mu$, and $\theta$:
\begin{eqnarray}\label{eq:FPorderzeroaux}
\fl\hspace{1cm}
-\nabla_\bR\cdot\int \bB \, u( F_{\spe 0}\ln F_{\spe 0}- F_{\spe 0})
\dd u\dd\mu\dd\theta\nonumber\\[5pt]
\fl\hspace{2cm} = -\int B\ln F_{\spe
0}\sum_{\spe'}\cT^{*}_{0,\spe} C_{\spe \spe'} [\cT^{-1*}_{\spe,0}
F_{\spe 0},\cT^{-1*}_{\spe',0}F_{\spe' 0}] \dd u\dd\mu\dd\theta.
\end{eqnarray}
Here it is convenient to define the flux-surface average of a function
$G(\psi,\Theta,\zeta)$, given by~\cite{dhaeseleer}
\begin{equation}
\fl\hspace{0.5cm}\langle G \rangle_\psi :=
\frac{\int_0^{2\pi}\int_0^{2\pi}\sqrt{g}\,
 G(\psi,\Theta,\zeta) \dd\Theta\dd\zeta}
{\int_0^{2\pi}\int_0^{2\pi}
\sqrt{g}\,
 \dd\Theta\dd\zeta}\, ,
\end{equation}
where
\begin{equation}\label{eq:sqrtg}
\sqrt{g}:=\frac{1}{\nabla_\bR\psi\cdot\left(
\nabla_\bR\Theta\times\nabla_\bR\zeta
\right)}
\end{equation}
is the square root of the determinant of the metric tensor in
coordinates $\{\psi,\Theta,\zeta\}$. It will also be useful to define
the volume enclosed by the flux surface labeled by $\psi$,
\begin{equation}\label{eq:defvolume}
V(\psi) = 
\int_0^\psi\dd\psi\int_0^{2\pi}\dd\Theta\int_0^{2\pi}\dd\zeta
\, \sqrt{g}\, .
\end{equation}
The flux-surface average of \eq{eq:FPorderzeroaux} yields
\begin{eqnarray}
\fl\left\langle
-
\sum_{\spe'}\int B \ln F_{\spe 0}\cT^{*}_{\spe,0} C_{\spe \spe'}
[\cT^{-1*}_{\spe,0} F_{\spe 0},\cT^{-1*}_{\spe',0}F_{\spe' 0}]
\dd u\dd\mu\dd\theta
\right\rangle_\psi = 0,
\end{eqnarray}
and after multiplying by $\tau_\spe$ and adding over $\sigma$:
\begin{eqnarray}\label{eq:entropyproduction}
\fl\left\langle
-\sum_{\sigma,\spe'}\tau_\spe
\int B \ln F_{\spe 0}\cT^{*}_{\spe,0} C_{\spe \spe'}
[\cT^{-1*}_{\spe,0} F_{\spe 0},\cT^{-1*}_{\spe',0}F_{\spe' 0}]
\dd u\dd\mu\dd\theta
\right\rangle_\psi = 0.
\end{eqnarray}
Observing that the Jacobian of $\cT_{\spe,0}$ at the point
$(\bR,u,\mu,\theta)$ is exactly $B(\bR)$, and using the formula for
the change of variables in an integral, we obtain:
\begin{eqnarray}\label{eq:entropyproduction2}
\fl\left\langle -\sum_{\sigma,\spe'}\tau_\spe
\int \ln
\left(\cT^{-1*}_{\spe,0}F_{\spe 0}\right) C_{\spe \spe'}
     [\cT^{-1*}_{\spe,0} F_{\spe 0},\cT^{-1*}_{\spe',0}F_{\spe' 0}] \dd^3
     v \right\rangle_\psi = 0.
\end{eqnarray}
This equation can be written as 
\begin{eqnarray}\label{eq:entropyproduction3}
\fl\Bigg\langle \sum_{\sigma,\spe'}\frac{\gamma_{\spe\spe'}}{2}
\int
\cT^{-1*}_{\spe,0}F_{\spe 0}\cT^{-1*}_{\spe',0}F_{\spe' 0}
\Big(\tau_\spe\nabla_\bv\ln\cT^{-1*}_{\spe,0}F_{\spe 0}
\nonumber\\[5pt]
\fl
-
\tau_{\spe'}\nabla_{\bv'}\ln\cT^{-1*}_{\spe',0}F_{\spe' 0}
\Big)\cdot\matW\cdot
\Big(\tau_\spe\nabla_\bv\ln\cT^{-1*}_{\spe,0}F_{\spe 0}
\nonumber\\[5pt]
\fl
-
\tau_{\spe'}\nabla_{\bv'}\ln\cT^{-1*}_{\spe',0}F_{\spe' 0}
\Big)
 \dd^3 v
\dd^3 v'
\Bigg\rangle_\psi = 0,
\end{eqnarray}
which is the entropy production in a flux surface. The only solution
to this equation is
\begin{eqnarray}
\fl \cT^{-1*}_{\spe,0} F_{\spe 0}({\boldr},{\bv},t)
=\frac{{n_{\spe}}({\boldr},t)}
{(2\pi {T_{\spe}}({\boldr},t))^{3/2}}
\exp\left(-\frac{(\bv -\tau_\spe^{-1}\bV(\boldr,t))^2}{2{T_{\spe}}
({\boldr},t)}\right),
\end{eqnarray}
where the temperature has to be the same for all the species (with the
exception of electrons if a subsidiary expansion in the mass ratio is
performed, or equivalently, if $\tau_e\sim\lambda_e\gg 1$ is
used). That is, in the previous equation, $T_\spe = T_{\spe'}$ for
every pair $\spe, \spe'$. Then,
\begin{eqnarray}\label{eq:F0}
&& \fl F_{\spe 0}(\bR,u,\mu,t)=\nonumber\\[5pt]
&& \fl \hspace{1cm}
\frac{n_{\spe}(\bR,t)}{(2\pi T_{\spe}(\bR,t))^{3/2}}
\exp\left(-\frac{\mu B(\bR) 
+ (u-\tau_\spe^{-1}V_{||}(\bR,t))^2/2}{T_{\spe}(\bR,t)}\right).
\end{eqnarray}
In the last expression we have made explicit the fact that the
component of $\bV(\bR,t)$ perpendicular to the magnetic field has to
be zero bacause otherwise $F_{\spe 0}$ would depend on the gyrophase;
that is, $\bV(\bR,t) = V_{||}(\bR,t)\bun(\bR,t)$. Now, take the
gyroaverage of \eq{eq:Vlasovorder0} and use
\eq{eq:Maxwelliansannihilatecoll_nondim} along with \eq{eq:F0} to
obtain
\begin{eqnarray}\label{eq:Vlasovorder0gyroav}
\fl
u\bun\cdot\Bigg[ \frac{1}{n_{\spe}}\nabla_\bR n_{\spe}
+
\left(\frac{\mu B + (u-\tau_\spe^{-1}V_{||})^2/2}{T_{\spe}} -\frac{3}{2}\right)
\frac{1}{T_{\spe}}
\nabla_\bR T_{\spe} 
\nonumber\\[5pt]
\fl\hspace{1cm}
+
\frac{1}{T_\spe}
\left(
\tau_\spe^{-1}(u-\tau_\spe^{-1}V_{||})\nabla_\bR V_{||}
+
Z_\spe\nabla_\bR\varphi_0
\right)\Bigg]
\nonumber\\[5pt]
\fl\hspace{1cm}
-\frac{V_{||}}{\tau_\spe T_\spe}
\bun\cdot
(\mu\nabla_\bR B
+ Z_\spe\nabla_\bR\varphi_0
)
 =
0.
\end{eqnarray}
Since this equation has to be satisfied for every $u$ and $\mu$, $V_{||}$ must vanish identically and $T_{\spe}$ must be a flux function. Then, from
\eq{eq:Vlasovorder0gyroav}, we infer that the combination
\begin{equation}
\eta_\spe = n_{\spe}
\exp\left(\frac{Z_\spe \varphi_0}{T_{\spe}}\right)
\end{equation}
is a function of $\psi$ and $t$ only, $\eta_\spe(\psi,t)$. The zeroth-order
long-wavelength quasineutrality equation (see \eq{eq:gyroPoissonlw2order0}
later on in this paper) gives
\begin{equation}
\sum_\sigma Z_\spe n_{\spe} = 0,
\end{equation}
or equivalently,
\begin{equation}
\sum_\sigma Z_\spe
\eta_{\spe}\exp\left(-\frac{Z_\spe
\varphi_0}{T_{\spe}}\right) = 0.
\end{equation}
Taking the parallel gradient of this equation, one shows that
$\varphi_0$ and $n_{\spe}$ are flux functions.

\subsection{Long-wavelength Fokker-Planck equation to $O(\epsilon_\spe)$}
\label{sec:FokkerPlancklong1}

The equation to order $\epsilon_\spe$ is fairly more complicated.
Apart from the material in \ref{sec:eqsofmotion}, we need the
long-wavelength limit of the pull-back of $F_{\spe 0}$ by $\cT_\spe^{-1}$
to order $\epsilon_\spe$. This is computed in \ref{sec:pullback} (see
\eq{eq:pullback_order1_Maxwellian}).  The result is
\begin{eqnarray}\label{eq:Vlasovorder1}
\fl
\left(
u\bun\cdot\nabla_\bR
- \mu\bun\cdot\nabla_\bR B\partial_u
\right) F_{\spe 1}^\lw
  - B\partial_\theta F_{\spe 2}^\lw \nonumber\\[5pt]
\fl\hspace{1cm}+
 \left(
\bv_\kappa + \bv_{\nabla B} + \bv_{E,\spe}^{(0)}
\right)\cdot\nabla_\bR F_{\spe 0}\nonumber\\[5pt]
\fl\hspace{1cm} + \left[
u\, \kappabf\cdot\left(\bv_{\nabla B} + \bv_{E,\spe}^{(0)}\right)
-Z_\spe\lambda_\spe\bun\cdot\nabla_\bR\varphi_1^{\lw}
\right]
\partial_u F_{\spe 0}\nonumber\\[5pt]
\fl\hspace{1cm}
= \sum_{\sigma'}\cT_{\spe,0}^*
C_{\spe\spe'}\Bigg[ \frac{1}{T_{\spe}} \left( \bv\cdot{\mathbf
V}^p_\spe + \left( \frac{v^2}{2T_{\spe}}-\frac{5}{2}
\right)\bv\cdot{\mathbf V}^T_{\spe}
\right)\cT^{-1*}_{\spe,0}F_{\spe 0}
\nonumber\\[5pt]
\fl\hspace{1cm}
+ \cT^{-1*}_{\spe,0}F_{\spe
1}^\lw ,\cT^{-1*}_{\spe',0}F_{\spe' 0} \Bigg]
 + \sum_{\sigma'}
\frac{\lambda_\spe}{\lambda_{\spe'}}\cT_{\spe,0}^*
C_{\spe\spe'} \Bigg[ \cT^{-1*}_{\spe,0}F_{\spe 0}, \frac{1}{
T_{\spe'}} \Bigg( \bv\cdot{\mathbf V}^p_{\spe'}
\nonumber\\[5pt]
\fl\hspace{1cm}
 + \left(
\frac{v^2}{2T_{\spe'}}-\frac{5}{2} \right)\bv\cdot{\mathbf
V}^T_{\spe'}
\Bigg)\cT^{-1*}_{\spe',0}F_{\spe' 0} +\,
\cT^{-1*}_{\spe',0}F_{\spe' 1}^\lw
\Bigg],
\end{eqnarray}
where
\begin{eqnarray}
 \bv_\kappa:=\frac{u^2}{B}\bun\times\kappabf,
\\[5pt]
 \bv_{\nabla B}:= \frac{\mu}{B}\bun\times\nabla_\bR B,
\\[5pt] 
 \bv_{E,\spe}^{(0)}:=
\frac{Z_\spe}{B}\bun\times\nabla_{\bR}\varphi_0,
\end{eqnarray}
and
\begin{equation}
\kappabf:= \bun\cdot\nabla_\bR\bun
\end{equation}
is the magnetic field curvature. The velocities ${\mathbf V}^p_{\spe}$ and
${\mathbf V}^T_{\spe}$ are defined by
\begin{equation}\label{eq:velocitiesPandT}
{\mathbf V}^p_\spe := \frac{1}{n_\spe B}\bun\times\nabla p_\spe, \quad
{\mathbf V}^T_\spe := \frac{1}{B}\bun\times\nabla T_\spe.
\end{equation}
Here, $p_\spe := n_\spe T_\spe$ is the pressure of species $\spe$. On
the right-hand side of \eq{eq:Vlasovorder1} we have employed
\eq{eq:flowcollnondim} to prove that the contribution of $\varphi_0$
appearing in \eq{eq:pullback_order1_Maxwellian} vanishes within the
collision operator. It is easy to find the equation for the
gyrophase-dependent piece of $F_{\spe 2}^\lw$:
\begin{eqnarray}
\fl  - B\partial_\theta (F_{\spe 2}^\lw -\langle F_{\spe 2}^\lw\rangle)
\nonumber\\[5pt]
\fl\hspace{1cm}
=
\sum_{\sigma'}\cT^{*}_{\spe,0}
C_{\spe\spe'}\Bigg[ \frac{1}{T_{\spe}} \Bigg( \bv\cdot{\mathbf
V}^p_\spe 
\nonumber\\[5pt]
\fl\hspace{1cm}
+ \Bigg( \frac{v^2}{2T_{\spe}}-\frac{5}{2}
\Bigg)\bv\cdot{\mathbf V}^T_{\spe}
\Bigg)\cT^{-1*}_{\spe,0}F_{\spe 0} ,\cT^{-1*}_{\spe',0}F_{\spe'
0} \Bigg]
\nonumber\\[5pt]
\fl\hspace{1cm} + \sum_{\sigma'}
\frac{\lambda_\spe}{\lambda_{\spe'}}\cT^{*}_{\spe,0}
C_{\spe\spe'} \Bigg[ \cT^{-1*}_{\spe,0}F_{\spe 0}, \frac{1}{
T_{\spe'}} \Bigg( \bv\cdot{\mathbf V}^p_{\spe'}
\nonumber\\[5pt]
\fl\hspace{1cm}
 + \left(
\frac{v^2}{2T_{\spe'}}-\frac{5}{2} \right)\bv\cdot{\mathbf
V}^T_{\spe'} \Bigg)\cT^{-1*}_{\spe',0}F_{\spe' 0}
\Bigg].
\end{eqnarray}

The gyroaverage of \eq{eq:Vlasovorder1} yields an equation
for $F_{\sigma 1}^\lw$ (recall from Section
\ref{sec:FokkerPlancklong0} that $F_{\spe 1}$ is
gyrophase-independent):
\begin{eqnarray}\label{eq:Vlasovorder1gyroav}
\fl
\left(
 u\bun\cdot\nabla_\bR
- \mu\bun\cdot\nabla_\bR B\partial_u
\right)
 F_{\spe 1}^\lw
  \nonumber\\[5pt]
\fl\hspace{1cm}+
 \left(
\bv_\kappa + \bv_{\nabla B} + \bv_{E,\spe}^{(0)}
\right)\cdot\nabla_\bR F_{\spe 0}\nonumber\\[5pt]
\fl\hspace{1cm} + \left[
u\, \kappabf\cdot\left(\bv_{\nabla B} + \bv_{E,\spe}^{(0)}\right)
-Z_\spe\lambda_\spe\bun\cdot\nabla_\bR\varphi_1^{\lw}
\right]
\partial_u F_{\spe 0}\nonumber\\[5pt]
\fl\hspace{1cm}
=
\sum_{\sigma'}\cT^{*}_{\spe,0}
C_{\spe\spe'}\left[ \cT^{-1*}_{\spe,0}F_{\spe 1}^\lw
,\cT^{-1*}_{\spe',0}F_{\spe' 0} \right]
\nonumber\\[5pt]
\fl\hspace{1cm}
 + \sum_{\sigma'}
\frac{\lambda_\spe}{\lambda_{\spe'}}\cT^{*}_{\spe,0}
C_{\spe\spe'} \left[ \cT^{-1*}_{\spe,0}F_{\spe 0},
\cT^{-1*}_{\spe',0}F_{\spe' 1}^\lw \right].
\end{eqnarray}

Up to this point our computations are valid for an arbitrary
time-independent magnetic field with nested flux surfaces. Now, we
particularize to the case of an equilibrium tokamak magnetic field:
\begin{equation}\label{eq:tokamakB}
\bB = I(\psi)\nabla_\bR\zeta + \nabla_\bR\zeta\times\nabla_\bR\psi.
\end{equation}
In \ref{sec:calculationsFPorder1} we show that, in this setting, equation
\eq{eq:Vlasovorder1gyroav} can be written as
\begin{eqnarray}\label{eq:Vlasovorder1gyroav4}
\fl
\left(u\bun\cdot\nabla_\bR - \mu\bun\cdot\nabla_\bR
B\partial_u
\right)
G_{\spe 1}^\lw
\nonumber\\[5pt]
\fl\hspace{1cm}
= \sum_{\sigma'}\cT_{\spe,0}^*
C_{\spe\spe'}\left[ \cT^{-1*}_{\spe,0}\left(G_{\spe
1}^\lw-\frac{Iu}{B}
\Upsilon_\spe
 F_{\spe 0}\right)
,\cT^{-1*}_{\spe',0}F_{\spe' 0} \right]
\nonumber\\[5pt]
\fl\hspace{1cm} + \sum_{\sigma'}
\frac{\lambda_\spe}{\lambda_{\spe'}}\cT_{\spe,0}^*
C_{\spe\spe'} \Bigg[ \cT^{-1*}_{0,\spe}F_{\spe 0},
\cT^{-1*}_{0,\spe'}\Bigg(G_{\spe' 1}^\lw- \frac{Iu}{B}
\Upsilon_{\spe'}
F_{{\spe'} 0}
\Bigg)
\Bigg],
\end{eqnarray}
where
\begin{eqnarray}\label{eq:defGspe1}
\fl G_{\spe 1}^\lw&:=& F_{\spe 1}^\lw
+
\Bigg\{
\frac{Z_\spe\lambda_\spe}{T_\spe}\varphi_1^\lw
+\frac{Iu}{B}
\left(
\frac{Z_\spe}{T_\spe}\partial_\psi\varphi_0
+\Upsilon_\spe
\right)
\Bigg\}F_{\spe 0},
\end{eqnarray}
and
\begin{eqnarray}\label{eq:defUpsilon}
\Upsilon_\spe := 
\partial_\psi \ln n_\spe+
\left(\frac{u^2/2 +\mu B}{T_\spe}-\frac{3}{2}\right)
\partial_\psi \ln T_\spe\, .
\end{eqnarray}
It is a remarkable fact that in terms of the functions $G_{\spe
  1}^\lw$ the first-order Fokker-Planck equations do not involve the
electrostatic potential. Equation \eq{eq:Vlasovorder1gyroav4} is in a form that makes it easy to compare with the results of neoclassical theory \cite{hinton76, helander02bk}.

\subsection{Short-wavelength Fokker-Planck and quasineutrality
  equations to $O(\epsilon_\spe)$}
\label{sec:FPandQuasineutSW}

Here, the equations for $F_{\spe 1}^\sw$ and
$\phi_{\spe 1}^\sw$ are given because they enter the second-order, long-wavelength
piece of the Fokker-Planck equation. Before presenting such
short-wavelength equations, we need to define a new operator $\modTinv$
acting on phase-space functions $F(\bR,u,\mu,\theta)$. Namely,
\begin{equation}
\fl\modTinv F(\boldr,\bv) :=
F\left(\boldr-\epsilon_\spe\cT_{\spe,0}^{-1*}\rhobf(\boldr,\bv),
\bv\cdot\bun(\boldr), \frac{v_\bot^2}{2B(\boldr)},\arctan
\left(
\frac{\bv\cdot\eun_2(\boldr)}{\bv\cdot\eun_1(\boldr)}
\right)\right).
\end{equation}
This operator is useful to write some expressions involving the short
wavelength pieces of the distribution function and the potential, for
which it is not possible to Taylor expand the dependence on $\boldr -
\epsilon_\spe \cT_{\spe, 0}^{-1 *} \rhobf (\boldr, \bv)$ around
$\boldr$.

The first-order, short-wavelength terms of
\eq{eq:FokkerPlancknondimgyro} yield
\begin{eqnarray}\label{eq:sworder1distfunction}
\fl\frac{1}{\tau_\spe}\partial_t F_{\spe 1}^\sw
+\left(u\bun\cdot\nabla_\bR-\mu\bun\cdot\nabla_\bR B\partial_u\right)
F_{\spe 1}^\sw
\nonumber\\[5pt]
\fl\hspace{0.5cm}
+
\left[
\frac{Z_\spe\lambda_\spe}{B}
\left(\bun\times\nabla_{\bR_\perp/\epsilon_\spe}
\langle
\phi_{\spe 1}^\sw
\rangle
\right)
\cdot\nabla_{\bR_\perp/\epsilon_\spe}F_{\spe 1}^\sw
\right]^\sw
\nonumber\\[5pt]
\fl\hspace{0.5cm}
+\left(
\frac{u^2}{B}\bun\times\kappabf
+\frac{\mu}{B}\bun\times\nabla_\bR B
+\frac{Z_\spe}{B}\bun\times\nabla_\bR\varphi_0
\right)
\cdot\nabla_{\bR_\perp/\epsilon_\spe}F_{\spe 1}^\sw
\nonumber\\[5pt]
\fl\hspace{0.5cm}
+\frac{Z_\spe\lambda_\spe}{B}
\left(\bun\times\nabla_{\bR_\bot/\epsilon_\spe}
\langle
\phi_{\spe 1}^\sw
\rangle
\right)
\cdot\nabla_\bR F_{\spe 0}
\nonumber\\[5pt]
\fl\hspace{0.5cm}
-Z_\spe\lambda_\spe
\Big(
\bun\cdot\nabla_\bR\langle\phi_{\spe 1}^\sw\rangle
+\frac{u}{B}
\bun\times(\bun\cdot\nabla_\bR\bun)
\cdot
\nabla_{\bR_\perp/\epsilon_\spe}
\langle\phi_{\spe 1}^\sw\rangle
\Big)
\partial_u F_{\spe 0}
\nonumber\\[5pt]
\fl\hspace{0.5cm}
=
\sum_{\spe^\prime} \left \langle \cT_{NP, \spe}^* C_{\sigma \sigma^\prime}
\left [\modTinv F_{\sigma 1}^\sw -
\frac{Z_\spe\lambda_\spe}{T_\spe}
\modTinv\tilde\phi_{\spe 1}^\sw \cT_{\spe,0}^{-1
*} F_{\spe 0}, \cT_{\spe',0}^{-1
*}F_{\spe' 0}
 \right ] \right \rangle\nonumber\\[5pt]
\fl\hspace{0.5cm}
 + \
\sum_{\spe^\prime} \frac{\lambda_\spe}{\lambda_{\spe'}}
\left \langle \cT_{NP, \spe}^*
C_{\sigma
\sigma^\prime} \left [ \cT_{\spe,0}^{-1 *}F_{\spe 0} ,
\modTinvprime F_{\sigma' 1}^\sw -
\frac{Z_{\spe'}\lambda_{\spe'}}{T_{\spe'}}
\modTinvprime\tilde\phi_{\spe'
1}^\sw \cT_{\spe',0}^{-1 *}F_{\spe' 0}
 \right ] \right \rangle.
\end{eqnarray}
As for the short-wavelength,
first-order quasineutrality equation:
\begin{eqnarray}\label{eq:quasinautralitySWorder1appendix}
\fl\sum_\spe \frac{Z_\spe}{\lambda_\spe}
&\int
B
\Bigg[-Z_\spe\lambda_\spe \phiwig_{\spe
      1}^\sw \left(\boldr
-\epsilon_\spe\rhobf(\boldr,\mu,\theta),\mu,\theta,t\right)
\frac{F_{\spe 0}(\boldr,u,\mu,t)}{T_{\spe}(\boldr,t)}\nonumber\\[5pt]
\fl&
 + F_{\spe 1}^\sw
\left(\boldr
-\epsilon_\spe\rhobf(\boldr,\mu,\theta),u,\mu,t
\right)
\Bigg]
  \dd u \dd \mu \dd \theta = 0.
\end{eqnarray}

\subsection{Long-wavelength Fokker-Planck equation to $O(\epsilon_\spe^2)$}
\label{sec:FokkerPlancklong2}

The second-order contribution to \eq{eq:FokkerPlancknondimgyro} is
cumbersome. In order to avoid lengthy calculations to those readers
interested in reaching quickly the final expressions and main results,
most of the manipulations in this subsection are deferred to the
appendices. 

The pieces of order $\epsilon_\spe^2$ in
\eq{eq:FokkerPlancknondimgyro} yield
\begin{eqnarray} \label{eq:FPsecondorder}
\fl
\left(
u \bun \cdot \nabla_\bR   - \mu \bun \cdot
\nabla_\bR B \, \partial_u
\right)
F^\lw_{\sigma 2}
 - B \partial_\theta F^\lw_{\sigma 3} +
\frac{\lambda_\spe^2}{\tau_\spe}\partial_{\epsilon_s^2 t} F_{\spe 0}
\nonumber\\[5pt]
\fl\hspace{0.5cm} + 
\left(
\bv_\kappa + \bv_{\nabla B} + \bv_{E,\spe}^{(0)}
\right)
 \cdot \nabla_\bR F^\lw_{\sigma 1} \nonumber\\[5pt]
\fl\hspace{0.5cm}
+\left[- Z_\spe \lambda_\spe
\bun \cdot \nabla_\bR \varphi_1^\lw
+u\, \kappabf\cdot\left(\bv_{\nabla B} + \bv_{E,\spe}^{(0)}\right)
\right]
\partial_u F^\lw_{\sigma 1}
\nonumber\\[5pt]
\fl\hspace{0.5cm} + \Bigg[
\bv_{E,\spe}^{(1)}-\frac{u}{B}
(\bun \cdot \nabla_\bR \times \bun)
\left(\bv_\kappa + \bv_{\nabla B} + \bv_{E,\spe}^{(0)}\right)
 \nonumber\\[5pt]
\fl\hspace{0.5cm} 
 - \frac{u \mu}{B}
 (\nabla_\bR \times \bK)_\bot
+ Z_\spe \lambda_\spe \partial_u\Psi_{\phi
B,\spe}^\lw \bun +
\partial_u\Psi_{B,\spe} \bun \Bigg]\cdot\nabla_\bR F_{\spe 0}
\nonumber\\[5pt]
\fl\hspace{0.5cm} - \Bigg \{ Z_\spe\lambda_\spe^2 \bun \cdot
\nabla_\bR \left [ \varphi_2^\lw + \frac{\mu}{2\lambda_\spe^2 B}
(\matI - \bun \bun) : \nabla_\bR \nabla_\bR \varphi_0 \right ] 
  \nonumber\\[5pt]
\fl\hspace{0.5cm}
+\bun \cdot \nabla_\bR \Psi_{B,\spe} + Z_\spe\lambda_\spe
\bun \cdot \nabla_\bR \Psi_{\phi
B}^\lw
 + Z_\spe^2 \lambda_\spe^2 \bun \cdot
\nabla_\bR \Psi_\phi^\lw
\nonumber\\[5pt]
\fl\hspace{0.5cm}
-u\,\kappabf\cdot\bv_{E,\spe}^{(1)}
+\Bigg[\frac{u^2}{B}\left(\bun\cdot\nabla_\bR\times\bun\right)
\kappabf
\nonumber\\[5pt]
\fl\hspace{0.5cm}
+ \mu\left(\left(\nabla_\bR\times\bK\right)\times\bun\right)\Bigg]
\cdot
\left(\bv_{\nabla B}+\bv_{E,\spe}^{(0)}\right)
\Bigg \} \partial_u F_{\spe 0} 
 \nonumber\\[5pt]
\fl\hspace{0.5cm}
 + 
\frac{Z_\spe \lambda_\spe}{B} \left[
\nabla_\bR \cdot \left(\bun\times\nabla_{\bR_\perp/\epsilon_\spe}
 \langle \phi_{\spe 1}^\sw \rangle
F_{\spe 1}^\sw\right)\right]^\lw
\nonumber\\[5pt]
\fl\hspace{0.5cm} - Z_\spe \lambda_\spe \partial_u \left[
\left( \bun\cdot\nabla_\bR 
\langle \phi_{\spe 1}^\sw \rangle
 +  \frac{u}{B}
\left(\bun \times \kappabf\right)\cdot
\nabla_{\bR_\perp/\epsilon_\spe}\langle \phi_{\spe 1}^\sw \rangle
\right)F_{\spe 1}^\sw \right]^\lw
\nonumber\\[5pt]
\fl\hspace{0.5cm} = \sum_{\spe'}
\left[\cT_{\spe,1}^*C_{\spe\spe'}^{(1)}\right]^\lw +\sum_{\spe'}
\cT_{\spe,0}^*C_{\spe\spe'}^{(2)\lw}.
\end{eqnarray}
Here,
\begin{eqnarray}
\fl\bv_{E,\spe}^{(1)} = \frac{Z_\spe\lambda_\spe}{B}
\bun\times\nabla_{\bR_\perp}\varphi_1^\lw,
\end{eqnarray}
\begin{eqnarray} \label{eq:PsincphiB}
\fl \Psi^\lw_{\phi B} = - \frac{3 \mu}{2 \lambda_\spe B^2}
\nabla_\bR B \cdot \nabla_{\bR} \varphi_0
- \frac{u^2}{\lambda_\spe B^2} (\bun \cdot \nabla_\bR \bun)
 \cdot \nabla_{\bR}
\varphi_0,
\end{eqnarray}
and
\begin{eqnarray} \label{eq:Psincphi}
\fl \Psi^\lw_\phi = 
 - \frac{1}{2\lambda_\spe^2 B^2} |\nabla_{\bR}
\varphi_0|^2 - \frac{1}{2B}\partial_\mu
\left[\langle (\phiwig_{\spe 1}^\sw)^2 \rangle
\right]^\lw.
\end{eqnarray}
As for the collision operator,
\begin{equation}
C_{\spe\spe'}^{(1)} =
 C_{\spe\spe'}^{(1)\lw}
 + C_{\spe\spe'}^{(1)\sw},
\end{equation}
\begin{eqnarray}\label{eq:T1C1lw}
\fl\left[\cT_{\spe,1}^* C_{\spe\spe'}^{(1)}\right]^\lw &=
\left(
\rhobf\cdot\nabla_\bR+
\hat{u}_1\partial_u
+
\hat{\mu}_1^\lw\partial_\mu +
\hat{\theta}_1^\lw
 \partial_\theta\right)
\cT_{\spe,0}^{*}C_{\spe\spe'}^{(1)\lw}\nonumber\\[5pt]
\fl
&+
\left[
\cT_{\spe,1}^*C_{\spe\spe'}^{(1)\sw}\right]^\lw,
\end{eqnarray}
where
\begin{eqnarray} \label{eq:C1nc}
\fl  C_{\sigma \sigma^\prime}^{(1)\lw}
=
 C_{\sigma \sigma^\prime} \Bigg[
\frac{1}{T_\spe} \bv\cdot \left (\bV_{\sigma}^p
+ \left ( \frac{v^2}{2T_\sigma} - \frac{5}{2} \right )
\bV^T_{\sigma} \right ) \cT_{\spe,0}^{-1 *}F_{\spe 0} \nonumber\\[5pt]
\fl\hspace{0.5cm} + \cT_{\spe,0}^{-1 *} \Bigg( G_{\spe 1} - \frac{I
u}{B}\Upsilon_\spe F_{\spe 0}
\Bigg),
\cT_{\spe',0}^{-1 *}F_{\spe' 0} \Bigg]
\nonumber\\[5pt]
\fl\hspace{0.5cm} + \
\frac{\lambda_\spe}{\lambda_{\spe'}}C_{\sigma
\sigma^\prime} \Bigg[ \cT_{\spe,0}^{-1 *}F_{\spe 0} ,
\frac{1}{T_{\spe'}} \bv\cdot \Bigg(\bV_{
\sigma'}^p
 + \left ( \frac{v^2}{2T_{\sigma'}}
 - \frac{5}{2} \Bigg) \bV^T_{ \sigma'} \right ) \cT_{\spe',0}^{-1 *}F_{\spe' 0}
\nonumber\\[5pt]
\fl\hspace{0.5cm} + \cT_{\spe',0}^{-1*}
 \left( G_{\spe' 1} - \frac{I u}{B} \Upsilon_{\spe'} F_{\spe' 0}
\right)
\Bigg],
\end{eqnarray}
and
\begin{eqnarray}
\fl  C_{\sigma \sigma^\prime}^{(1)\sw} = C_{\sigma \sigma^\prime}
\left [\modTinv F_{\sigma 1}^\sw -
\frac{Z_\spe\lambda_\spe}{T_\spe}
\modTinv\tilde\phi_{\spe 1}^\sw \cT_{\spe,0}^{-1
*} F_{\spe 0}, \cT_{\spe',0}^{-1
*}F_{\spe' 0}
 \right ]\nonumber\\[5pt]
\fl\hspace{0.5cm} + \
\frac{\lambda_\spe}{\lambda_{\spe'}}C_{\sigma
\sigma^\prime} \left [ \cT_{\spe,0}^{-1 *}F_{\spe 0} ,
\modTinvprime F_{\sigma' 1}^\sw -
\frac{Z_{\spe'}\lambda_{\spe'}}{T_{\spe'}}
\modTinvprime\tilde\phi_{\spe'
1}^\sw \cT_{\spe',0}^{-1 *}F_{\spe' 0}
 \right ].
\end{eqnarray}
The left-hand side of \eq{eq:FPsecondorder} is written by employing
again the Poisson brackets obtained in \ref{sec:eqsofmotion}. The
first-order coordinate transformation that enters explicitly the
expression of the collision operator is computed in detail in
\ref{sec:pullback}; in particular, $\hat{u}_1,\hat{\mu}_1^\lw,
\hat{\theta}_1^\lw$ are defined in
\eq{eq:totalchangecoorfirstorderCorrections} and
\eq{eq:totalchangecoorfirstorderCorrectionsLW}. In \ref{sec:Cturb} we
calculate the last term of \eq{eq:T1C1lw}, $[\cT_{\spe, 1}^{*} C_{\spe
  \spe^\prime}^{(1)\sw}]^\lw$, and its gyroaverage. Finally,
$C_{\spe\spe'}^{(2)\lw}$ is calculated in \ref{sec:C2nc}.

In \ref{sec:calculationsFPorder2} we prove that the
gyroaverage of \eq{eq:FPsecondorder} can be rearranged so that it
reads
\begin{eqnarray} \label{eq:eqH2sigma}
\fl\Big( u \bun \cdot &  \nabla_\bR  - \mu \bun \cdot \nabla_\bR
B\partial_u \Big) G_{\spe 2}^\lw +
\frac{\lambda_\spe^2}{\tau_\spe} \partial_{\epsilon_s^2 t} F_{\spe 0}
\nonumber\\\fl & - \bun \cdot \nabla_\bR \Theta
\partial_\psi \Bigg \{
\frac{Z_\spe \lambda_\spe}{\bB \cdot \nabla_\bR
\Theta} \left[ F_{\sigma 1}^\sw
(\nabla_{\bR_\perp/\epsilon_\spe} \langle \phi_{\spe 1}^\sw
\rangle \times \bun) \cdot \nabla_\bR \psi
\right]^\lw \nonumber\\ \fl & +
 \frac{1}{\bun \cdot \nabla_\bR
\Theta} \left \langle
 \left(\frac{Iu}{B} +
\rhobf \cdot \nabla_\bR \psi \right ) \sum_{\sigma^\prime}
\cT_{\spe,0}^*C_{\sigma \sigma^\prime}^{(1)\lw} \right \rangle
\Bigg \} \nonumber\\
\fl &
- \partial_u \Bigg \{ \Bigg[
 Z_\spe\lambda_\spe
 F_{\sigma 1}^\sw
 \Big(  \bun\cdot\nabla_\bR
\langle\phi_{\spe
 1}^\sw\rangle
\nonumber\\
\fl &
 + \frac{\mu}{u B} \partial_\Theta B(\bun \times
\nabla_\bR \Theta) \cdot
\nabla_{\bR_\perp/\epsilon_\spe}\langle\phi_{\spe 1}^\sw\rangle
 \nonumber\\
\fl &+ 
\frac{u}{B} [\bun \times (\bun \cdot
\nabla_\bR \bun)] \cdot
\nabla_{\bR_\perp/\epsilon_\spe}\langle\phi_{\spe 1}^\sw\rangle 
 \Big)  \Bigg]^\lw \nonumber\\
\fl & - \left \langle \frac{I}{B}
\left ( \mu \partial_\psi B  +
Z_\spe
 \partial_\psi\varphi_0  \right )
\sum_{\sigma^\prime} \cT_{\spe,0}^*C_{\sigma \sigma^\prime}^{(1)\lw} \right
\rangle \Bigg \} \nonumber\\\fl & + \partial_\mu \Bigg \langle
 \frac{1}{B}
\rhobf \cdot \nabla_\bR \psi \left ( \mu 
\partial_\psi B + Z_\spe \partial_\psi\varphi_0
 \right ) \sum_{\sigma^\prime}
\cT_{\spe,0}^*C_{\sigma \sigma^\prime}^{(1)\lw}
 \Bigg \rangle
\nonumber\\
\fl & = - \sum_{\sigma^\prime} \partial_u \Bigg \langle
 \Bigg [ \frac{Z_\spe \lambda_\spe
 \varphi_1^\lw}{u}
 +
\mu \bun \cdot \nabla_\bR \times \bun \nonumber\\
\fl & - \frac{1}{u}\rhobf
\cdot \left ( \mu \partial_\Theta B \nabla_{\bR} \Theta 
 + u^2 \bun \cdot \nabla_\bR \bun \right )
\Bigg ]  \cT_{\spe,0}^*C_{\sigma \sigma^\prime}^{(1)\lw}
 \Bigg \rangle  \nonumber\\
\fl &
+
\sum_{\sigma^\prime} \partial_\mu \Bigg \langle
\Bigg[ \frac{u \mu}{B} \bun \cdot \nabla_\bR
\times
\bun
\nonumber\\
\fl &
 - \frac{1}{B}\rhobf \cdot \left ( \mu \partial_\Theta B
 \nabla_{\bR} \Theta
 + u^2 \bun \cdot \nabla_\bR
\bun \right ) \Bigg ] \cT_{\spe,0}^*C_{\sigma \sigma^\prime}^{(1)\lw}
\Bigg\rangle
\nonumber\\
\fl &+\sum_{\sigma^\prime}\left[ \left\langle
\cT_{\spe, 1}^* C_{\sigma
\sigma^\prime}^{(1)\sw}\right\rangle
\right]^\lw
+\sum_{\sigma^\prime} \left \langle 
\cT_{\spe,0}^*C_{\sigma \sigma^\prime}^{(2)\lw}  \right
\rangle,
\end{eqnarray}
where $G_{\spe 2}^\lw$ is defined in \eq{eq:defG2again3}. Note that
the first-order, short-wavelength pieces of the distribution function
and electrostatic potential, $F_{\spe 1}^\sw$ and $\phi_{\spe 1}^\sw$,
enter equation~\eq{eq:eqH2sigma}. The equations needed to determine
them are given in subsection \ref{sec:FPandQuasineutSW}. Observe also
that the time derivative of $F_{\spe 0}$ appears in \eq{eq:eqH2sigma},
something that has very important consequences. We will learn that
equation \eq{eq:eqH2sigma} has non-trivial solvability conditions that
involve the time evolution of certain moments of $F_{\spe 0}$.

The function $G_{\spe 2}^\lw$ is defined to make comparisons with
neoclassical theory easier. It is also useful to obtain the
solvability conditions in subsection \ref{sec:transportEqs} with less
algebra.

\section{Long-wavelength quasineutrality equation}
\label{sec:longwavePoisson}

In this section we obtain the quasineutrality equation,
\eq{eq:gyroQuasineutrality}, at long-wavelengths. For convenience, we
repeat here equation \eq{eq:gyroQuasineutrality}:
\begin{eqnarray}\label{eq:gyroQuasineutralityagain}
\fl\sum_\spe Z_\spe  \int B_{||,\spe}^*
F_\spe
\delta\Big(\pi^{\boldr}\Big(\cT_{\spe}(\bR, u, \mu, \theta,
t)\Big)-\boldr\Big)\dd ^3 R\, \dd u\, \dd\mu\, \dd\theta = 0,
\end{eqnarray}
with $\pi^\boldr(\boldr,\bv):=\boldr$.

At long wavelengths we can simply expand the argument of the Dirac
delta function around $\bR-\boldr$. Using that
\begin{eqnarray}
\fl\pi^{\boldr}{\cal T}_{\spe}(\bR, u, \mu, \theta,
t) =  \bR +\epsilon_\spe\rhobf +\epsilon_\spe^2
\left(\bR_{\spe,2}+\mu_{\spe,1}\partial_\mu\rhobf
+\theta_{\spe,1}\partial_\theta\rhobf\right)
 + O(\epsilon_\spe^3),
\end{eqnarray}
that the first-order term of $B_{||,\spe}^*$ is odd in $u$, that
$F_{\spe 0}$ is even in $u$, that $F_{\spe 1}^\lw$ does not depend on
$\theta$, and integrating over $\bR $, it is straightforward to obtain
\begin{eqnarray}\label{eq:gyroPoissonlw2order0}
\fl\sum_\spe
Z_\spe n_\spe(\boldr,t) = 0
\end{eqnarray}
to order $\epsilon_s^0$,
\begin{eqnarray}\label{eq:gyroPoissonlw2order1}
\fl\sum_\spe
\frac{Z_\spe}{\lambda_\spe}
\int B(\boldr)F_{\spe 1}^\lw(\boldr,u,\mu,t)
\dd u\dd\mu\dd\theta = 0
\end{eqnarray}
to order $\epsilon_s$, and
\begin{eqnarray}\label{eq:gyroPoissonlw2}
\fl\sum_\spe
\frac{Z_\spe}{\lambda_\spe^2}
\Bigg[
\int
(BF_{\spe 2}^\lw -\mu\bun\cdot\nabla_\boldr\times\bK \, F_{\spe 0}
+u\bun\cdot\nabla_\boldr\times\bun \, F_{\spe 1}^\lw)
\dd u\dd\mu\dd\theta\nonumber\\[5pt]
\fl\hspace{0.5cm}
-\nabla_\boldr\cdot\int(\bR_{\spe,2}^\lw
+\mu_{\spe,1}^\lw
\partial_\mu\rhobf
+\theta_{\spe,1}^\lw
\partial_\theta\rhobf)BF_{\spe 0}\,\dd u\dd\mu\dd\theta
\nonumber\\[5pt]
\fl\hspace{0.5cm}
+\frac{1}{2}\nabla_\boldr\nabla_\boldr:
\int\rhobf\rhobf BF_{\spe 0}\, \dd u\dd\mu\dd\theta
\Bigg]
\end{eqnarray}
to order $\epsilon_s^2$. Here everything is evaluated at
$\bR=\boldr$. In writing the arguments of some functions we have
stressed that they are evaluated at $\bR=\boldr$, e.g. $n_\spe(\boldr)$,
but we should not forget that $n_\spe$, for example, depends only on
$\psi$ in flux coordinates. Note
that to be formally correct we need a unique, species-independent
expansion parameter, and we have chosen $\epsilon_s$ as indicated in
Section \ref{sec:secondordergyrokinetics}. In
\ref{sec:computations_quasineutralitylw} we show that
\eq{eq:gyroPoissonlw2} can be transformed into
\begin{eqnarray}\label{eq:gyroPoissonlw4}
\fl\sum_\spe
\frac{Z_\spe}{\lambda_\spe^2}
\Bigg[
\int(BF_{\spe 2}^\lw
+u\bun\cdot(\nabla_\boldr \times\bun)F_{\spe 1}^\lw)\dd u\dd\mu\dd\theta
\nonumber\\[5pt]
\fl\hspace{1cm}
-\bun\cdot(\nabla_\boldr \times\bK)\frac{n_{\spe}T_{\spe}}{B^2}
+
\nabla_\boldr\cdot\left(\frac{3}{2}
\frac{\nabla_{\boldr_\bot} B}{B^3}n_{\spe}T_{\spe}\right)
\nonumber\\[5pt]
\fl\hspace{1cm}
+\frac{1}{2}\nabla_\boldr\nabla_\boldr:
\left(
\left({\matI}-\bun\bun
\right)
\frac{n_{\spe}T_{\spe}}{B^2}
\right)\nonumber\\[5pt]
\fl\hspace{1cm}
+
\nabla_\boldr\cdot\left(
(\bun\cdot\nabla_\boldr \bun)
\frac{n_{\spe}T_{\spe}}{B^2}
\right)
+
\nabla_\boldr\cdot\left(
\frac{Z_\spe n_{\spe}}{B^2}\nabla_\boldr\varphi_0
\right)
\Bigg]
=0.
\end{eqnarray}
Observe that the above expressions for the long-wavelength
quasineutrality equation are completely general, i.e. we have not
particularized for tokamak geometry. We proceed to do it next by writing \eq{eq:gyroPoissonlw2order1} and
\eq{eq:gyroPoissonlw4} in terms of the functions $G_{\spe 1}^\lw$ and
$G_{\spe 2}^\lw$, defined in \eq{eq:defGspe1} and
\eq{eq:defG2again3}. This is obvious for the first-order piece of the
quasineutrality equation, yielding
\begin{eqnarray}\label{eq:gyroPoissonlw2order1aux2}
\sum_\spe \frac{Z_\spe}{\lambda_\spe} &\Bigg( \int B(\boldr) G_{\spe
  1}^\lw(\boldr,u,\mu,t)\dd u\dd\mu\dd\theta
\nonumber\\[5pt]
 &-\frac{Z_\spe
  \lambda_\spe}{T_\spe}n_\spe(\boldr,t) \varphi_1^\lw(\boldr,t)\Bigg) =
0,
\end{eqnarray}
and in \ref{sec:computations_quasineutralitylw} it is shown that the
result for the second-order piece is
\begin{eqnarray}\label{eq:gyroPoissonlw4aux3}
\fl\sum_\spe&
\frac{Z_\spe}{\lambda_\spe^2}
\int B
\Bigg[
G_{\spe 2}^\lw + \frac{u}{B}\bun\cdot\nabla_\boldr\times\bun\, G_{\spe 1}^\lw
\nonumber\\
\fl&
+
\left(\frac{Z_\spe\lambda_\spe \varphi_1^\lw}{u} \partial_u
-
\frac{Iu}{B}\partial_\psi\right) G_{\sigma 1}^\lw
\nonumber\\
\fl&
- \left[ \frac{Z_\spe\lambda_\spe }{
u \bB \cdot
\nabla_\boldr \Theta} F_{\sigma 1}^\sw (\bun\times
\nabla_{\boldr_\perp/\epsilon_\spe}
\langle \phi_{\spe 1}^\sw \rangle) \cdot \nabla_\boldr
\Theta \right]^\lw\nonumber\\
\fl& +
 \left \langle \frac{1}{ u \bun \cdot
  \nabla_\boldr \Theta} \rhobf \cdot 
\nabla_\boldr\Theta
\sum_{\sigma^\prime}\cT_{\spe,0}^* C_{\sigma
    \sigma^\prime}^{(1)\lw} \right \rangle
\nonumber\\
\fl&
+ \frac{Z_\spe^2\lambda_\spe^2}{2T_\spe^2}
\left[ \left \langle
(\phiwig_{\spe 1}^\sw)^2 \right \rangle \right]^\lw F_{\spe
0}
\Bigg]
\dd u\dd\mu\dd\theta
\nonumber\\
\fl&+
\sum_\spe
\frac{Z_\spe}{\lambda_\spe^2}n_\spe T_\spe\Bigg\{
\frac{Z_\spe\lambda_\spe^2}{T_\sigma^2}
\left(
\frac{Z_\spe}{2 T_\sigma}(\varphi_1^\lw)^2
-
\varphi_2^\lw
\right)
\nonumber\\[5pt]
\fl&
+
\frac{R^2}{2}
\Bigg[
\left(\frac{Z_\spe}{T_\sigma}
\partial_\psi\varphi_0 
+
\partial_\psi \ln n_\spe\right)^2
\nonumber\\[5pt]
\fl&
+\left(
\partial_\psi \ln T_\spe
\right)^2
+
2
\partial_\psi \ln n_\spe
\partial_\psi \ln T_\spe
\nonumber\\[5pt]
\fl&
+\partial_\psi^2\ln n_\spe + \frac{Z_\spe}{T_\spe}\partial_\psi^2\varphi_0
+
\partial_\psi^2\ln T_\spe
\Bigg]
\Bigg\}=0,
\end{eqnarray}
where $R$ is the major
radius coordinate, i.e. it is the distance to the axis of symmetry of
the tokamak.

\section{Indeterminacy of the long-wavelength radial electric field}
\label{sec:indeterminacyRadialElectricField}

With the results of Sections \ref{sec:FokkerPlancklong} and
\ref{sec:longwavePoisson} at hand it is reasonably easy to prove that
in a tokamak $\varphi_0(\psi)$ is not determined by second-order
Fokker-Planck and quasineutrality equations. In order to be as clear
as possible, we divide the argument into three steps. In subsection
\ref{sec:indeterminacyquasineutrality} we show that the
quasineutrality equation gives no information about the radial
electric field, even though naively one would have expected to use this
equation to solve for it. In subsection \ref{sec:transportEqs} we
learn that \eq{eq:eqH2sigma} possesses non-trivial solvability
conditions and work them out. They are transport equations
for the lowest order density and temperature functions. In subsection  
\ref{sec:timeevolutionquasineutrality} we prove that these solvability
conditions do not yield new equations for the
radial electric field. The proof amounts to explicitly
showing that the turbulent tokamak is intrinsically ambipolar.

\subsection{Quasineutrality equation and long-wavelength radial
  electric field}
\label{sec:indeterminacyquasineutrality}

It is obvious from
equations~\eq{eq:Vlasovorder1gyroav4} and \eq{eq:eqH2sigma} that if
$G_{\spe j}^\lw$, $j=1,2$ are solutions of the first and second-order
Fokker-Planck equations, then so are $G_{\spe j}^\lw + h_{\spe j}$,
$j=1,2$, where
\begin{eqnarray}\label{eq:kernel}
 h_{\spe j} &=&
\left[
\frac{n_{\spe  j}}{n_\spe}
+\left(\frac{\mu B+u^2/2}{T_\spe}-\frac{3}{2}\right)
\frac{T_{\spe j}}{T_\spe}
\right]
F_{\spe 0},
\end{eqnarray}
for an arbitrary set of flux functions $\{n_{\spe j}(\psi,t), T_{\spe
  j}(\psi,t)\}_\spe$, with the only restriction $T_{\spe
  j}/\lambda_\spe^j =T_{\spe' j}/\lambda_{\spe'}^j$, for all
$\sigma,\sigma'$ (the temperature of the electrons is allowed to be
different if we expand in the mass ratio, that is, if we use
$\lambda_e \sim \tau_e \gg 1$). In other words, the operator acting on
$G_{\spe 1}^\lw$ in \eq{eq:Vlasovorder1gyroav4} and on $G_{\spe
  2}^\lw$ in \eq{eq:eqH2sigma} has a kernel given by \eq{eq:kernel}
with an obvious interpretation: it consists of corrections of order
$\epsilon_\spe^j$ to the zeroth-order particle densities, $n_\spe$,
and temperatures, $T_\spe$. Therefore, in order to have a unique
solution for the Fokker-Planck equation, one needs to prescribe a
condition that eliminates the freedom
introduced by the existence of a non-zero kernel. An example of such a
condition is given by imposing, for $j=1,2$,
\begin{eqnarray}\label{eq:gaugefixingExample}
\left\langle\int B G_{\sigma j}\dd u\dd\mu\dd\theta\right\rangle_\psi = 0, \mbox{ for every $\sigma$, and}
\nonumber\\
\left\langle\sum_\sigma\frac{1}{\lambda_\sigma^j} \int B \left(u^2/2+\mu
  B\right)G_{\sigma j}
\dd u\dd\mu\dd\theta\right\rangle_\psi = 0.
\end{eqnarray}
Of course, even though this is a natural choice, there are
infinitely many different possibilities.

Assume that $\textsf{G}_{\spe j}$, $j=1,2$ (note the different font)
are particular solutions of \eq{eq:Vlasovorder1gyroav4} and
\eq{eq:eqH2sigma} (not necessarily satisfying \eq{eq:gaugefixingExample}). Then,
any solution of \eq{eq:Vlasovorder1gyroav4} and \eq{eq:eqH2sigma} is
of the form $\textsf{G}_{\spe j}^\lw + h_{\spe j}$, $j=1,2$. When
introduced in \eq{eq:gyroPoissonlw2order1aux2} and
\eq{eq:gyroPoissonlw4aux3} we find:
\begin{eqnarray}\label{eq:quasineutralityWithCorrectionsOrder1}
\fl
\sum_\spe \frac{Z_\spe}{\lambda_\spe} n_{\spe 1}+
\sum_\spe \frac{Z_\spe}{\lambda_\spe} \Bigg( \int B(\boldr) \textsf{G}_{\spe
  1}^\lw(\boldr,u,\mu,\theta,t)\dd u\dd\mu\dd\theta\nonumber\\[5pt]
\fl\hspace{1cm}
-\frac{Z_\spe
  \lambda_\spe}{T_\spe}n_\spe(\boldr) \varphi_1^\lw(\boldr,t)\Bigg) =
0,
\end{eqnarray}
\begin{eqnarray}\label{eq:quasineutralityWithCorrectionsOrder2}
\fl
\sum_\spe
\frac{Z_\spe}{\lambda_\spe^2}n_{\spe 2}+
\sum_\spe
\frac{Z_\spe}{\lambda_\spe^2}
\int B
\Bigg[
\textsf{G}_{\spe 2}^\lw +
\frac{u}{B}\bun\cdot\nabla_\boldr
\times\bun\, \textsf{G}_{\spe 1}^\lw
\nonumber\\[5pt]
\fl\hspace{1cm}
+
\left(\frac{Z_\spe\lambda_\spe \varphi_1^\lw}{u} \partial_u
-
\frac{Iu}{B}\partial_\psi\right) \textsf{G}_{\sigma 1}^\lw
\nonumber\\
[5pt]
\fl\hspace{1cm}
- \left[ \frac{Z_\spe\lambda_\spe }{
u \bB \cdot
\nabla_\boldr\Theta} F_{\sigma 1}^\sw (\bun\times
\nabla_{\boldr_\perp/\epsilon_\spe}
\langle \phi_{\spe 1}^\sw \rangle) \cdot \nabla_\boldr
\Theta \right]^\lw
\nonumber\\
[5pt]
\fl\hspace{1cm} + \left \langle \frac{1}{ u \bun \cdot
  \nabla_\boldr \Theta}
 \rhobf \cdot \nabla_\boldr\Theta
\sum_{\sigma^\prime} \mathcal{T}^*_{\spe,0}
C_{\sigma
    \sigma^\prime}^{(1)\lw} \right \rangle
\nonumber\\
[5pt]
\fl\hspace{1cm} 
+ \frac{Z_\spe^2\lambda_\spe^2}{2T_\spe^2}
\left[ \left \langle
(\phiwig_{\spe 1}^\sw)^2 \right \rangle \right]^\lw F_{\spe
0}
\Bigg]
\dd u\dd\mu\dd\theta
\nonumber\\
[5pt]
\fl\hspace{1cm}+
\sum_\spe
\frac{Z_\spe}{\lambda_\spe^2}n_\spe T_\spe\Bigg\{
\frac{Z_\spe\lambda_\spe^2}{T_\sigma^2}
\left(
\frac{Z_\spe}{2 T_\sigma}(\varphi_1^\lw)^2
-
\varphi_2^\lw
\right)
\nonumber\\[5pt]
\fl\hspace{1cm}
+
\frac{R^2}{2}
\Bigg[
\left(\frac{Z_\spe}{T_\sigma}
\partial_\psi\varphi_0 
+
\partial_\psi \ln n_\spe\right)^2
\nonumber\\
[5pt]
\fl\hspace{1cm} 
+
\left(
\partial_\psi \ln T_\spe
\right)^2
+
2
\partial_\psi \ln n_\spe \partial_\psi \ln T_\spe
\nonumber\\[5pt]
\fl\hspace{1cm}
+\partial_\psi^2\ln n_\spe + \frac{Z_\spe}{T_\spe}\partial_\psi^2\varphi_0
+
\partial_\psi^2\ln T_\spe
\Bigg]
\Bigg\}=0.
\end{eqnarray}
Even though $\varphi_0$ enters this equation to second-order, it
cannot be determined. The first and second-order pieces of the
long-wavelength quasineutrality equation simply give constraints on
the corrections $n_{\spe 1}$ and $n_{\spe 2}$. Each function $n_{\spe
  j}$ will be determined by a transport equation that appears as a
solvability condition for a higher order long-wavelength piece of the
Fokker-Planck equation, just as a transport equation for $n_\spe$ is
derived in subsection \ref{sec:transportEqdensity} as a solvability
condition for equation \eq{eq:eqH2sigma}. Note that we cannot choose
the value of $n_{\spe j}$; we can only decide which piece of $G_{\spe
  j}$ we call $n_{\spe j}$ and which piece we leave within
$\textsf{G}_{\spe j}$. The density corrections $n_{\spe j}$ cannot be
set to zero arbitrarily because we need them to satisfy the
solvability conditions of the higher order pieces of the Fokker-Planck
equation.\footnote{Our procedure here has diverged from the canonical
  Chapman-Enskog approach, where the density of the lowest order
  Maxwellian is not broken into pieces of different orders. Instead,
  in the Chapman-Enskog theory, the conservation equation for particle
  density, obtained from the solvability conditions (see subsection
  \ref{sec:transportEqs}), contains terms of different orders. Our
  procedure is different in that the conservation equation will be
  split into different orders, each giving an equation for a piece
  $n_{\spe j}$. In this way we highlight that the quasineutrality
  equation to $O(\epsilon_s^2)$ does not allow to solve for
  $\varphi_0$ because this would require knowing $n_{\spe 1}$ and
  $n_{\spe 2}$, which are determined from higher order pieces of the
  long-wavelength Fokker-Planck equation.}
  
  We cannot calculate $\varphi_0$ from the
  quasineutrality equation to this order, but the first and
  second-order pieces of the long-wavelength poloidal electric field
  can be found, respectively, from
  \eq{eq:quasineutralityWithCorrectionsOrder1} and
  \eq{eq:quasineutralityWithCorrectionsOrder2}. This can be viewed by
  acting on the latter equations with $\bun\cdot\nabla_\boldr$ and
  employing that $n_{\spe 1}$ and $n_{\spe 2}$ are only functions of
  $\psi$. Not surprisingly, $\varphi_1^\lw$ and $\varphi_2^\lw$ are
  determined up to an arbitrary, additive function of $\psi$, that can
  be absorbed by redefining the corrections $n_{\spe 1}$ and $n_{\spe
    2}$. Without loss of generality, we fix the ambiguity by taking
\begin{equation}\label{eq:fixvarphi1}
\left\langle\varphi_1^\lw\right\rangle_\psi = 0
\end{equation}
and
\begin{equation}\label{eq:fixvarphi2}
\left\langle\varphi_2^\lw\right\rangle_\psi = 0.
\end{equation}

In Subsection \ref{sec:transportEqs} we explain that the second-order
Fokker-Planck equation possesses some solvability conditions. We have
to show that their fulfillment does not impose any additional
conditions that give $\varphi_0(\psi,t)$, and we do so in subsection
\ref{sec:timeevolutionquasineutrality}.

\subsection{Transport equations}
\label{sec:transportEqs}

Some of the benefits of writing the Fokker-Planck equation precisely
in the form \eq{eq:eqH2sigma} will be appreciated in this subsection,
where we show that time evolution equations for the lowest order
density and temperature functions $n_\spe$ and $T_\spe$ are obtained
as solvability conditions for the second-order, long-wavelength
Fokker-Planck equation. That is, we prove that if a solution for
$G_{\spe 2}^\lw$ (equivalently, for $F_{\spe 2}^\lw$) exists, then
\eq{eq:eqH2sigma} imposes certain constraints among lower-order
quantities (solvability conditions). These conditions turn out to be
transport equations for density and energy. In
\ref{app:solvabilityconditions} we prove that these transport equations
are indeed the only solvability conditions obtained from the
Fokker-Planck equation up to order $\epsilon_s^2$.

\subsubsection{Transport equation for density.}
\label{sec:transportEqdensity}

Go back to \eq{eq:eqH2sigma}, multiply by $\tau_\spe
B/\lambda_\spe^2$, integrate over $u,\mu,\theta$ and take the
flux-surface average:
\begin{eqnarray}\label{eq:transportEq_density}
\fl
\partial_{\epsilon_s^2 t} n_{\spe}(\psi,t)
= \frac{1}{V'(\psi)}\partial_\psi
\Bigg\langle
V'(\psi)
\int\dd u\dd\mu\dd\theta\
 \Bigg \{
\nonumber\\[5pt]
\fl\hspace{1cm}
\left[ F_{\sigma 1}^\sw (\nabla_{\bR_\perp/\epsilon_\spe}
\langle \phi_{\spe 1}^\sw
\rangle \times \bun) \cdot \nabla_\bR \psi \right]^\lw
\nonumber\\[5pt]
\fl
\hspace{1cm}
+\frac{B}{Z_\spe \lambda_\spe}
  \left \langle
 \left(\frac{Iu}{B} +
\rhobf \cdot \nabla_\bR \psi \right ) \sum_{\sigma^\prime}
C_{\sigma \sigma^\prime}^{(1)\lw} \right \rangle
\Bigg \}\Bigg\rangle_\psi,
\end{eqnarray}
where $V'(\psi)$ is the derivative of the function $V(\psi)$, defined
in \eq{eq:defvolume}, which gives the volume enclosed by the flux
surface with label $\psi$. We have also used that for the tokamak the
square root of the determinant of the metric tensor (recall
\eq{eq:sqrtg}) is
\begin{equation}
\sqrt{g} = \frac{1}{\bB\cdot\nabla_\bR\Theta} \, .
\end{equation}
Equation \eq{eq:transportEq_density} is a transport equation for each
lowest-order particle density function $n_\spe$. Note that $\varphi_0$
and $\varphi_1^\lw$ do not appear.

\subsubsection{Transport equation for energy.}
\label{sec:transportEqEnergy}

Now, we do something similar for the total energy. Multiply
\eq{eq:eqH2sigma} by
$(\tau_\spe/\lambda_\spe^2)B(u^2/2 + \mu B)$,
integrate over $u,\mu,\theta$, and take the flux-surface
average. Then,
\begin{eqnarray}\label{eq:transportEq_speEnergy}
  \fl \partial_{\epsilon_s^2 t}
  \left(\frac{3}{2}n_\spe (\psi,t) T_\spe (\psi,t) \right) 
  =
  \nonumber\\
  \fl \hspace{0.5cm}
  \frac{1}{V^\prime(\psi)} \partial_\psi \Bigg\langle
  V^\prime (\psi) \int \left(u^2/2 + \mu B\right)
  \Bigg\{
  \nonumber\\
  \fl \hspace{0.5cm}
    \left[ F_{\sigma 1}^\sw
      (\nabla_{\bR_\perp/\epsilon_\spe} \langle \phi_{\spe 1}^\sw
      \rangle \times \bun) \cdot \nabla_\bR \psi \right]^\lw
  \nonumber\\ \fl \hspace{0.5cm} 
  +\frac{B}{Z_\spe \lambda_\spe} \left \langle
    \left(\frac{Iu}{B} +
      \rhobf \cdot \nabla_\bR \psi \right ) \sum_{\sigma^\prime}
    \cT_{\spe,0}^*C_{\sigma \sigma^\prime}^{(1)\lw} \right \rangle \Bigg\}
  \dd u\dd\mu\dd\theta\Bigg\rangle_\psi
  \nonumber\\
  \fl \hspace{0.5cm} -\Bigg\langle
  \int B \Bigg [ F_{\sigma 1}^\sw
  \Big(  u \bun\cdot\nabla_\bR
  \langle\phi_{\spe
    1}^\sw\rangle
  \nonumber\\
  \fl \hspace{0.5cm}
  + \frac{\mu}{B} (\bun \times
  \nabla_\bR B) \cdot
  \nabla_{\bR_\perp/\epsilon_\spe}\langle\phi_{\spe 1}^\sw\rangle
  \nonumber\\
  \fl \hspace{0.5cm} + 
  \frac{u^2}{B} [\bun \times (\bun \cdot
  \nabla_\bR \bun)] \cdot
  \nabla_{\bR_\perp/\epsilon_\spe}\langle\phi_{\spe 1}^\sw\rangle 
  \Big)  \Bigg]^\lw \dd u \dd \mu \dd \theta \Bigg \rangle_\psi
  \nonumber\\
  \fl \hspace{0.5cm}
+ 
\frac{1}{\lambda_\spe} 
 \partial_\psi\varphi_0 \Bigg
 \langle
 \int B \left ( \frac{I u}{B} + \rhobf \cdot \nabla_\bR \psi \right )
  \sum_{\sigma^\prime} \cT_{\spe,0}^*C_{\sigma \sigma^\prime}^{(1)\lw}
  \dd u\dd\mu\dd\theta\Bigg
\rangle_\psi
  \nonumber\\
  \fl \hspace{0.5cm} +
  \Bigg\langle
  \frac{\tau_\spe}{\lambda_\spe^2}
  \int
  B
  \left(u^2/2+\mu B\right)
  \sum_{\sigma^\prime}\Bigg[ \left[\left\langle
      \cT_{\spe, 1}^* C_{\sigma
        \sigma^\prime}^{(1)\sw}
    \right\rangle
  \right]^\lw
  \nonumber\\
  \fl \hspace{0.5cm} +\left \langle 
    \cT_{\spe,0}^*C_{\sigma \sigma^\prime}^{(2)\lw}  \right
  \rangle
  \Bigg]
  \dd u\dd\mu\dd\theta\Bigg\rangle_\psi,
\end{eqnarray}
which is a transport equation for the energy density of species
$\spe$. The term containing $\varphi_1^\lw$ in \eq{eq:eqH2sigma} does
not contribute to \eq{eq:transportEq_speEnergy} because the collision
operator conserves the total number of particles of each
species. Equation \eq{eq:transportEq_speEnergy} gives an equation for
the temperature of each species. Unless we expand in the mass ratio
$\lambda_e \sim \tau_e \gg 1$, and allow different temperatures for
electrons and ions, this equation still contains the function $G_{\spe
  2}^\lw$ in $C_{\spe\spe^\prime}^{(2)\lw}$ and cannot be considered a
solvability condition. It is possible to prove that for $\lambda_e
\sim \tau_e \gg 1$, the equations for the electron and ion
temperatures do not contain $G_{\spe 2}^\lw$ and consequently are
independent solvability conditions that determine $T_i$ and
$T_e$. However, in general, the only way to eliminate $G_{\spe 2}^\lw$
is summing over all species. We obtain
\begin{eqnarray}\label{eq:transportEq_totalEnergyAux}
  \fl \partial_{\epsilon_s^2 t} \left( \sum_\spe 
\frac{3}{2}n_\spe (\psi,t) T_\spe (\psi,t) \right) 
  =
  \nonumber\\
  \fl \hspace{0.5cm}
  \frac{1}{V^\prime(\psi)} \partial_\psi \Bigg\langle
  V^\prime (\psi) \int \left(u^2/2 + \mu B\right)
  \sum_\spe \Bigg\{
  \nonumber\\
  \fl \hspace{0.5cm}
 \left[ F_{\sigma 1}^\sw
    (\nabla_{\bR_\perp/\epsilon_\spe} \langle \phi_{\spe 1}^\sw
    \rangle \times \bun) \cdot \nabla_\bR \psi \right]^\lw
  \nonumber\\ \fl \hspace{0.5cm}
  + \frac{B}{Z_\spe \lambda_\spe} \left \langle
    \left(\frac{Iu}{B} +
      \rhobf \cdot \nabla_\bR \psi \right ) \sum_{\sigma^\prime}
    \cT_{\spe,0}^*C_{\sigma \sigma^\prime}^{(1)\lw} \right \rangle \Bigg\}
  \dd u\dd\mu\dd\theta\Bigg\rangle_\psi
  \nonumber\\
  \fl \hspace{0.5cm} -\Bigg\langle \sum_\spe
  \int B \Bigg [ F_{\sigma 1}^\sw
  \Big(  u \bun\cdot\nabla_\bR
  \langle\phi_{\spe
    1}^\sw\rangle
  \nonumber\\
  \fl \hspace{0.5cm}
  + \frac{\mu}{B} (\bun \times
  \nabla_\bR B) \cdot
  \nabla_{\bR_\perp/\epsilon_\spe}\langle\phi_{\spe 1}^\sw\rangle
  \nonumber\\
  \fl \hspace{0.5cm} + 
  \frac{u^2}{B} [\bun \times (\bun \cdot
  \nabla_\bR \bun)] \cdot
  \nabla_{\bR_\perp/\epsilon_\spe}\langle\phi_{\spe 1}^\sw\rangle 
  \Big)  \Bigg]^\lw \dd u \dd \mu \dd \theta \Bigg \rangle_\psi
  \nonumber\\
  \fl \hspace{0.5cm} +
  \Bigg\langle
  \sum_{\spe,\spe'} \frac{\tau_\spe}{\lambda_\spe^2}
  \int
  B
  \left(u^2/2+\mu B\right)
   \left[\left\langle
      \cT_{\spe, 1}^* C_{\sigma
        \sigma^\prime}^{(1)\sw}
    \right\rangle
  \right]^\lw
  \dd u\dd\mu\dd\theta\Bigg\rangle_\psi.
\end{eqnarray}
Here we have used the conservation of momentum and energy by the
collision operator to find
\begin{eqnarray} \label{eq:cancellationcollisions}
\fl \sum_{\spe, \spe^\prime} \Bigg \langle \frac{1}{\lambda_\spe} 
 \int B \left ( \frac{I u}{B} + \rhobf \cdot \nabla_\bR \psi \right )
\cT_{\spe,0}^*C_{\sigma \sigma^\prime}^{(1)\lw}
\dd u\dd\mu\dd\theta\Bigg\rangle_\psi = 0,
\nonumber\\
\fl \sum_{\spe, \spe^\prime} \Bigg\langle
\frac{\tau_\spe}{\lambda_\spe^2}
\int
B
\left(u^2/2+\mu B\right)
\left \langle 
\cT_{\spe,0}^*C_{\sigma \sigma^\prime}^{(2)\lw}  \right
\rangle
\dd u\dd\mu\dd\theta\Bigg\rangle_\psi = 0.
\end{eqnarray}
These expressions are easily deduced from the conservation properties
of the collision operator \eq{eq:propcollnondim} by realizing that the
summations of collision operators in \eq{eq:cancellationcollisions}
can always be decomposed in binomials of the form
\begin{eqnarray}
\fl \int B \left ( \frac{I u}{B} + \rhobf \cdot \nabla_\bR \psi \right )
( \cT_{\spe,0}^*C_{\sigma \sigma^\prime} [ \lambda_\spe^{-1} g_\spe, \cT_{\spe^\prime, 0}^{-1*} F_{\spe^\prime 0} ] \nonumber\\+ \cT_{\spe^\prime,0}^*C_{\spe^\prime \spe} [  \cT_{\spe^\prime, 0}^{-1*} F_{\spe^\prime 0}, \lambda_{\spe}^{-1} g_\spe]  )
\dd u\dd\mu\dd\theta = 0,
\nonumber\\
\fl \int B \left ( u^2/2 + \mu B \right )
( \tau_\spe \cT_{\spe,0}^*C_{\sigma \sigma^\prime} [ \lambda_\spe^{-1} g_\spe, \lambda_{\spe^\prime}^{-1} g_{\spe^\prime} ] \nonumber\\+ \tau_{\spe^\prime} \cT_{\spe^\prime,0}^*C_{\spe^\prime \spe} [  \lambda_{\spe^\prime}^{-1} g_{\spe^\prime}, \lambda_\spe^{-1} g_\spe]  )
\dd u\dd\mu\dd\theta = 0,
\nonumber\\
\fl \int B \left ( u^2/2 + \mu B \right )
( \tau_\spe \cT_{\spe,0}^*C_{\sigma \sigma^\prime} [ \lambda_\spe^{-2} g_\spe, \cT_{\spe^\prime, 0}^{-1*} F_{\spe^\prime 0} ] \nonumber\\+ \tau_{\spe^\prime} \cT_{\spe^\prime,0}^*C_{\spe^\prime \spe} [  \cT_{\spe^\prime, 0}^{-1*} F_{\spe^\prime 0}, \lambda_{\spe}^{-2} g_\spe]  )
\dd u\dd\mu\dd\theta = 0,
\end{eqnarray}
where $g_{\spe} (\boldr, \bv, t)$ and $g_{\spe^\prime} (\boldr, \bv,
t)$ are just placeholders for the functions that appear in the
collision operators. We have not written these functions explicitly
because their particular form is unimportant for the cancellations. To
show that the summation of collision operators in
\eq{eq:cancellationcollisions} gives these binomials it is important
to keep track of the factors $\lambda_{\spe}^{-1}$ that multiply the
collision operator arguments.

Equation \eq{eq:transportEq_totalEnergyAux} can be simplified by
further cancellations. \ref{app:cancellationenergy} contains the proof
that
\begin{eqnarray}\label{eq:cancellationenergy}
\fl -\Bigg\langle \sum_\spe
 \int B \Bigg [ F_{\sigma 1}^\sw
 \Big(  u \bun\cdot\nabla_\bR
\langle\phi_{\spe
 1}^\sw\rangle
 \nonumber\\
\fl \hspace{0.5cm}
+ \frac{\mu}{B} (\bun \times
\nabla_\bR B) \cdot
\nabla_{\bR_\perp/\epsilon_\spe}\langle\phi_{\spe 1}^\sw\rangle
 \nonumber\\
\fl \hspace{0.5cm} + 
\frac{u^2}{B} [\bun \times (\bun \cdot
\nabla_\bR \bun)] \cdot
\nabla_{\bR_\perp/\epsilon_\spe}\langle\phi_{\spe 1}^\sw\rangle 
 \Big)  \Bigg]^\lw \dd u \dd \mu \dd \theta \Bigg \rangle_\psi
\nonumber\\
\fl \hspace{0.5cm} +
 \Bigg\langle
\sum_{\spe,\spe'} \frac{\tau_\spe}{\lambda_\spe^2}
\int
B
\left(u^2/2+\mu B\right)
 \left[\left\langle
\cT_{\spe, 1}^* C_{\sigma
\sigma^\prime}^{(1)\sw}
\right\rangle
 \right]^\lw
\dd u\dd\mu\dd\theta\Bigg\rangle_\psi
\nonumber\\
\fl \hspace{0.5cm}
 = O(\epsilon_s),
\end{eqnarray}
and therefore the final expression for the total energy transport
equation is
\begin{eqnarray}\label{eq:transportEq_totalEnergy}
\fl \partial_{\epsilon_s^2 t}
 \left( \sum_\spe \frac{3}{2}n_\spe (\psi,t) T_\spe (\psi,t) \right) 
=
\nonumber\\
\fl \hspace{0.5cm}
\frac{1}{V^\prime(\psi)} \partial_\psi \Bigg\langle
V^\prime (\psi) \int \left(u^2/2 + \mu B\right)
\sum_\spe \Bigg\{
\nonumber\\
\fl \hspace{0.5cm}
\left[ F_{\sigma 1}^\sw
(\nabla_{\bR_\perp/\epsilon_\spe} \langle \phi_{\spe 1}^\sw
\rangle \times \bun) \cdot \nabla_\bR \psi \right]^\lw
 \nonumber\\ \fl \hspace{0.5cm}
+\frac{B}{Z_\spe \lambda_\spe} \left \langle
 \left(\frac{Iu}{B} +
\rhobf \cdot \nabla_\bR \psi \right ) \sum_{\sigma^\prime}
\cT_{\spe,0}^*C_{\sigma \sigma^\prime}^{(1)\lw} \right \rangle \Bigg\}
\dd u\dd\mu\dd\theta\Bigg\rangle_\psi.
\end{eqnarray}

\subsection{Time evolution of the lowest-order quasineutrality condition:
intrinsic ambipolarity of the turbulent tokamak}
\label{sec:timeevolutionquasineutrality}

The zeroth-order piece of the long-wavelength quasineutrality equation
imposes the well-known condition on the lowest order particle
densities, equation \eq{eq:gyroPoissonlw2order0}:
\begin{equation}\label{eq:gyroPoissonlw2order0again}
\sum_\spe Z_\spe n_{\spe}(\psi,t) = 0.
\end{equation}
On the other hand, we have obtained as a solvability condition of the
long-wavelength second-order Fokker-Planck equation a time evolution
equation for each function $n_\spe$,
\eq{eq:transportEq_density}. Thus, we can deduce a time evolution
equation for $\sum_\spe Z_\spe n_\spe$. It is
important to find out whether \eq{eq:gyroPoissonlw2order0again} is
automatically preserved by the time evolution or, on the contrary, its
preservation implies additional constraints on low-order
quantities. In principle, it might have happened that imposing
$\partial_t\sum_\spe Z_\spe n_\spe = 0$ implied a new equation
involving the long-wavelength radial electric field. In this
subsection we show that this is not the case in a tokamak.

The contribution of the last term in \eq{eq:transportEq_density} to
$\partial_t \sum_\spe Z_\spe n_\spe$ vanishes due to the momentum
conservation properties of the collision operator (see
\eq{eq:cancellationcollisions}). As a result, we recover that, in
neoclassical theory, $\partial_t\sum_\spe Z_\spe n_\spe\equiv 0$. The
contribution of the first term on the right side of
\eq{eq:transportEq_density} to $\partial_t \sum_\spe Z_\spe n_\spe$
also vanishes,
\begin{eqnarray} \label{eq:turbulentintrinsicambipolarity}
\sum_\spe Z_\spe \Bigg\langle \int B
\left[ F_{\sigma 1}^\sw (\nabla_{\bR_\perp/\epsilon_\spe}
\langle \phi_{\spe 1}^\sw
\rangle \times \bun) \cdot \nabla_\bR \psi \right]^\lw \dd u \dd\mu\dd\theta \Bigg\rangle_\psi = 0.
\end{eqnarray}
To prove this property we need the short-wavelength quasineutrality
equation to first order in the expansion parameter,
given in
\eq{eq:quasinautralitySWorder1appendix}. We repeat it
here for convenience:
\begin{eqnarray}\label{eq:quasinautralitySWorder1}
\fl\sum_\spe \frac{Z_\spe}{\lambda_\spe}
&\int B
\Bigg[-Z_\spe\lambda_\spe \phiwig_{\spe
      1}^\sw \left(\boldr
-\epsilon_\spe\rhobf(\boldr,\mu,\theta),\mu,\theta,t\right)
\frac{F_{\spe 0}(\boldr,u,\mu,t)}{T_{\spe}(\boldr,t)}\nonumber\\[5pt]
\fl&
 + F_{\spe 1}^\sw
\left(\boldr
-\epsilon_\spe\rhobf(\boldr,\mu,\theta),u,\mu,t
\right)
\Bigg]
  \dd u \dd \mu \dd \theta = 0.
\end{eqnarray}
Acting on \eq{eq:quasinautralitySWorder1} with
$\varphi_1^\sw(\boldr,t)\nabla_{\boldr_\perp/\epsilon_s}$, taking the
coarse-grain average, and observing that
\begin{equation}
  \varphi_1^\sw(\boldr,t) =
  \phi_{\spe 1}^\sw
(\boldr-\epsilon_\spe\rhobf(\boldr,\mu,\theta),\mu,\theta,t)
 + O(\epsilon_\spe),
\end{equation}
we obtain
\begin{eqnarray}\label{eq:quasinautralitySWorder1prime}
\fl
\sum_\spe Z_\spe \int B
\left[
\phi_{\spe 1}^\sw\nabla_{\boldr_\perp/\epsilon_\spe}
\left(-\frac{Z_\spe\lambda_\spe\phiwig_{\spe
      1}^\sw}{T_{\spe}} F_{\spe 0} + F_{\spe 1}^\sw\right)
\right]^\lw
  \dd u \dd \mu \dd \theta = O(\epsilon_\spe).
\end{eqnarray}
In this expression the functions $\phi_{\spe 1}^\sw$ and $F_{\spe
  1}^\sw$ are evaluated at $\bR = \boldr - \epsilon_\spe \rhobf
(\boldr, \mu, \theta)$, but after the coarse grain average we can
Taylor expand and, to lowest order, they can be evaluated at $\bR =
\boldr$. This leads to
\begin{eqnarray}\label{eq:quasinautralitySWorder1_2}
\fl -\sum_\spe Z_\spe \int B
\nabla_{\boldr_\perp/\epsilon_\spe}
\left[
\frac{Z_\spe\lambda_\spe (\phiwig_{\spe
      1}^\sw)^2}{2 T_{\spe}} F_{\spe 0}
\right]^\lw \dd u \dd\mu \dd\theta
\nonumber\\[5pt]
\hspace{1cm}
+ \sum_\spe Z_\spe \int B
\left[
F_{\spe 1}^\sw\nabla_{\boldr_\perp/\epsilon_\spe}
\langle\phi_{\spe 1}\rangle^\sw
\right]^\lw
  \dd u \dd \mu \dd \theta = O(\epsilon_\spe).
\end{eqnarray} 
The fact that
\begin{equation}
\nabla_{\boldr_\perp/\epsilon_\spe} g^\lw = O(\epsilon_\spe)
\end{equation}
whenever $g = O(1)$ implies
\begin{eqnarray}\label{eq:quasinautralitySWorder1_3}
\sum_\spe Z_\spe\int B
\left[
F_{\spe 1}^\sw\nabla_{\boldr_\perp/\epsilon_\spe}\langle\phi_{\spe 1}^\sw\rangle
\right]^\lw
  \dd u \dd \mu \dd \theta = O(\epsilon_\spe),
\end{eqnarray}
whence we immediately infer equation
\eq{eq:turbulentintrinsicambipolarity}, giving
\begin{equation}
\partial_t\sum_\spe Z_\spe n_\spe\equiv 0,
\end{equation}
identically, at the relevant order. Consequently, we have proven that
the well-known neoclassical intrinsic ambipolarity property of the
tokamak still holds in gyrokinetic theory.

\section{Discussion of results and conclusions}
\label{sec:conclusions}

At the moment, the problem of extending the standard set of
gyrokinetic equations, and therefore computer simulations, to
transport time scales is an active research topic. Focusing on
toroidal angular momentum transport in tokamaks in electrostatic
gyrokinetics, the issue has been recently raised by Parra and
Catto~\cite{parra08,parra09b,parra09c,parra10b,parra10c,parra10a};
they argue that calculating momentum transport correctly requires
knowledge of the distribution function and electrostatic potential up
to second order in the expansion parameter, the gyroradius over the
macroscopic length scale. An intimately related result of this series
of works is that in a tokamak the system consisting of second-order
Fokker-Planck and quasineutrality equations does not determine the
long-wavelength radial electric field. A method to correctly compute
the radial transport of toroidal angular momentum (and therefore the
radial electric field) when the second-order pieces of the
distribution function and the electrostatic potential are known is
given in reference \cite{parra10a, parra11d}.

Using the recent derivation of the second-order
  gyrokinetic equations~\cite {ParraCalvo2011} in general magnetic
  geometry, we have worked out the long-wavelength limit of the
  Fokker-Planck and quasineutrality equations in a tokamak, a
  necessary first step towards the formulation of a set of equations
  to compute the radial transport of toroidal angular momentum without
  having to resort to subsidiary expansions such as the expansion in
  $B_p/B\ll 1$ of references \cite{parra10a, parra11d}. Specifically, we have
  obtained (see the main text for notation and details):
\begin{enumerate}
\item[(i)] The long-wavelength Fokker-Planck equations to second
  order, \eq{eq:Vlasovorder1gyroav4} and \eq{eq:eqH2sigma}, that give
  $G_{\spe 1}^\lw$ and $G_{\spe 2}^\lw$, and therefore the
  long-wavelength component of the distribution functions.
\item[(ii)] The quasineutrality equation up to second-order
  \eq{eq:gyroPoissonlw2order0}, \eq
  {eq:quasineutralityWithCorrectionsOrder1}, and
  \eq{eq:quasineutralityWithCorrectionsOrder2}, that determines the
  first and second-order pieces of the long-wavelength poloidal
  electric field. Equivalently, and under conditions
  \eq{eq:fixvarphi1} and \eq {eq:fixvarphi2}, the quasineutrality equation
  determines $\varphi_1^\lw$ and $\varphi_2^\lw$.
\item[(iii)] Transport equations for density
  \eq{eq:transportEq_density} and energy
  \eq{eq:transportEq_totalEnergy}.
\item[(iv)]  Equations \eq {eq:sworder1distfunction} and
  \eq{eq:quasinautralitySWorder1appendix}, that give the
  short-wavelength component of the distribution functions, $F_{\spe
    1}^\sw$, and electrostatic potential, $\phi_{\spe 1}^\sw$. They
  are needed because they enter equation \eq{eq:eqH2sigma}.
\end{enumerate}
In order to provide a model for toroidal angular momentum
transport in tokamaks one still needs to derive explicit
equations for the short-wavelength components of the distribution
functions and electrostatic potential to second order. This will be
the subject of a future publication.

In addition, in this paper, we have given a complete
  proof that the long-wavelength tokamak radial electric field cannot
  be determined by simply using Fokker-Planck and quasineutrality
  equations accurate to second order in the gyrokinetic expansion
  parameter. In other words, we have proven that gyrokinetics does not
  spoil the well-known neoclassical intrinsic ambipolarity property of
  the tokamak.


  \ack This research was supported in part by grant ENE2009-07247,
  Ministerio de Ciencia e Innovaci\'on (Spain), and by US DoE grant
  DE-SC008435.

\appendix

\section{Gyrokinetic equations of motion}
\label{sec:eqsofmotion}

Here, the gyrokinetic equations of motion corresponding
to the Poisson bracket \eq{eq:poissonbracket} and the gyrokinetic
Hamiltonian given in equations~\eq{eq:Hgyro0}, \eq{eq:Hgyro1}, and
\eq{eq:Hgyro2}, are explicitly written:
\begin{eqnarray} \label{eq:dRdt}
\fl 
\frac{\dd\bR}{\dd t}&= 
\{\bR ,\overline{H}_\spe\}_\bZ  = 
\nonumber \\
\fl &\left ( u + Z_\spe\lambda_\spe
\epsilon^2_\spe \partial_u\Psi_{\phi B,\spe} +
\epsilon^2_\spe \partial_u\Psi_{B,\phi} \right )
\frac{\bB_\spe^*}{B_{||,\spe}^*}
\nonumber \\
\fl &
+ \frac{1}{B_{||,\spe}^*} \bun \times \Bigg (
\epsilon_\spe \mu \nabla_\bR B 
+ Z_\spe\lambda_\spe \epsilon_\spe
\nabla_{\bR_\perp/ \epsilon_\spe}
 \langle\phi_\spe\rangle 
 \nonumber \\
\fl &
 +
{Z_\spe^2\lambda_\spe^2 \epsilon^2_\spe}
 \nabla_{\bR_\bot/
\epsilon_\spe} \Psi_{\phi,\spe}
 + Z_\spe\lambda_\spe \epsilon^2_\spe
\nabla_{\bR_\bot/ \epsilon_\spe}
 \Psi_{\phi B,\spe}\nonumber \\ \fl &
 +
Z_\spe^2\lambda_\spe^2 \epsilon_\spe^3 \nabla_\bR \Psi_{\phi,\spe}
+ Z_\spe\lambda_\spe \epsilon_\spe^3 \nabla_\bR \Psi_{\phi B,\spe}
 + \epsilon_\spe^3 \nabla_\bR \Psi_{B,\spe} \Bigg ),
\end{eqnarray}
\begin{eqnarray} \label{eq:dudt}
\fl
\frac{\dd u}{\dd t}&=
\{u ,\overline{H}_\spe\}_\bZ =
\nonumber \\
\fl &
 - \frac{\mu}{B^*_{||,\spe}} \bB_\spe^*
\cdot \nabla_\bR B - Z_\spe\lambda_\spe \epsilon_\spe \bun \cdot \nabla_\bR
\langle\phi_\spe\rangle  \nonumber \\
\fl &
- Z_\spe^2\lambda_\spe^2 \epsilon_\spe^2
 \bun \cdot \nabla_\bR
\Psi_{\phi,\spe}
- Z_\spe\lambda_\spe \epsilon_\spe^2
 \bun \cdot \nabla_\bR
\Psi_{\phi B,_\spe}
 \nonumber \\[3pt]
\fl &
- \epsilon_\spe^2 \bun \cdot
\nabla_\bR \Psi_{B,\spe}
 - \frac{1}{B_{||,\spe}^*} \Big[ u \bun \times (\bun
\cdot \nabla_\bR \bun)
 \nonumber \\[3pt]
\fl &
 - \epsilon_\spe \mu (\nabla_\bR \times \bK)_\bot
\Big] \cdot \Bigg ( Z_\spe\lambda_\spe \epsilon_\spe
\nabla_{\bR_\bot/ \epsilon_\spe} \langle\phi_\spe\rangle
 \nonumber \\[3pt]
\fl & +
Z_\spe^2\lambda_\spe^2 \epsilon_\spe^2
 \nabla_{\bR_\bot/
\epsilon_\spe} \Psi_{\phi,\spe}
 + Z_\spe\lambda_\spe \epsilon_\spe^2
\nabla_{\bR_\bot/\epsilon_\spe} \Psi_{\phi B,_\spe}  \nonumber \\
\fl &
+
Z_\spe^2\lambda_\spe^2 \epsilon_\spe^3 \nabla_\bR \Psi_{\phi,\spe}
+ Z_\spe\lambda_\spe
\epsilon_\spe^3 \nabla_\bR \Psi_{\phi B,\spe}
 +
\epsilon_\spe^3 \nabla_\bR \Psi_{B,\spe} \Bigg) ,
\end{eqnarray}
\begin{eqnarray} \label{eq:dmudt}
\fl \frac{\dd \mu}{\dd t} = \{\mu,\overline{H}_\spe\}_\bZ = 0,
\end{eqnarray}
\begin{eqnarray} \label{eq:dthetadt}
\fl
\frac{\dd \theta}{\dd t}&=
\{\theta ,\overline{H}_\spe\}_\bZ  =
 \nonumber \\
\fl &
 -
\frac{1}{\epsilon_\spe} B - Z_\spe\lambda_\spe
\partial_\mu\langle\phi_\spe\rangle
 - Z_\spe^2\lambda_\spe^2
\epsilon_\spe \partial_\mu\Psi_{\phi,\spe}
 \nonumber \\
\fl &
- 
Z_\spe\lambda_\spe\epsilon_\spe \partial_\mu\Psi_{\phi B,\spe}
 - \epsilon_\spe \partial_\mu\Psi_{B,\spe} 
 \nonumber \\
\fl &
 - \frac{\bB_\spe^* \cdot
\bK}{B_{||,\spe}^*} \Bigg( u + Z_\spe\lambda_\spe \epsilon_\spe^2
\partial_u\Psi_{\phi B,\spe}
+ \epsilon_\spe^2 \partial_u\Psi_{B,\spe} \Bigg)
 \nonumber \\[3pt]
\fl &
- \frac{1}{B_{||,\spe}^*} (\bK \times
\bun) \cdot \Big ( \epsilon_\spe \mu \nabla_\bR B 
 +
Z_\spe\lambda_\spe \epsilon_\spe
\nabla_{\bR_\perp/
\epsilon_\spe} \langle\phi_\spe\rangle
 \nonumber \\[3pt]
\fl &
 + Z_\spe^2\lambda_\spe^2 \epsilon_\spe^2
\nabla_{\bR_\bot/\epsilon_\spe} \Psi_{\phi,\spe} 
+
Z_\spe\lambda_\spe \epsilon_\spe^2
 \nabla_{\bR_\bot/\epsilon_\spe}
 \Psi_{\phi B,\spe}
 \nonumber \\[3pt]
\fl &
+ Z_\spe^2\lambda_\spe^2 \epsilon_\spe^3 \nabla_\bR
\Psi_{\phi,\spe}
  + Z_\spe\lambda_\spe \epsilon_\spe^3 \nabla_\bR
\Psi_{\phi B,\spe} + \epsilon_\spe^3 \nabla_\bR \Psi_{B,\spe} \Big ).
\end{eqnarray}

\section{Some basic properties of the collision operator}
\label{sec:collisionoperator}

We recall (see, for example, reference \cite{Lenard1960}) that the
collision operator \eq{eq:collisionoperator} satisfies, for every
$\spe,\spe'$, the conservation properties
\begin{eqnarray}\label{eq:propcoll}
\int C_{\spe\spe'}\dd^3 v = 0
\nonumber\\[5pt]
\int m_\spe\bv C_{\spe\spe'}\dd^3 v = -\int m_{\spe'}\bv C_{\spe'\spe}\dd^3 v
\nonumber\\[5pt]
\int \frac{1}{2}m_\spe\bv^2 C_{\spe\spe'}\dd^3 v
= -\int \frac{1}{2} m_{\spe'}\bv^2 C_{\spe'\spe}\dd^3 v,
\end{eqnarray}
and the statistical equilibrium condition
\begin{equation}\label{Maxwellianannihilatecoll}
C_{\spe\spe'}[f_{M\spe},f_{M\spe'}] = 0
\end{equation}
when both distribution functions are Maxwellian,
\begin{eqnarray}\label{eq:maxwelliancoll}
&&{f_{M\spe}}({\boldr},{\bv})
=n_{\spe}(\boldr)
\left(
\frac{m_\spe}{2\pi T_{\spe}(\boldr)}
\right)^{3/2}
\exp\left(-\frac{m_\spe {\bv}^2}{2T_{\spe}
(\boldr)}\right),\nonumber\\[5pt]
&&{f_{M\spe'}}({\boldr},{\bv})
=n_{\spe'}(\boldr)
\left(
\frac{m_{\spe'}}{2\pi T_{\spe'}(\boldr)}
\right)^{3/2}
\exp\left(-\frac{m_{\spe'}\bv^2}{2T_{\spe'}(\boldr)}\right),
\end{eqnarray}
with $T_\spe(\boldr) = T_{\spe'}(\boldr)$ at every point. These are the
only solutions to the equations \eq{Maxwellianannihilatecoll}. The
easiest way to see this is noting that the entropy production,
\begin{equation}
-\sum_{\spe,\spe'}
\int
\ln f_\spe
C_{\spe\spe'}[f_\spe,f_{\spe'}]
\dd^3v  \, ,
\end{equation}
is non-negative and vanishes only when $f_\spe$ and $f_{\spe'}$ are
Maxwellians with the same temperature.

Another well-known property, derived from
\eq{Maxwellianannihilatecoll}, is
\begin{equation}\label{eq:flowcoll}
C_{\spe\spe'}\left[\frac{m_\spe}{T_\spe}\bv f_{M\sigma},f_{M\sigma'}\right]
+C_{\spe\spe'}\left[f_{M\sigma},\frac{m_{\spe'}}{T_{\spe'}}\bv f_{M\sigma'}\right]
\equiv 0.
\end{equation}
This property implies that displacing both Maxwellians by the same
average velocity gives another solution of \eq{Maxwellianannihilatecoll}.

It is useful to have the explicit translation of these properties into
our non-dimensional variables. With the definition
\eq{eq:collisionoperatornondim} we have
\begin{eqnarray}\label{eq:propcollnondim}
\int \underline{C_{\spe\spe'}}\dd^3 \underline{v} = 0
\nonumber\\[5pt]
\int  \underline{\bv}\, \underline{C_{\spe\spe'}}\,\dd^3
\underline{v} = -\int \underline{\bv} \,
\underline{C_{\spe'\spe}}\,\dd^3 \underline{v}
\nonumber\\[5pt]
\int\frac{1}{2}\,\tau_\spe \, \underline{\bv}^2 \,
\underline{C_{\spe\spe'}} \, \dd^3 \underline{v} =
-\int \frac{1}{2}\, \tau_{\spe'} \, \underline{\bv}^2 \,
\underline{C_{\spe'\spe}} \, \dd^3 \underline{v}.
\end{eqnarray}
Also,
\begin{equation}\label{eq:Maxwelliansannihilatecoll_nondim}
\underline{C_{\spe\spe'}} [\underline{f_{M\spe}},\underline{f_{M\spe'}}]
\equiv
0
\end{equation}
when
\begin{eqnarray}\label{eq:maxwelliancollnondim}
&&\underline{f_{M\spe}}(\underline{\boldr},\underline{\bv})
=\frac{\underline{n_{\spe}}(\underline{\boldr})}
{(2\pi \underline{T_{\spe}}(\underline{\boldr}))^{3/2}}
\exp\left(-\frac{\underline{\bv}^2}{2\underline{T_{\spe}}
(\underline{\boldr})}\right),\nonumber\\[5pt]
&&\underline{f_{M\spe'}}(\underline{\boldr},\underline{\bv})
=\frac{\underline{n_{\spe'}}(\underline{\boldr})}
{(2\pi \underline{T_{\spe'}}(\underline{\boldr}))^{3/2}}
\exp\left(-\frac{\underline{\bv}^2}{2\underline{T_{\spe'}}
(\underline{\boldr})}\right),
\end{eqnarray}
and $\underline{T_{\spe}}(\underline{\boldr})= 
\underline{T_{\spe'}} (\underline{\boldr})$ at every
point. Finally,
\begin{equation}\label{eq:flowcollnondim}
\underline{C_{\spe\spe'}}
\left[\frac{1}{\tau_{\spe}\underline{T_\spe}}\underline{\bv}\
 \underline{f_{M\sigma}},
\underline{f_{M\sigma'}}\right]
+\underline{C_{\spe\spe'}}\left[\underline{f_{M\sigma}},
\frac{1}{\tau_{\spe'}\underline{T_{\spe'}}}\underline{\bv}\
 \underline{f_{M\sigma'}}\right]
\equiv 0.
\end{equation}

\section{Gyrokinetic transformation to first order}
\label{sec:pullback}

In this appendix we provide explicit expressions for the gyrokinetic
transformation $(\boldr,\bv)=\cT_\spe(\bR,u,\mu,\theta,t)$ to
order $\epsilon_\spe$. Define
\begin{eqnarray}\label{eq:defcoordinates0}
v_{||}&:=\bv\cdot\bun(\boldr),\\[5pt]
\mu_0&:=\frac{(\bv-v_{||}\bun(\boldr))^2}{2B(\boldr)},\\[5pt]
\theta_0&:=\arctan
\left(
\frac{\bv\cdot\eun_2(\boldr)}{\bv\cdot\eun_1(\boldr)}
\right),
\end{eqnarray}
and let us compute $(\boldr,v_{||},\mu_0,\theta_0)$
as a function of $(\bR,u,\mu,\theta)$ to first order in
$\epsilon_\spe$. From the definition \eq{eq:changeNonPert} we find
$(\boldr,v_{||},\mu_0,\theta_0)$ as a function of
$(\bR_g,v_{||g},\mu_g,\theta_g)$:
\begin{eqnarray} \label{0gtransformation}
\fl\boldr = \bR_g + \epsilon_\spe\rhobf_g,\nonumber\\[5pt]
\fl
v_{||} = v_{||g}
+ \epsilon_\spe B_g(\rhobf_g\times\bun_g)\rhobf_g:\nabla_{\bR_g}\bun_g
+ O(\epsilon_\spe^2),\nonumber\\[5pt]
\fl\mu_0 = \mu_g - \epsilon_\spe\left(
\frac{\mu_g}{B_g}\rhobf_g\cdot\nabla_{\bR_g}B_g +
v_{||g}(\rhobf_g\times\bun_g)\rhobf_g:\nabla_{\bR_g}\bun_g
\right)
\nonumber\\[5pt]
\fl\hspace{1cm}
+ O(\epsilon_\spe^2),
\nonumber\\[5pt]
\fl\theta_0 = \theta_g +\epsilon_\spe\left(
\rhobf_g\cdot\nabla_{\bR_g}\eun_{2g}\cdot\eun_{1g}
-\frac{v_{||g}}{2\mu_g}\rhobf_g\cdot
\nabla_{\bR_g}\bun_g\cdot\rhobf_g
\right)
\nonumber\\[5pt]
\fl\hspace{1cm}+ O(\epsilon_\spe^2),
\end{eqnarray}
where a subindex $g$ stresses that the quantity is evaluated at
$(\bR_g,v_{||g},\mu_g,\theta_g)$. Using \eq{newvar}, \eq{u1},
\eq{mu1}, \eq{theta1}, and the identities
\begin{eqnarray}
(\rhobf\times\bun)\rhobf
:\nabla_{\bR}\bun = \rhobf(\rhobf\times\bun)
:\nabla_{\bR}\bun - \frac{2\mu}{B}\bun\cdot\nabla_\bR\times\bun,
\nonumber\\[5pt]
\rhobf\rhobf + (\rhobf\times\bun)(\rhobf\times\bun)
= \frac{2\mu}{B}(\matI-\bun\bun),
\end{eqnarray}
we arrive at
\begin{eqnarray}\label{eq:totalchangecoorfirstorder}
\fl \boldr &= \bR + \epsilon_\spe\rhobf
+ O(\epsilon_\spe^2)
,
\nonumber\\[5pt]
\fl v_{||} &= u
+ \epsilon_\spe\hat{u}_1
+ O(\epsilon_\spe^2),\nonumber\\[5pt]
\fl \mu_0 &= \mu + \epsilon_\spe\hat{\mu}_1
+ O(\epsilon_\spe^2),\nonumber\\[5pt]
\fl \theta_0 &= \theta
+\epsilon_\spe\hat{\theta}_1
+ O(\epsilon_\spe^2),
\end{eqnarray}
where
\begin{eqnarray}\label{eq:totalchangecoorfirstorderCorrections}
\fl \hat{u}_1 &= 
u\bun\cdot\nabla_\bR\bun\cdot\rhobf+
\frac{B}{4}[\rhobf(\rhobf\times\bun)
+(\rhobf\times\bun)\rhobf]
:\nabla_{\bR}\bun
\nonumber\\[5pt]
\fl&-\mu\bun\cdot\nabla_\bR\times\bun,
\nonumber\\[5pt]
\fl\hat{\mu}_1 &=
-\frac{\mu}{B}\rhobf\cdot\nabla_{\bR}B
-\frac{u}{4}
\left(
\rhobf(\rhobf\times\bun)
+(\rhobf\times\bun)\rhobf
\right):\nabla_{\bR}\bun
\nonumber\\[5pt]
\fl&+\frac{u\mu}{B}\bun\cdot\nabla_\bR\times\bun
-\frac{u^2}{B}\bun\cdot\nabla_\bR\bun\cdot\rhobf
-\frac{Z_\spe\lambda_\spe}{B}\tilde\phi_{\spe 1},
\nonumber\\[5pt]
\fl\hat{\theta}_1&= 
(\rhobf\times\bun)
\cdot
\Bigg(
\nabla_\bR\ln B + \frac{u^2}{2\mu B}\bun\cdot\nabla_\bR\bun
\nonumber\\[5pt]
\fl&-\bun\times\nabla_\bR\eun_2\cdot\eun_1
\Bigg)
-\frac{u}{8\mu}
\left(
\rhobf\rhobf - (\rhobf\times\bun)(\rhobf\times\bun)
\right):\nabla_\bR\bun
\nonumber\\[5pt]
\fl&
+\frac{u}{2B^2}\bun\cdot\nabla_\bR B
+\frac{Z_\spe\lambda_\spe}{B}\partial_\mu\tilde\Phi_{\spe 1}
.
\end{eqnarray}
It is useful to have the long-wavelength limit of the previous
expressions at hand. Employing
\eq{eq:potentiallw2}, and \eq{eq:potentiallw3} we get:
\begin{eqnarray}\label{eq:totalchangecoorfirstorderCorrectionsLW}
\fl \hat{u}_1^\lw &= \hat{u}_1
\nonumber\\[5pt]
\fl\hat{\mu}_1^\lw &=
-\frac{\mu}{B}\rhobf\cdot\nabla_{\bR}B
-\frac{u}{4}
\left(
\rhobf(\rhobf\times\bun)
+(\rhobf\times\bun)\rhobf
\right):\nabla_{\bR}\bun
\nonumber\\[5pt]
\fl&+\frac{u\mu}{B}\bun\cdot\nabla_\bR\times\bun
-\frac{u^2}{B}\bun\cdot\nabla_\bR\bun\cdot\rhobf
-\frac{Z_\spe}{B}\rhobf\cdot\nabla_\bR\varphi_0,
\nonumber\\[5pt]
\fl\hat{\theta}_1^\lw&= 
(\rhobf\times\bun)
\cdot
\Bigg(
\nabla_\bR\ln B + \frac{u^2}{2\mu B}\bun\cdot\nabla_\bR\bun
\nonumber\\[5pt]
\fl&-\bun\times\nabla_\bR\eun_2\cdot\eun_1
\Bigg)
-\frac{u}{8\mu}
\left(
\rhobf\rhobf - (\rhobf\times\bun)(\rhobf\times\bun)
\right):\nabla_\bR\bun
\nonumber\\[5pt]
\fl&
+\frac{u}{2B^2}\bun\cdot\nabla_\bR B
+\frac{Z_\spe}{2\mu B}\left(\rhobf\times\bun\right)\cdot\nabla_\bR\varphi_0
.
\end{eqnarray}

Next, we proceed to calculate the long-wavelength limit of
$\cT_\spe^{-1*}F_{\spe 0}$ to first order in $\epsilon_\spe$, needed
to write \eq{eq:Vlasovorder1} in Section
\ref{sec:FokkerPlancklong1}. Inverting
\eq{eq:totalchangecoorfirstorder} to first order, and recalling
\eq{eq:totalchangecoorfirstorderCorrectionsLW} and the relations
$\partial_u F_{\spe0 } = -(u/T_\spe) F_{\spe0 }$, $\partial_\mu
F_{\spe0 } = -(B/T_\spe) F_{\spe0 }$, one finds that
\begin{eqnarray}\label{eq:pullback_order1_Maxwellian}
\fl \left[\cT^{-1*}_\spe F_{\spe 0}\right]^\lw &=& \cT^{-1*}_{\spe,0}F_{\spe 0}
+
\frac{\epsilon_\spe}{ T_{\spe}}
\Bigg[
\bv\cdot{\mathbf V}^p_\spe +
\left(
\frac{v^2}{2T_{\spe}}-\frac{5}{2}
\right)\bv\cdot{\mathbf V}^T_\spe
\nonumber\\[5pt]
\fl&+&
\frac{Z_\spe}{B}
\bv\cdot(\bun\times\nabla_\boldr\varphi_0)
\Bigg]\cT^{-1*}_{\spe,0}F_{\spe 0},
\end{eqnarray}
with ${\mathbf V}^p_\spe$ and ${\mathbf V}^T_\spe$ defined in
\eq{eq:velocitiesPandT}.

\section{Calculations for the Fokker-Planck equation to 
$O(\epsilon_\spe)$}
\label{sec:calculationsFPorder1}

In what follows we detail the calculations that recast
\eq{eq:Vlasovorder1gyroav} into \eq{eq:Vlasovorder1gyroav4} when the
magnetic field has the form \eq{eq:tokamakB}. First, rewrite
\eq{eq:Vlasovorder1gyroav} as
\begin{eqnarray}\label{eq:Vlasovorder1gyroavappendixaux}
\fl
\left(
 u\bun\cdot\nabla_\bR
- \mu\bun\cdot\nabla_\bR B\partial_u
\right)
 F_{\spe 1}^\lw
  \nonumber\\[5pt]
\fl\hspace{1cm}+
 \left(
\bv_\kappa + \bv_{\nabla B} + \bv_{E,\spe}^{(0)}
\right)\cdot
\Big(\nabla_\bR F_{\spe 0} + \frac{\mu F_{\spe 0}}{T_\spe}\nabla_\bR B
\nonumber\\[5pt]
\fl\hspace{1cm}
+\frac{Z_\spe F_{\spe 0}}{T_\spe}\nabla_\bR\varphi_0\Big)
+\frac{Z_\spe\lambda_\spe}{T_\spe} u\bun\cdot\nabla_\bR\varphi_1^\lw F_{\spe0}
\nonumber\\[5pt]
\fl\hspace{1cm}
=
\sum_{\sigma'}\cT^{*}_{\spe,0}
C_{\spe\spe'}\left[ \cT^{-1*}_{\spe,0}F_{\spe 1}^\lw
,\cT^{-1*}_{\spe',0}F_{\spe' 0} \right]
\nonumber\\[5pt]
\fl\hspace{1cm}
 + \sum_{\sigma'}
\frac{\lambda_\spe}{\lambda_{\spe'}}\cT^{*}_{\spe,0}
C_{\spe\spe'} \left[ \cT^{-1*}_{\spe,0}F_{\spe 0},
\cT^{-1*}_{\spe',0}F_{\spe' 1}^\lw \right].
\end{eqnarray}

Denote by $R$ the
cylindrical coordinate giving the distance to the axis of the
torus, and by $\zun$ the unit vector in the toroidal direction. The identities
\begin{eqnarray}
\fl B^2 = \frac{I^2+|\nabla_\bR\psi|^2}{R^2},\\[5pt]
\fl\bun\times\nabla_\bR\psi = I\bun - RB\zun,\\[5pt]
\fl\nabla_\bR\cdot\zun = 0,\\[5pt]
\fl[\bun\times(\bun\cdot\nabla_\bR\bun)]\cdot\nabla_\bR\psi
=\nonumber\\
\fl\hspace{0.5cm}
\left(\nabla_\bR\times\bun\right)\cdot\nabla_\bR\psi
=\nabla_\bR\cdot(I\bun) = \bB\cdot\nabla_\bR\left(\frac{I}{B}\right),\\[5pt]
\fl\bun\times\nabla_\bR\varphi_0 =
\partial_\psi\varphi_0(I\bun-RB\zun),
\end{eqnarray}
and
\begin{eqnarray}
\fl (\bun\times\nabla_\bR B)\cdot\nabla_\bR\psi = -I\bun\cdot\nabla_\bR B
\end{eqnarray}
are useful to write \eq{eq:Vlasovorder1gyroavappendixaux} as
\begin{eqnarray}\label{eq:Vlasovorder1gyroav2}
\fl
\left(
u\bun\cdot\nabla_\bR
- \mu\bun\cdot\nabla_\bR B\partial_u
\right) F_{\spe 1}^\lw
\nonumber\\[5pt]
\fl
\hspace{1cm}
\left(
u^2\bun\cdot\nabla_\bR\left(\frac{I}{B}\right)
-\frac{I \mu}{B}\bun\cdot\nabla_\bR B
\right)
\Bigg(
\Upsilon_\spe 
\nonumber\\[5pt]
\fl
\hspace{1cm}
+\frac{Z_\spe}{T_\spe}
\partial_\psi\varphi_0
\Bigg) F_{\spe 0}
+\frac{Z_\spe\lambda_\spe u}{T_\spe}\bun\cdot\nabla_\bR\varphi_1^\lw F_{\spe 0}
\nonumber\\[5pt]
\fl
\hspace{1cm}
=
\sum_{\sigma'}\cT_{\spe,0}^* C_{\spe\spe'}\left[
\cT^{-1*}_{\spe,0}F_{\spe 1}^\lw
,\cT^{-1*}_{\spe',0}F_{\spe' 0}
\right]\nonumber\\[5pt]
\fl\hspace{1cm}
+
\sum_{\sigma'}
\frac{\lambda_\spe}{\lambda_{\spe'}}\cT_{\spe,0}^* C_{\spe\spe'}
\left[
\cT^{-1*}_{\spe,0}F_{\spe 0},
\cT^{-1*}_{\spe',0}F_{\spe' 1}^\lw
\right],
\end{eqnarray}
where $\Upsilon_\spe$ is defined in \eq{eq:defUpsilon}.
Equation \eq{eq:Vlasovorder1gyroav2} is equivalent to
\begin{eqnarray}\label{eq:Vlasovorder1gyroav3}
\fl (u\bun\cdot\nabla_\bR
- \mu\bun\cdot\nabla_\bR B\partial_u)
 \Bigg[
F_{\spe 1}^\lw
\nonumber\\[5pt]
\hspace{1cm}
\fl +
\Bigg\{
\frac{Z_\spe\lambda_\spe}{T_\spe}\varphi_1^\lw
+\frac{Iu}{B}
\left(\frac{Z_\spe}{T_\spe}\partial_\psi\varphi_0+\Upsilon_\spe\right)
\Bigg\}F_{\spe 0}
\Bigg]
\nonumber\\[5pt]
\hspace{1cm}
\fl=
\sum_{\sigma'}\cT_{\spe,0}^* C_{\spe\spe'}\left[
\cT^{-1*}_{\spe,0}F_{\spe 1}^\lw
,\cT^{-1*}_{\spe',0}F_{\spe' 0}
\right]
\nonumber\\[5pt]
\fl
\hspace{1cm}
+
\sum_{\sigma'}
\frac{\lambda_\spe}{\lambda_{\spe'}}\cT_{\spe,0}^* C_{\spe\spe'}
\left[
\cT^{-1*}_{\spe,0}F_{\spe 0},
\cT^{-1*}_{\spe',0}F_{\spe' 1}^\lw
\right].
\end{eqnarray}

The definition of the new function $G_{\spe 1}^\lw$ given in
\eq{eq:defGspe1} seems appropriate. Employing that the collision
operator vanishes when acting on Maxwellians with the same
temperature, and using property \eq{eq:flowcollnondim}, the dependence
on $\varphi_0$ and $\varphi_1^\lw$ is removed and
equation~\eq{eq:Vlasovorder1gyroav4} is obtained.

\section{Computation of the turbulent piece of the collision operator}
\label{sec:Cturb}

We have to calculate $[
\cT_{\spe,1}^*C_{\spe\spe'}^\sw]^\lw$ appearing in
\eq{eq:T1C1lw} and this appendix is devoted to that end. Then,
\begin{eqnarray}\label{eq:turbpiececoll}
\fl\left[ \cT_{\spe,1}^*C_{\spe\spe'}^\sw
\right]^\lw=
\nonumber\\[5pt]
\fl\hspace{0.5cm}
\Bigg[
\frac{Z_\spe\lambda_\spe}{B}
\Bigg(-\frac{1}{B}\left(\bun\times
\nabla_{\bR_\perp/\epsilon_\spe}\tilde\Phi_\spe^\sw\right)
\cdot\nabla_{\bR_\perp/\epsilon_\spe}
-\tilde\phi_{\spe 1}^\sw\partial_\mu
+\partial_\mu\tilde\Phi_{\spe 1}^\sw\partial_\theta
\Bigg)
\nonumber\\[5pt]
\fl\hspace{0.5cm}
\Bigg\{
\cT_{NP, \spe}^* C_{\sigma \sigma^\prime}
\left [\modTinv F_{\sigma 1}^\sw -
\frac{Z_\spe\lambda_\spe}{T_\spe}
\modTinv\tilde\phi_{\spe 1}^\sw \cT_{\spe,0}^{-1
*} F_{\spe 0}, \cT_{\spe',0}^{-1
*}F_{\spe' 0}
 \right ]\nonumber\\[5pt]
\fl\hspace{0.5cm}
+ \
\frac{\lambda_\spe}{\lambda_{\spe'}}
\cT_{NP, \spe}^*
C_{\sigma
\sigma^\prime} \Bigg[ \cT_{\spe,0}^{-1 *}F_{\spe 0} ,
\modTinvprime F_{\sigma' 1}^\sw
\nonumber\\[5pt]
\fl\hspace{0.5cm}
 -
\frac{Z_{\spe'}\lambda_{\spe'}}{T_{\spe'}}
\modTinvprime\tilde\phi_{\spe'
1}^\sw \cT_{\spe',0}^{-1 *}F_{\spe' 0}
 \Bigg]
\Bigg\}
\Bigg]^\lw.
\end{eqnarray}
But the first term on the right side of \eq{eq:turbpiececoll} does not
contribute in the long-wavelength limit because, for any $g = O(1)$,
\begin{eqnarray}
\fl\left[\left(\bun\times
\nabla_{\bR_\perp/\epsilon_\spe}\tilde\Phi_\spe^\sw\right)
\cdot\nabla_{\bR_\perp/\epsilon_\spe} g^\sw\right]^\lw =
\nonumber\\[5pt]
\fl\hspace{1cm}
-
\bun\cdot \nabla_{\bR_\perp/\epsilon_\spe}\times
\left[
g^\sw \nabla_{\bR_\perp/\epsilon_\spe}\tilde\Phi_\spe^\sw
\right]^\lw
=
 O(\epsilon_s).
\end{eqnarray}
Therefore,
\begin{eqnarray}\label{eq:turbpiececoll2}
\fl\left[ \cT_{\spe,1}^*C_{\spe\spe'}^\sw
\right]^\lw=
\Bigg[
\frac{Z_\spe\lambda_\spe}{B}
\Bigg(
-\tilde\phi_{\spe 1}^\sw\partial_\mu
+\partial_\mu\tilde\Phi_{\spe 1}^\sw\partial_\theta
\Bigg)
\nonumber\\[5pt]
\fl\hspace{0.5cm}
\Bigg\{
\cT_{NP, \spe}^* C_{\sigma \sigma^\prime}
\left [\modTinv F_{\sigma 1}^\sw -
\frac{Z_\spe\lambda_\spe}{T_\spe}
\modTinv\tilde\phi_{\spe 1}^\sw \cT_{\spe,0}^{-1
*} F_{\spe 0}, \cT_{\spe',0}^{-1
*}F_{\spe' 0}
 \right ]\nonumber\\[5pt]
\fl\hspace{0.5cm} + \
\frac{\lambda_\spe}{\lambda_{\spe'}}
\cT_{NP, \spe}^*
C_{\sigma
\sigma^\prime} \Bigg[ \cT_{\spe,0}^{-1 *}F_{\spe 0} ,
\modTinvprime F_{\sigma' 1}^\sw
\nonumber\\[5pt]
\fl\hspace{0.5cm}
 -
\frac{Z_{\spe'}\lambda_{\spe'}}{T_{\spe'}}
\modTinvprime\tilde\phi_{\spe'
1}^\sw \cT_{\spe',0}^{-1 *}F_{\spe' 0}
 \Bigg]
\Bigg\}
\Bigg]^\lw.
\end{eqnarray}
As for its gyroaverage,
\begin{eqnarray}\label{eq:turbpiececollgyroave}
\fl
\Big\langle
\Big[ &\cT_{\spe,1}^*C_{\spe\spe'}^\sw
\Big]^\lw
\Big\rangle
=-
\partial_\mu
\Bigg\langle\Bigg[
\frac{Z_\spe\lambda_\spe}{B}
\tilde\phi_{\spe 1}^\sw
\nonumber\\[5pt]
\fl&
\Bigg\{
\cT_{NP, \spe}^* C_{\sigma \sigma^\prime}
\left [\modTinv F_{\sigma 1}^\sw -
\frac{Z_\spe\lambda_\spe}{T_\spe}
\modTinv\tilde\phi_{\spe 1}^\sw \cT_{\spe,0}^{-1
*} F_{\spe 0}, \cT_{\spe',0}^{-1
*}F_{\spe' 0}
 \right ]
\nonumber\\[5pt]
\fl& + \
\frac{\lambda_\spe}{\lambda_{\spe'}}
\cT_{NP, \spe}^*
C_{\sigma
\sigma^\prime} \Bigg[ \cT_{\spe,0}^{-1 *}F_{\spe 0} ,
\modTinvprime F_{\sigma' 1}^\sw
\nonumber\\[5pt]
\fl&
 -
\frac{Z_{\spe'}\lambda_{\spe'}}{T_{\spe'}}
\modTinvprime\tilde\phi_{\spe'
1}^\sw \cT_{\spe',0}^{-1 *}F_{\spe' 0}
 \Bigg]
\Bigg\}
\Bigg]^\lw
\Bigg\rangle.
\end{eqnarray}
In order to get the last expression we have integrated by parts in
$\theta$ and $\mu$.

\section{Computation of the last term of \eq{eq:FPsecondorder}}
\label{sec:C2nc}

\begin{eqnarray}\label{eq:C2lw}
\fl  C_{\spe \spe'}^{(2)\lw} &=
C_{\sigma \sigma^\prime}
 \Big[\cT^{-1*}_{\spe,0} F_{\sigma 2}^\lw +
[\cT_{\sigma,1}^{-1 *} F_{\sigma 1}^\lw]^\lw 
 \nonumber\\
\fl&
+
[\cT_{\sigma,1}^{-1 *} F_{\sigma 1}^\sw]^\lw +
[\cT_{\sigma,2}^{-1 *} F_{\spe 0}]^\lw,
\cT^{-1*}_{\spe',0}F_{\spe' 0} \Big] \nonumber\\
\fl&+\left(\frac{\lambda_\spe}{\lambda_{\spe'}}\right)^2C_{\sigma \sigma^\prime}
\Big[
\cT^{-1*}_{\spe,0}F_{\spe 0},\cT^{-1*}_{\spe',0} F_{\sigma' 2}^\lw
 \nonumber\\
\fl&
+
[\cT_{\sigma',1}^{-1 *} F_{\sigma' 1}^\lw]^\lw +
[\cT_{\sigma',1}^{-1 *} F_{\sigma' 1}^\sw]^\lw +
[\cT_{\sigma',2}^{-1 *} F_{\spe' 0}]^\lw
\Big] \nonumber\\
\fl& +
\frac{\lambda_\spe}{\lambda_{\spe'}}C_{\sigma\sigma^\prime}
\left[\cT^{-1*}_{\spe,0} F_{\sigma 1}^\lw
+ [\cT_{\sigma,1}^{-1 *} F_{\sigma 0}]^\lw,
\cT^{-1*}_{\spe',0}F_{\sigma^\prime 1}^\lw +
[\cT_{\sigma',1}^{-1 *} F_{\sigma' 0}]^\lw
\right ] \nonumber\\
\fl& +
\frac{\lambda_\spe}{\lambda_{\spe'}}
\Bigg[
C_{\sigma\sigma^\prime} \Bigg[
\modTinv F_{\spe 1}^\sw -
\frac{Z_\spe\lambda_\spe
}{T_\spe}
\modTinv \phiwig_{\spe 1}^\sw\cT_{\spe,0}^{-1 *}F_{\spe 0}
, 
\modTinvprime F_{\spe' 1}^\sw
\nonumber\\[5pt]
\fl&
-
\frac{Z_{\spe'}\lambda_{\spe'} 
}{T_{\spe'}}
\modTinvprime\phiwig_{\spe' 1}^\sw\cT_{\spe',0}^{-1 *}F_{\spe' 0}
\Bigg] \Bigg]^\lw,
\end{eqnarray}
with
\begin{eqnarray}\label{eq:T1F1lw}
\fl [\cT_{\sigma,1}^{-1 *} F_{\spe 1}^\lw]^\lw =  -
\cT_{\spe,0}^{-1 *}\Bigg\{
\rhobf\cdot\nabla_\bR\nonumber\\[5pt]
\fl\hspace{0.5cm}
 +
\Bigg(u\bun\cdot\nabla_\bR\bun\cdot\rhobf
+\frac{B}{4}[\rhobf(\rhobf\times\bun)+(\rhobf\times\bun)\rhobf]:\nabla_\bR\bun
\nonumber\\[5pt]
\fl\hspace{0.5cm}
-\mu\bun\cdot\nabla_\bR\times\bun
\Bigg)\partial_u
+
\Bigg(-\frac{\mu}{B}\rhobf\cdot\nabla_\bR B
\nonumber\\[5pt]
\fl\hspace{0.5cm}
-\frac{u}{4}[\rhobf(\rhobf\times\bun)+(\rhobf\times\bun)\rhobf]:\nabla_\bR\bun
+\frac{u\mu}{B}\bun\cdot\nabla_\bR\times\bun \nonumber\\[5pt]
\fl\hspace{0.5cm}
- \frac{u^2}{B}\bun\cdot\nabla_\bR\bun\cdot\rhobf
- \frac{Z_\spe}{B}\rhobf\cdot\nabla\varphi_0
\Bigg)\partial_\mu
\Bigg\}
F_{\spe 1}^\lw,
\end{eqnarray}
\begin{eqnarray}\label{eq:T1F1sw}
\fl [\cT_{\sigma,1}^{-1 *} F_{\sigma 1}^\sw]^\lw =
- \cT_{\spe,0}^{-1 *}
\Bigg[
\Bigg( \frac{Z_\spe\lambda_\spe}{B^2}
 (\nabla_{\bR_\perp/\epsilon_\spe}
\Phiwig_{\sigma 1}^\sw \times \bun ) \cdot \nabla_{\bR_\perp/\epsilon_\spe}
\nonumber\\[5pt]
\fl\hspace{0.5cm}
-
\frac{Z_\spe\lambda_\spe \phiwig_{\spe 1}^\sw}{B}
\frac{\partial}{\partial \mu}  \Bigg) F_{\spe 1}^\sw
\Bigg]^\lw.
\end{eqnarray}
Here to obtain \eq{eq:T1F1lw} and \eq{eq:T1F1sw} we have used the
results in \ref{sec:pullback}. The term $[\cT_{\sigma,2}^{-1 *}
F_{\spe 0}]^\lw$ is calculated in
\ref{sec:secondordertransfMaxwellian}. Now, let us write the
gyroaverage of \eq{eq:C2lw}:
\begin{eqnarray}\label{eq:C2lwgyro}
\fl\Big\langle \cT_{\spe, 0}^{*}C_{\spe \spe'}^{(2)\lw}\Big\rangle =
\cT_{\spe, 0}^{*} C_{\sigma \sigma^\prime}
 \Big[\cT_{\spe, 0}^{-1*}\left\langle F_{\sigma 2}^\lw \right\rangle +
\cT_{\spe, 0}^{-1*}\left\langle 
\cT_{\spe, 0}^{*} [\cT_{\sigma,1}^{-1 *} F_{\sigma 1}^\lw]^\lw\right\rangle 
\nonumber\\\hspace{0.5cm}
\fl
 +
\cT_{\spe, 0}^{-1*}\left\langle \cT_{\spe, 0}^{*}
[\cT_{\sigma,2}^{-1 *} F_{\spe 0}]^\lw\right\rangle ,
\cT^{-1*}_{\spe',0}F_{\spe' 0} \Big] 
+\left(\frac{\lambda_\spe}{\lambda_{\spe'}}\right)^2
\cT_{\spe, 0}^{*}C_{\sigma \sigma^\prime}
\Big[
\cT^{-1*}_{\spe,0}F_{\spe 0},
\nonumber\\\hspace{0.5cm}
\fl
\cT_{\spe^\prime, 0}^{-1*}
 \left\langle F_{\sigma' 2}^\lw\right\rangle +
\cT_{\spe^\prime, 0}^{-1*}
\left\langle
\cT_{\spe^\prime, 0}^{*}[\cT_{\sigma',1}^{-1 *} F_{\sigma' 1}^\lw]^\lw\right\rangle +
\cT_{\spe^\prime, 0}^{-1*}\left\langle\cT_{\spe^\prime, 0}^{*}
[\cT_{\sigma',2}^{-1 *} F_{\spe' 0}]^\lw\right\rangle
\Big] \nonumber\\\hspace{0.5cm}
\fl +
\frac{\lambda_\spe}{\lambda_{\spe'}}
\left\langle
\cT_{\spe, 0}^{*}C_{\sigma\sigma^\prime}
\left[\cT^{-1*}_{\spe,0} F_{\sigma 1}^\lw
+ [\cT_{\sigma,1}^{-1 *} F_{\sigma 0}]^\lw,
\cT^{-1*}_{\spe',0}F_{\sigma^\prime 1}^\lw +
[\cT_{\sigma',1}^{-1 *} F_{\sigma' 0}]^\lw
\right]
\right\rangle
\nonumber\\\hspace{0.5cm}
\fl +
\frac{\lambda_\spe}{\lambda_{\spe'}}
\Bigg
\langle
\cT_{\spe, 0}^{*} \Bigg[C_{\sigma\sigma^\prime} \Bigg[ 
\modTinv F_{\spe 1}^\sw -
\frac{Z_\spe\lambda_\spe
}{T_\spe}
\modTinv \phiwig_{\spe 1}^\sw\cT_{\spe,0}^{-1 *}F_{\spe 0}
, 
\modTinvprime F_{\spe' 1}^\sw \nonumber\\\hspace{0.5cm}
\fl
-
\frac{Z_{\spe'}\lambda_{\spe'} 
}{T_{\spe'}}
\modTinvprime\phiwig_{\spe' 1}^\sw\cT_{\spe',0}^{-1 *}F_{\spe' 0}
\Bigg] \Bigg]^\lw
\Bigg
\rangle,
\end{eqnarray}
where we have used that $\left\langle
\cT_{\spe, 0}^{*}[\cT_{\sigma,1}^{-1 *} F_{\sigma
    1}^\sw]^\lw\right\rangle = 0$. Here,
\begin{eqnarray}\label{eq:T1F1lwAveraged}
\fl \left\langle
\cT_{\spe, 0}^{*}[\cT_{\sigma,1}^{-1 *} F_{\spe 1}^\lw]^\lw
\right\rangle = 
\mu\bun\cdot\nabla_\bR\times\bun
\left(
\partial_u
-\frac{u}{B} 
\partial_\mu
\right)
F_{\spe 1}^\lw,
\end{eqnarray}
and 
\begin{eqnarray}\label{gyroaverageT2F0}
\fl\left\langle\cT_{\spe, 0}^{*} [\cT_{\sigma,2}^{-1 *} F_{\spe 0}]^\lw\right\rangle
 =\nonumber\\
\fl\hspace{1cm} \frac{\mu}{2B} (\matI - \bun \bun) : \Bigg [
\nabla_\bR \nabla_\bR \ln n_\spe + \left ( \frac{u^2/2 + \mu B}{T_\spe} -
\frac{3}{2} \right ) \nabla_\bR \nabla_\bR \ln T_\spe \Bigg ] F_{\spe 0}
\nonumber\\
\fl\hspace{1cm} 
- \frac{\mu}{B} \frac{Z_\spe}{T_\spe^2}
\nabla_\bR \varphi_0 \cdot \nabla_\bR T_\spe F_{\spe 0} -
\frac{\mu}{2B} \frac{u^2/2 + \mu B}{T_\spe^3} |\nabla_\bR T_\spe|^2
F_{\spe 0} \nonumber\\
\fl\hspace{1cm} 
+ \frac{\mu}{2B} \Bigg |
\frac{\nabla_\bR n_\spe}{n_\spe} + \frac{Z_\spe \nabla_\bR
\varphi_0}{T_\spe} + \left ( \frac{u^2/2 + \mu B}{T_\spe} - \frac{3}{2}
\right ) \frac{\nabla_\bR T_\spe}{T_\spe} \Bigg |^2 F_{\spe 0}
\nonumber\\
\fl\hspace{1cm} 
- \frac{\mu}{2B^2} \nabla_{\bR_\bot} B \cdot \left (
\frac{\nabla_\bR n_\spe}{n_\spe} + \frac{Z_\spe \nabla_\bR
\varphi_0}{T_\spe} + \left ( \frac{u^2/2 + \mu B}{T_\spe} - \frac{3}{2}
\right ) \frac{\nabla_\bR T_\spe}{T_\spe} \right ) F_{\spe 0}
\nonumber\\
\fl\hspace{1cm} 
+ \frac{Z_\spe^2\lambda_\spe^2}{2T_\spe^2}
\left[ \left \langle
(\phiwig_{\spe 1}^\sw)^2 \right \rangle \right]^\lw F_{\spe
0} + \frac{1}{T_\spe} \Bigg [ - \frac{Z_\spe^2}{2B^2}
|\nabla_\bR \varphi_0|^2 \nonumber\\
\fl\hspace{1cm} 
- \frac{Z_\spe^2\lambda_\spe^2}{2B}
\partial_\mu \left[ \left
\langle (\phiwig_{\spe 1}^\sw)^2 \right \rangle \right]^\lw
- \frac{3 Z_\spe \mu}{2 B^2} \nabla_{\bR_\bot} B \cdot
\nabla_\bR \varphi_0
\nonumber\\
\fl\hspace{1cm} 
 - \frac{Z_\spe u^2}{B^2} \bun
\cdot \nabla_\bR \bun \cdot \nabla_\bR \varphi_0  + \Psi_{B, \spe}
\nonumber\\
\fl\hspace{1cm} 
+ 
\frac{Z_\spe
\mu}{B} (\matI - \bun \bun): \nabla_\bR \nabla_\bR \varphi_0 \Bigg
] F_{\spe 0}.
\end{eqnarray}
This last result has been obtained by gyroaveraging \eq{T2F0}.

\section{Second-order inverse transformation of a Maxwellian}
\label{sec:secondordertransfMaxwellian}

The calculation of $C_{\spe\spe'}^{(2)\lw}$ in \ref{sec:C2nc} requires
$[\mathcal{T}^{-1\ast}_{\spe, 2} F_{\spe 0}]^\lw$.  We
start by using that $F_{\spe 0}$ is a Maxwellian that depends on $\bR$
and $u^2/2 + \mu B(\bR)$, giving
\begin{eqnarray}
\fl \left [ \mathcal{T}^{-1\ast}_{\spe, 2} F_{\spe 0} \right ]^\lw
=
\nonumber\\
\fl\hspace{0.5cm}
 \frac{1}{2B^2} (\bv \times \bun) (\bv \times \bun) : \Bigg [
\nabla_\boldr \nabla_\boldr \ln n_\spe + \left ( \frac{v^2}{2T_\spe} -
\frac{3}{2} \right ) \nabla_\boldr \nabla_\boldr \ln T_\spe 
\nonumber\\
\fl\hspace{0.5cm}
-
\frac{v^2}{2T_\spe^3} \nabla_\boldr T_\spe \nabla_\boldr T_\spe
+
\Bigg ( \frac{\nabla_\boldr n_\spe}{n_\spe} + \left ( \frac{v^2}{2T_\spe}
- \frac{3}{2} \right ) \frac{\nabla_\boldr T_\spe}{T_\spe} \Bigg ) \Bigg
( \frac{\nabla_\boldr n_\spe}{n_\spe} 
\nonumber\\
\fl\hspace{0.5cm}
+ \left ( \frac{v^2}{2T_\spe} -
\frac{3}{2} \right ) \frac{\nabla_\boldr T_\spe}{T_\spe} \Bigg ) \Bigg ]
\cT_{\spe,0}^{-1*}F_{\spe 0}
\nonumber\\
\fl\hspace{0.5cm}
+ {\bR}_{02}^\lw \cdot \left ( \frac{\nabla_\boldr
n_\spe}{n_\spe} + \left ( \frac{v^2}{2T_\spe} - \frac{3}{2} \right
) \frac{\nabla_\boldr T_\spe}{T_\spe} \right ) \cT_{\spe,0}^{-1*}F_{\spe 0}
\nonumber\\
\fl\hspace{0.5cm}
-
\frac{1}{B} H_{01}^\lw (\bv \times \bun) \cdot \Bigg (
\frac{\nabla_\boldr n_\spe}{n_\spe} + \left ( \frac{v^2}{2T_\spe} -
\frac{5}{2} \right ) \frac{\nabla_\boldr T_\spe}{T_\spe} \Bigg )
\frac{\cT_{\spe,0}^{-1*} F_{\spe0}}{T_\spe}
\nonumber\\
\fl\hspace{0.5cm}
+ \frac{1}{2} \left [ H_{01}^2
\right ]^\lw \frac{\cT_{\spe,0}^{-1*}F_{\spe 0}}{T_\spe^2} - H_{02}^\lw
\frac{\cT_{\spe,0}^{-1*}F_{\spe 0}}{T_\spe},
\end{eqnarray}
where the functions $\bR_{02}$, $H_{01}$ and $H_{02}$ are given by
\begin{eqnarray}
\bR = \boldr + \frac{\epsilon_\spe}{B} \bv \times \bun +
\epsilon_\spe^2 \bR_{02}   + O(\epsilon_\spe^3)
\end{eqnarray}
and
\begin{eqnarray}
\frac{u^2}{2} + \mu B(\bR) = \frac{v^2}{2} + \epsilon_\spe H_{01}
+ \epsilon_\spe^2 H_{02}+ O(\epsilon_\spe^3).
\end{eqnarray}
In what follows we calculate $\bR_{02}$, $H_{01}$ and $H_{02}$. 

To compute $\bR_{02}$ we use that
\begin{equation}
\boldr = \bR_g + \epsilon_\spe \rhobf (\bR_g, \mu_g, \theta_g).
\end{equation}
Employing the results in \eq{0gtransformation} it is easy to see that
\begin{eqnarray}
\fl \boldr = \bR_g - \frac{\epsilon_\spe}{B} \bv \times \bun +
\epsilon_\spe^2 \mathcal{T}_{\spe, 0}^{-1\ast} \Bigg [ - \rhobf
\cdot \nabla_\bR \rhobf 
\nonumber\\[5pt]
\fl\hspace{1cm}+ \left(
\frac{\mu}{B}\rhobf\cdot\nabla_\bR B + u(\rhobf \times\bun )\rhobf
:\nabla_\bR \bun \right) \partial_\mu \rhobf
\nonumber\\[5pt]
\fl\hspace{1cm}
- \left(
\rhobf \cdot\nabla_\bR \eun_2 \cdot\eun_1 -\frac{u}{2\mu}\rhobf
\cdot \nabla_\bR\bun \cdot\rhobf \right)
\partial_\theta \rhobf \Bigg ] + O(\epsilon_\spe^3).
\end{eqnarray}
Using $\nabla_\bR \rhobf = - (2B)^{-1} \nabla_\bR B \rhobf -
(\nabla_\bR \bun \cdot \rhobf) \bun + (\nabla_\bR \eun_2 \cdot
\eun_1) \rhobf \times \bun$, $\partial_\mu \rhobf = (2\mu)^{-1}
\rhobf$ and $\partial_\theta \rhobf = - \rhobf \times \bun$, we
obtain
\begin{eqnarray}
\fl \boldr = \bR_g - \frac{\epsilon_\spe}{B} \bv \times \bun +
\frac{\epsilon_\spe^2}{B^2} \Bigg[ \frac{1}{B} (\bv \times
\bun)(\bv \times \bun) \cdot \nabla_\boldr B  \nonumber\\
\fl\hspace{1cm}
+ (\bv \times \bun) \cdot
\nabla_\boldr \bun \cdot (\bv \times \bun) \bun
+ v_{||} (\bv
\times \bun) \cdot \nabla_\boldr \bun \times \bun \Bigg ]
\nonumber\\
\fl\hspace{1cm}
+O(\epsilon_\spe^3).
\end{eqnarray}
Finally, since $\bR_g = \bR + \epsilon_\spe^2 \bR_2 +
  O(\epsilon_\spe^3)$ with $\bR_2$ given in \eq{R2}, we obtain
\begin{eqnarray}
\fl \bR_{02} = \frac{1}{B} \Bigg[ \left ( v_{||} \bun +
\frac{1}{4} \bv_\bot \right ) \bv \times \bun 
\nonumber\\
\fl\hspace{1cm}
+ \bv \times \bun
\left ( v_{||} \bun + \frac{1}{4} \bv_\bot \right ) \Bigg]
\dotcross \nabla_\boldr \left ( \frac{\bun}{B} \right ) \nonumber\\
\fl\hspace{1cm}
+
\frac{v_{||}}{B^2} \bv_\bot \cdot \nabla_\boldr \bun + \frac{v_{||}}{B^2}
\bun \bun \cdot \nabla_\boldr \bun \cdot \bv_\bot \nonumber\\
\fl\hspace{1cm}
+
\frac{\bun}{8B^2} [ \bv_\bot \bv_\bot - (\bv_\bot \times \bun)
(\bv_\bot \times \bun) ]: \nabla_\boldr \bun 
\nonumber\\
\fl\hspace{1cm}
+\frac{Z_\spe\lambda_\spe}{ B^2} \bun \times
\modTinv\nabla_{(\bR_\bot/\epsilon_\spe)} \Phiwig_\spe
\nonumber\\
\fl\hspace{1cm}
+ \frac{v_\bot^2}{2B^3} \bun \bun \cdot \nabla_\boldr B -
\frac{v_\bot^2}{4B^3} \nabla_{\boldr_\bot} B,
\end{eqnarray}
where ${\mathbf a}{\mathbf b}\dotcross\matM = {\mathbf
  a}\times({\mathbf b}\cdot\matM)$. The long-wavelength component is
\begin{eqnarray}
  \fl \bR_{02}^\lw = \frac{1}{B} \Bigg[ \left ( v_{||} \bun +
    \frac{1}{4} \bv_\bot \right ) \bv \times \bun
  \nonumber\\
  \fl\hspace{1cm}
  + \bv \times \bun
  \left ( v_{||} \bun + \frac{1}{4} \bv_\bot \right ) \Bigg]
  \dotcross \nabla_\boldr \left ( \frac{\bun}{B} \right )
 \nonumber\\
\fl\hspace{1cm}+
  \frac{v_{||}}{B^2} \bv_\bot \cdot \nabla_\boldr \bun + \frac{v_{||}}{B^2}
  \bun \bun \cdot \nabla_\boldr \bun \cdot \bv_\bot \nonumber\\\fl\hspace{1cm}+
  \frac{\bun}{8B^2} [ \bv_\bot \bv_\bot - (\bv_\bot \times \bun)
  (\bv_\bot \times \bun) ]: \nabla_\boldr \bun \nonumber\\\fl\hspace{1cm}+
  \frac{v_\bot^2}{2B^3} \bun \bun \cdot \nabla_\boldr B -
  \frac{v_\bot^2}{4B^3} \nabla_{\boldr_\bot} B.
\end{eqnarray}

To obtain $H_{01}$ and $H_{02}^\lw$, we use that the expressions of
the Hamiltonian in the two different sets of variables are related
(with some abuse of notation) by
\begin{eqnarray}\label{eq:relationbetweenHamiltonians}
\fl \frac{v^2}{2} + Z_\spe\lambda_\spe \epsilon_\spe \varphi
(\boldr,\bv,t)
 = 
 \nonumber\\[5pt]
\fl\hspace{1cm}
\frac{u^2}{2} + \mu B(\bR) 
+ Z_\spe\lambda_\spe
\epsilon_\spe \langle \phi_\spe \rangle(\bR,\mu,t)
 \nonumber\\[5pt]
\fl\hspace{1cm}
+ Z_\spe^2\lambda_\spe^2
\epsilon_\spe^2 \Psi_{\phi,\spe} + Z_\spe\lambda_\spe \epsilon_\spe^2
\Psi_{\phi B,\spe} + \epsilon_\spe^2 \Psi_{B,\spe}
 \nonumber\\[5pt]
\fl\hspace{1cm}
-
\frac{Z_\spe\lambda_\spe \epsilon_\spe^2}{B}
\partial_t \Phiwig_\spe
 + O(\epsilon_\spe^3).
\end{eqnarray}
Let us give a more detailed explanation of the last equation. As shown
in reference \cite{ParraCalvo2011}, and to the order of interest, the
Hamiltonian in gyrokinetic coordinates, $\overline{H}_\spe$, is the
Hamiltonian in cartesian coordinates, $H^{\bX}_\spe$, after a change
of coordinates and the addition of the partial derivative with respect
to time of a gauge function. This function is $-
  S_{NP,\spe}-\epsilon_\spe^2 S_{P,\spe}^{(2)}$, where $S_{NP,\spe}$
  (which does not depend on time) and $S_{P,\spe}^{(2)}$ are given in
  equations (81) and (108) of reference \cite{ParraCalvo2011},
  respectively. As a result,
\begin{equation}
  \overline{H}_\spe = 
\cT_\spe^* H_\spe^{\bX} -\epsilon_\spe^2 \partial_t S_{P,\spe}^{(2)}
+ O(\epsilon_\spe^3),
\end{equation}
This is the origin of the last term in \eq{eq:relationbetweenHamiltonians}.

The function $\langle \phi_\spe
\rangle (\bR, \mu, t)$ is
\begin{eqnarray}
\fl\langle
\phi_\spe\rangle (\bR, \mu, t) 
= \langle \phi_\spe
 \rangle (\bR_g, \mu_g, t) 
\nonumber\\[5pt]
\fl\hspace{0.5cm}
 - \epsilon_\spe\left( \bR_2 \cdot
\nabla_{\bR_{g\perp}/\epsilon_\spe } \langle \phi_\spe \rangle +
 \mu_1 \partial_{\mu_g} \langle\phi_\spe
  \rangle \right)
\nonumber\\[5pt]
\fl\hspace{0.5cm}
 + O(\epsilon_\spe^2).
\end{eqnarray}
Here it is worth distinguishing between long-wavelength and
short-wavelength pieces. For the long-wavelength potential,
\begin{eqnarray}
\fl \langle \phi_\spe^\lw \rangle (\bR_g, \mu_g, t) =
\frac{1}{\epsilon_\spe \lambda_\spe} \varphi_0(\bR_g, t) +
\varphi_1^\lw(\bR_g, t)
\nonumber\\[5pt]
\fl\hspace{0.5cm}
 + \frac{\epsilon_\spe \mu_g}{2 \lambda_\spe
B (\bR_g)} \left(\matI - \bun (\bR_g) \bun (\bR_g)\right):
 \nabla_{\bR_g} \nabla_{\bR_g}
\varphi_0(\bR_g, t) + O(\epsilon_\spe^2).
\end{eqnarray}
This has to be written in $(\boldr, \bv)$
variables. Employing the results in
  \eq{0gtransformation} gives
\begin{eqnarray}
\fl \langle \phi_\spe^\lw \rangle (\bR_g, \mu_g, t) =
\frac{1}{\epsilon_\spe \lambda_\spe} \varphi_0  +
\varphi_1^\lw  + \frac{1}{\lambda_\spe B} (\bv \times
\bun) \cdot \nabla_\boldr \varphi_0
\nonumber\\[5pt]
\fl\hspace{0.5cm}
 + \frac{\epsilon_\spe}{B} (\bv \times
\bun) \cdot \nabla_\boldr \varphi_1^\lw 
\nonumber\\[5pt]
\fl\hspace{0.5cm}
-
\frac{1}{\lambda_\spe} \mathcal{T}_{\spe, 0}^{-1\ast} 
\Bigg [ - \rhobf
\cdot \nabla_\bR \rhobf + \left(
\frac{\mu}{B}\rhobf\cdot\nabla_\bR B + u(\rhobf \times\bun )\rhobf
:\nabla_\bR \bun \right) \partial_\mu \rhobf
\nonumber\\[5pt]
\fl\hspace{0.5cm}- \left(
\rhobf \cdot\nabla_\bR \eun_2 \cdot\eun_1 -\frac{u}{2\mu}\rhobf
\cdot \nabla_\bR\bun \cdot\rhobf \right)
\partial_\theta \rhobf \Bigg ] \cdot \nabla_\boldr \varphi_0 
\nonumber\\[5pt]
\fl\hspace{0.5cm}
+
\frac{\epsilon_\spe}{2\lambda_\spe B^2} \left [ (\bv \times \bun)
(\bv \times \bun) + \frac{v_\bot^2}{2} (\matI - \bun \bun) \right
]: \nabla_\boldr \nabla_\boldr \varphi_0 +O(\epsilon_\spe^2),
\end{eqnarray}
where on the right-hand side everything is evaluated at $\boldr$. With
this result, we find that to lowest order
\begin{eqnarray}
\fl \frac{u^2}{2} + \mu B(\bR) &= \frac{v^2}{2} -
\frac{Z_\spe\epsilon_\spe}{B}
 (\bv \times \bun(\boldr)) \cdot
\nabla_\boldr \varphi_0(\boldr,t) \nonumber\\[5pt]
\fl&+ Z_\spe\lambda_\spe \epsilon_\spe \modTinv
\phiwig_{\spe 1}^\sw(\boldr,\bv,t) + O(\epsilon_\spe^2),
\end{eqnarray}
giving
\begin{eqnarray}
\fl H_{01}^\lw(\boldr,\bv,t) = - \frac{Z_\spe}{B}
 (\bv \times \bun(\boldr)) \cdot
\nabla_\boldr \varphi_0(\boldr,t)
\end{eqnarray}
and
\begin{eqnarray}
\fl \left [ H_{01}^2 \right ]^\lw(\boldr,\bv,t)
 &= \frac{Z_\spe^2}{B^2}
\left [ (\bv \times \bun(\boldr)) \cdot \nabla_\boldr
 \varphi_0(\boldr,t) \right ]^2
\nonumber\\[5pt]
\fl& +
Z_\spe^2\lambda_\spe^2 
\mathcal{T}^{-1\ast}_{\spe,0}
\left[
(\phiwig_{\spe 1}^\sw)^2 \right]^\lw(\boldr,\bv,t).
\end{eqnarray}
Going to higher order, we find
\begin{eqnarray}
\fl H_{02}^\lw(\boldr,\bv,t)
 = - \frac{Z_\spe\lambda_\spe}{B} (\bv \times \bun) \cdot
\nabla_\boldr \varphi_1^\lw
 -
Z_\spe \bR_{02}^\lw \cdot \nabla_\boldr
\varphi_0
\nonumber\\
\fl\hspace{0.5cm}
 - Z_\spe^2\lambda_\spe^2 \mathcal{T}^{-1\ast}_{\spe,0}
\Psi_{\phi,\spe}^\lw - Z_\spe\lambda_\spe \mathcal{T}^{-1\ast}_{\spe,0}
\Psi_{\phi B,\spe}^\lw
\nonumber\\
\fl\hspace{0.5cm}
 - \mathcal{T}^{-1\ast}_{\spe,0}
\Psi_{B,\spe} - 
\frac{Z_\spe}{2B^2} \left [ (\bv
\times \bun) (\bv \times \bun) + \frac{v_\bot^2}{2} (\matI - \bun
\bun) \right ]: \nabla_\boldr \nabla_\boldr \varphi_0
\nonumber\\
\fl\hspace{0.5cm}
+
\frac{Z_\spe^2\lambda_\spe^2}{B}
\mathcal{T}^{-1\ast}_{\spe,0} \left[
\left((\nabla_{\bR_\bot/\epsilon_\spe}  {\Phiwig_\spe^\sw}) \times
\bun \right) \cdot
\nabla_{\bR_\bot/\epsilon_\spe}
\langle  {\phi_\spe^\sw} \rangle \right]^\lw
\nonumber\\
\fl\hspace{0.5cm}
-
\frac{Z_\spe^2\lambda_\spe^2}{B}\mathcal{T}^{-1\ast}_{\spe,0} \left[
  {\phiwig_{\spe 1}^\sw}\partial_\mu 
\langle  {\phi_{\spe 1}^\sw} \rangle
 \right]^\lw .
\end{eqnarray}
Note that  $[ (\nabla_{(\bR_\bot/\epsilon_\spe)}
\Phiwig_\spe^\sw \times \bun ) \cdot \nabla_{\bR_\bot/\epsilon_\spe}
 \langle \phi_\spe^\sw \rangle ]^\lw =
O(\epsilon_\spe)$ and can be neglected, giving
\begin{eqnarray}
\fl H_{02}^\lw(\boldr,\bv,t)
 = - \frac{Z_\spe\lambda_\spe}{B} (\bv \times \bun) \cdot
\nabla_\boldr\varphi_1^\lw - 
 {Z_\spe} \bR_{02}^\lw \cdot \nabla_\boldr
\varphi_0 
\nonumber\\[5pt]
\fl\hspace{0.5cm}
- Z_\spe^2\lambda_\spe^2 \mathcal{T}^{-1\ast}_{\spe,0}
\Psi_{\phi,\spe}^{\lw} - Z_\spe\lambda_\spe \mathcal{T}^{-1\ast}_{\spe,0}
\Psi_{\phi B,\spe}^{\lw}- \mathcal{T}^{-1\ast}_{\spe,0}
\Psi_{B,\spe}
\nonumber\\[5pt]
\fl\hspace{0.5cm}
 - 
 {\frac{Z_\spe}{2B^2}} \left [ (\bv
\times \bun) (\bv \times \bun) + \frac{v_\bot^2}{2} (\matI - \bun
\bun) \right ]: \nabla_\boldr \nabla_\boldr \varphi_0\nonumber\\[5pt]
\fl\hspace{0.5cm}
-
\frac{Z_\spe^2\lambda_\spe^2}{B}
\mathcal{T}^{-1\ast}_{\spe,0}\left[
{\phiwig_{\spe 1}^\sw}\partial_\mu 
\langle  {\phi_{\spe 1}^\sw}\rangle
\right]^\lw.
\end{eqnarray}

Combining all these results we obtain
\begin{eqnarray} \label{T2F0}
\fl \left [ \mathcal{T}^{-1\ast}_{\spe, 2} F_{\spe 0} \right ]^\lw
= \frac{1}{2B^2} (\bv \times \bun) (\bv \times \bun) : \Bigg [
\nabla_\boldr \nabla_\boldr \ln n_\spe + 
 {\frac{Z_\spe}{T_\spe}} \nabla_\boldr
\nabla_\boldr \varphi_0
\nonumber\\\fl\hspace{0.5cm}
 - 
 {\frac{Z_\spe}{T_\spe^2}} ( \nabla_\boldr
\varphi_0 \nabla_\boldr T_\spe 
+ \nabla_\boldr T_\spe \nabla_\boldr
\varphi_0 ) + \left ( \frac{v^2}{2T_\spe} - \frac{3}{2} \right )
\nabla_\boldr \nabla_\boldr \ln T_\spe
\nonumber\\\fl\hspace{0.5cm} - \frac{v^2}{2T_\spe^3} 
\nabla_\boldr T_\spe
\nabla_\boldr
 T_\spe \Bigg ] \cT_{\spe,0}^{-1*}F_{\spe 0} 
+ \frac{1}{2B^2} \Bigg
[ (\bv \times \bun) \cdot \Bigg 
( \frac{\nabla_\boldr n_\spe}{n_\spe}
\nonumber\\\fl\hspace{0.5cm} +
 {\frac{Z_\spe\nabla_\boldr \varphi_0}{T_\spe}}
 + \left (
\frac{v^2}{2T_\spe} - \frac{3}{2} \right ) \frac{\nabla_\boldr
T_\spe}{T_\spe} \Bigg ) \Bigg ]^2 \cT_{\spe,0}^{-1*}F_{\spe 0} \nonumber\\\fl\hspace{0.5cm}+
\bR_{02}^\lw \cdot \left ( \frac{\nabla_\boldr n_\spe}{n_\spe} +
 {\frac{Z_\spe \nabla_\boldr \varphi_0}{T_\spe}} 
+ \left (
\frac{v^2}{2T_\spe} - \frac{3}{2} \right ) \frac{\nabla_\boldr
T_\spe}{T_\spe} \right ) \cT_{\spe,0}^{-1*}F_{\spe 0} 
\nonumber\\\fl\hspace{0.5cm}+
\frac{Z_\spe^2\lambda_\spe^2}{2T_\spe^2}
\mathcal{T}^{-1\ast}_{\spe,0}\left[
(\phiwig_{\spe 1}^\sw)^2 \right]^\lw
F_{\spe 0} 
+ \frac{1}{T_\spe} \Bigg [ \frac{Z_\spe\lambda_\spe}{B} (\bv \times
\bun) \cdot \nabla_\boldr \varphi_1^\lw
\nonumber\\\fl\hspace{0.5cm}
 + Z_\spe^2\lambda_\spe^2
\mathcal{T}^{-1\ast}_{\spe,0} \Psi_{\phi,\spe}^\lw 
+
Z_\spe\lambda_\spe \mathcal{T}^{-1\ast}_{\spe,0} \Psi_{\phi B,\spe}^{\lw}
\nonumber\\\fl\hspace{0.5cm}
+ \mathcal{T}^{-1\ast}_{\spe,0} \Psi_{B,\spe} + 
 {\frac{Z_\spe
v_\bot^2}{4B^2} }(\matI - \bun \bun): \nabla_\boldr \nabla_\boldr \varphi_0
\nonumber\\\fl\hspace{0.5cm}+ \frac{Z_\spe^2\lambda_\spe^2}{B}
\mathcal{T}^{-1\ast}_{\spe,0}\left[
{\phiwig_{\spe 1}^\sw}\partial_\mu 
\langle  {\phi_{\spe 1}^\sw}
\rangle
\right]^\lw
\Bigg ] \cT_{\spe,0}^{-1*}F_{\spe 0}.
\end{eqnarray}

\section{Calculations for the Fokker-Planck equation to 
$O(\epsilon_\spe^2)$}
\label{sec:calculationsFPorder2}

Start with \eq{eq:FPsecondorder}. Employing the definition of
$G_{\sigma 1}$ in \eq{eq:defGspe1}, $\partial_\zeta F_{\spe 1}^\lw
\equiv 0$, and using the identities
\begin{equation}
\fl (\bun \times \nabla_\bR \varphi_0) \cdot \nabla_\bR
G_{\spe 1}^\lw = \partial_\psi\varphi_0  I \bun
\cdot \nabla_\bR G_{\spe 1}^\lw,
\end{equation}
\begin{eqnarray}
\fl (\bun \times \nabla_\bR B) \cdot \nabla_\bR G_{\spe 1}^\lw =
\partial_\psi B  I \bun \cdot \nabla_\bR
G_{\spe 1}^\lw 
\nonumber\\[5pt]
\fl\hspace{1cm}
- \partial_\psi G_{\spe 1}^\lw I\, \bun
\cdot \nabla_\bR B,
\end{eqnarray}
\begin{eqnarray}
\fl [\bun \times (\bun \cdot \nabla_\bR \bun) ] \cdot \nabla_\bR
G_{\spe 1}^\lw =
\nonumber\\[5pt]
\fl\hspace{1cm}
 (\nabla_\bR \times \bun) \cdot \nabla_\bR
G_{\spe 1}^\lw - (\bun \cdot \nabla_\bR \times \bun) \bun \cdot
\nabla_\bR G_{\spe 1}^\lw=\nonumber\\[5pt]
\fl\hspace{1cm}
\bB
\cdot \nabla_\bR  \left (
\frac{I}{B}
\partial_\psi G_{\spe 1}^\lw  \right ) \nonumber\\[5pt]
\fl\hspace{1cm} - \bB \cdot \nabla_\bR \Theta
\partial_\psi  \left ( \frac{1}{\bB \cdot \nabla_\bR
\Theta} I \bun \cdot \nabla_\bR G_{\spe 1}^\lw \right )
 \nonumber\\[5pt]
\fl\hspace{1cm}  -
(\bun \cdot \nabla_\bR \times \bun) \bun \cdot \nabla_\bR
G_{\spe 1}^\lw,
\end{eqnarray}
\begin{eqnarray}
\fl \bun \times (\bun \cdot \nabla_\bR \bun) \cdot \left ( \mu
\nabla_\bR B + Z_\spe
 \nabla_{\bR_\perp} \varphi_0
\right ) =
\nonumber\\[5pt]
\fl\hspace{1cm} 
 \nabla_\bR \cdot \left ( \mu \bun \times \nabla_\bR B +
Z_\spe\bun \times
 \nabla_{\bR_\perp}   \varphi_0
\right ) \nonumber\\
\fl\hspace{1cm} 
 - \mu (\bun \cdot \nabla_\bR \times \bun)
\bun \cdot \nabla_\bR B 
\nonumber\\
\fl\hspace{1cm} 
= \bB \cdot \nabla_\bR \left [ \frac{I}{B}
\left ( \mu \partial_\psi B  +
Z_\spe
 \partial_\psi\varphi_0  \right )
\right ] \nonumber\\
\fl\hspace{1cm} 
 - \bB \cdot \nabla_\bR \Theta
\partial_\psi  \left ( \frac{I}{\bB \cdot \nabla_\bR
\Theta} \mu \bun \cdot \nabla_\bR B \right )
\nonumber\\
\fl\hspace{1cm}
 - \mu (\bun \cdot
\nabla_\bR \times \bun) \bun \cdot \nabla_\bR B,
\end{eqnarray}
\begin{eqnarray}
\fl
Z_\spe\lambda_\spe\bun\cdot\nabla_\bR\varphi_1^\lw
\partial_u
\left(\frac{Z_\spe\lambda_\spe\varphi_1^\lw}{T_\spe}F_{\spe 0}\right)=
\nonumber\\
\fl\hspace{1cm}
-
\left(
u\bun\cdot\nabla_\bR-\mu\bun\cdot\nabla_\bR B\partial_u
\right)
\left[
\frac{1}{2}
\left(
\frac{Z_\spe\lambda_\spe\varphi_1^\lw}{T_\spe}
\right)^2
F_{\spe 0}
\right],
\end{eqnarray}
\begin{eqnarray}
\fl \frac{1}{B}\Bigg[ - Z_\spe
 \nabla_\bR \varphi_0 \times \bun +
\mu \bun \times \nabla_\bR B 
\nonumber\\
\fl\hspace{1cm} 
+
u^2 \bun \times (\bun \cdot \nabla_\bR \bun)
\Bigg] \cdot \nabla_\bR \left(
\frac{Z_\spe\lambda_\spe\varphi_1^\lw}{T_\spe}
F_{ \sigma 0}
\right)
\nonumber\\
\fl\hspace{1cm}
-
Z_\spe\lambda_\spe\bun\cdot\nabla_\bR\varphi_1^\lw
\partial_u
\Bigg[
\frac{Iu}{B}
\Bigg(\frac{Z_\spe}{ T_\sigma}
\partial_\psi\varphi_0  + \Upsilon_\spe
\Bigg)
F_{\spe 0}
\Bigg]
\nonumber\\
\fl\hspace{1cm}
-\frac{u}{B}
\left[\bun\times\left(\bun\cdot\nabla_\bR\bun\right)
\right]
\cdot
\big(\mu\nabla_\bR B
\nonumber\\
\fl\hspace{1cm}
 +Z_\spe\nabla_\bR\varphi_0\big)
\partial_u
\left(
\frac{Z_\spe\lambda_\spe\varphi_1^\lw}{T_\spe}
F_{ \sigma 0}
\right) = 
\nonumber\\
\fl\hspace{1cm}
-\frac{Z_\spe\lambda_\spe I}{B}
\bun\cdot\nabla_\bR\varphi_1^\lw\Upsilon_\spe F_{\spe 0}
\nonumber\\
\fl\hspace{1cm}
+\frac{Z_\spe\lambda_\spe }{BT_\spe}F_{\spe 0}
\big[
\mu \bun \times \nabla_\bR B  
+
u^2 \bun \times (\bun \cdot \nabla_\bR \bun)
\big]
\cdot\nabla_\bR\varphi_1^\lw
\nonumber\\
\fl\hspace{1cm}
+
\left(
u\bun\cdot\nabla_\bR-\mu\bun\cdot\nabla_\bR B\partial_u
\right)
\Bigg[
\frac{Z_\spe\lambda_\spe I u }{T_\spe B}
\varphi_1^\lw F_{ \sigma 0}
\Bigg(\frac{Z_\spe}{ T_\sigma}
\partial_\psi\varphi_0
\nonumber\\
\fl\hspace{1cm}
  + \Upsilon_\spe -\frac{\partial_\psi T_\spe}{T_\spe}
\Bigg)
\Bigg],
\end{eqnarray}
\begin{eqnarray}
\fl \Bigg[ - \frac{Z_\spe}{B}
 \nabla_{\bR} \varphi_0 \times \bun +
\frac{\mu}{B} \bun \times \nabla_\bR B 
\nonumber\\
\fl\hspace{1cm} 
+
\frac{u^2}{B} \bun \times (\bun \cdot \nabla_\bR \bun)
\Bigg]
 \cdot \nabla_\bR \left ( \frac{Iu}{B} \right )
\nonumber\\
\fl\hspace{1cm} 
- \frac{u}{B} [\bun \times (\bun \cdot \nabla_\bR
\bun)] \cdot \left ( \mu \nabla_\bR B
+ Z_\spe \nabla_{\bR} \varphi_0 \right )
\partial_u \left ( \frac{Iu}{B}
\right ) \nonumber \\
\fl\hspace{1cm} 
 =
 - \frac{u}{B} (\bun \cdot
\nabla_\bR \times \bun) \left [ u \bun \cdot \nabla_\bR \left (
\frac{Iu}{B} \right ) - \mu \bun \cdot \nabla_\bR B
\partial_u \left ( \frac{Iu}{B}
\right ) \right ],
\end{eqnarray}
and
\begin{eqnarray}
\fl \frac{1}{B}\Bigg[ - Z_\spe
 \nabla_\bR \varphi_0 \times \bun +
\mu \bun \times \nabla_\bR B
\nonumber\\
\fl\hspace{1cm} 
+
u^2 \bun \times (\bun \cdot \nabla_\bR \bun)
\Bigg] \cdot \nabla_\bR \left(
\Bigg(\frac{Z_\spe}{ T_\sigma}
\partial_\psi\varphi_0  + \Upsilon_\spe
\Bigg)
F_{ \sigma 0}
\right)
\nonumber\\
\fl\hspace{1cm}
-\frac{u}{B}
\left[\bun\times\left(\bun\cdot\nabla_\bR\bun\right)
\right]
\cdot
\big(\mu\nabla_\bR B
\nonumber\\
\fl\hspace{1cm}
 +Z_\spe\nabla_\bR\varphi_0\big)
\partial_u
\left(
\Bigg(\frac{Z_\spe}{ T_\sigma}
\partial_\psi\varphi_0  + \Upsilon_\spe
\Bigg)
F_{ \sigma 0}
\right)
 =
\nonumber\\
\fl\hspace{1cm}
\left [ u \bun \cdot \nabla_\bR \left (
\frac{Iu}{B} \right ) - \mu \bun \cdot \nabla_\bR B
\partial_u \left ( \frac{Iu}{B}
\right ) \right ]\times
\nonumber\\
\fl\hspace{1cm}
\Bigg[
\partial_\psi
\left(\frac{Z_\spe}{T_\spe}\partial_\psi\varphi_0\right)
-\frac{Z_\spe}{T_\spe^2}
\partial_\psi\varphi_0
\partial_\psi T_\spe
+\partial_\psi\Upsilon_\spe
\nonumber\\
\fl\hspace{1cm}
-\frac{\mu}{T_\spe^2}\partial_\psi B\partial_\psi T_\spe
+
\Bigg(\frac{Z_\spe}{ T_\sigma}
\partial_\psi\varphi_0  + \Upsilon_\spe
\Bigg)^2
\Bigg]F_{\spe 0},
\end{eqnarray}
we obtain
\begin{eqnarray}\label{eq:auxEqToBeSimplified}
\fl \frac{1}{B}\Bigg[ - Z_\spe
 \nabla_\bR \varphi_0 \times \bun +
\mu \bun \times \nabla_\bR B 
\nonumber\\
\fl\hspace{1cm} 
+
u^2 \bun \times (\bun \cdot \nabla_\bR \bun)
\Bigg] \cdot \nabla_\bR F_{ \sigma 1}^\lw \nonumber\\
\fl\hspace{1cm} 
- \Bigg\{
Z_\spe\lambda_\spe \bun \cdot \nabla_\bR \varphi_1^\lw +
\frac{u}{B} [\bun \times (\bun \cdot \nabla_\bR \bun)]
\nonumber\\
\fl\hspace{1cm} 
\cdot \left ( \mu \nabla_\bR B + Z_\spe
\nabla_\bR \varphi_0 \right ) \Bigg\} \partial_u
F_{\sigma 1}^\lw \nonumber\\
\fl\hspace{1cm} 
= \frac{1}{B}
\left ( Z_\spe \partial_\psi
\varphi_0 + \mu \partial_\psi B 
\right ) I \bun \cdot \nabla_\bR G_{\sigma 1}^\lw 
\nonumber\\
\fl\hspace{1cm} 
-
\frac{I\mu}{B} \partial_\psi G_{\sigma 1}^\lw
 \bun \cdot \nabla_\bR B \nonumber\\
\fl\hspace{1cm}  - \bun \cdot \nabla_\bR
\Theta \partial_\psi  \left ( \frac{I u^2}{\bB
\cdot \nabla_\bR \Theta}  \bun \cdot
\nabla_\bR G_{\sigma 1}^\lw \right )
\nonumber\\
\fl\hspace{1cm} 
 + u \bun \cdot \nabla_\bR \left (
\frac{I u}{B} \partial_\psi G_{\sigma 1}^\lw \right )
\nonumber\\
\fl\hspace{1cm} 
- Z_\spe\lambda_\spe \bun \cdot
\nabla_\bR \varphi_1^\lw \partial_u G_{\sigma 1}^\lw
\nonumber\\
\fl\hspace{1cm} 
 - u
\bun \cdot \nabla_\bR \left [ \frac{I}{B} \left ( \mu
\partial_\psi B  + Z_\spe
\partial_\psi\varphi_0  \right ) \right ]
\partial_u G_{\sigma 1}^\lw \nonumber\\
\fl\hspace{1cm} 
+ \bun \cdot
\nabla_\bR \Theta \partial_\psi  \left (
\frac{Iu\mu}{\bB \cdot \nabla_\bR \Theta}
\bun \cdot \nabla_\bR B \right ) \partial_u
G_{\sigma 1}^\lw \nonumber\\
\fl\hspace{1cm} 
- \frac{u}{B}
(\bun \cdot \nabla_\bR \times \bun) \left ( u \bun \cdot
\nabla_\bR - \mu \bun \cdot \nabla_\bR B \partial_u \right )
\Bigg \{ G_{\sigma 1}^\lw 
\nonumber\\
\fl\hspace{1cm} 
- \frac{Iu}{B}
F_{\sigma 0} \left(\frac{Z_\spe}{ T_\sigma}
\partial_\psi\varphi_0  + \Upsilon_\spe
\right)
\Bigg \} \nonumber\\
\fl\hspace{1cm} 
+\frac{Z_\spe\lambda_\spe I}{B}
\bun \cdot \nabla_\bR \varphi_1^\lw F_{\sigma 0}
\Upsilon_\spe
\nonumber\\
\fl\hspace{1cm} 
-\frac{Z_\spe\lambda_\spe}{T_\spe B} F_{\spe 0}
\left[
\mu \bun \times \nabla_\bR B +
u^2 \bun \times (\bun \cdot \nabla_\bR \bun)
\right]\cdot \nabla_\bR \varphi_1^\lw
\nonumber\\
\fl\hspace{1cm} 
- \left ( u \bun
\cdot \nabla_\bR - \mu \bun \cdot \nabla_\bR B
\partial_u \right ) \Bigg\{
\frac{1}{2} \left (
\frac{Z_\spe\lambda_\spe \varphi_1^\lw}{T_\spe} \right )^2 F_{\sigma 0}
\nonumber\\
\fl\hspace{1cm} 
+\frac{Z_\spe\lambda_\spe Iu}{T_\spe B}\varphi_1^\lw F_{\spe 0}
\Bigg[ \frac{Z_\spe}{T_\sigma}
\partial_\psi\varphi_0 
+
\Upsilon_\spe
-\frac{1}{T_\sigma} \partial_\psi T_\sigma
\Bigg]
\nonumber\\
\fl\hspace{1cm} 
+\frac{1}{2} \left ( \frac{Iu}{B} \right )^2
F_{\spe 0}
\Bigg[
\partial_\psi\Upsilon_\spe
-\frac{\mu\partial_\psi B}{T_\spe}\partial_\psi\ln T_\spe
\nonumber\\
\fl\hspace{1cm} 
+ \partial_\psi  \left (
\frac{Z_\spe}{T_\spe} \partial_\psi\varphi_0  \right )
 -
\frac{Z_\spe}{T_\sigma^2}
\partial_\psi \varphi_0
\partial_\psi T_\sigma 
\nonumber\\
\fl\hspace{1cm} 
+\Bigg(
\frac{Z_\spe}{T_\spe} \partial_\psi \varphi_0 +\Upsilon_\spe
\Bigg )^2
\Bigg]
\Bigg\}.
\end{eqnarray}
To simplify this expression we use
\begin{eqnarray}
\fl \frac{1}{B} \left ( Z_\spe
\partial_\psi\varphi_0  + \mu \partial_\psi
B \right ) I \bun \cdot \nabla_\bR G_{\sigma 1}^\lw
=\nonumber\\[5pt]
\fl\hspace{1cm}
\partial_u \left [ \frac{I}{B}
\left(
Z_\spe
\partial_\psi\varphi_0 + \mu \partial_\psi B 
\right) u \bun \cdot
\nabla_\bR G_{\sigma 1}^\lw \right ] \nonumber\\[5pt]
\fl\hspace{1cm}
- \frac{Iu}{B}
\left ( Z_\spe
\partial_\psi\varphi_0  + \mu \partial_\psi B  \right )
 \bun \cdot \nabla_\bR \left (
\partial_u G_{\sigma 1}^\lw  \right ),
\end{eqnarray}
\begin{eqnarray}
\fl u \bun \cdot \nabla_\bR \left ( \frac{I u}{B}
\partial_\psi G_{\sigma 1}^\lw  \right ) -
\frac{I}{B} \partial_\psi G_{\sigma 1}^\lw 
\mu \bun \cdot \nabla_\bR B =
\nonumber\\[5pt]
\fl\hspace{1cm}
 \left ( u \bun \cdot \nabla_\bR -
\mu \bun \cdot \nabla_\bR B \partial_u \right )
\left ( \frac{I u}{B} \partial_\psi
G_{\sigma 1}^\lw \right )
\nonumber\\
\fl\hspace{1cm}
+ \frac{Iu}{B} \partial_{\psi}\partial_\mu G_{\sigma 1}^\lw
 \mu \bun \cdot \nabla_\bR B,
\end{eqnarray}
\begin{eqnarray}
\fl - Z_\spe\lambda_\spe \bun \cdot \nabla_\bR \varphi_1^\lw
\partial_u G_{\sigma 1}^\lw 
=
\nonumber\\[5pt]
\fl\hspace{1cm}
- \left ( u \bun \cdot
\nabla_\bR - \mu \bun \cdot \nabla_\bR B \partial_u \right ) \left( \frac{Z_\spe\lambda_\spe \varphi_1^\lw}{u}
\partial_u G_{\sigma 1}^\lw  \right ) \nonumber\\
\fl\hspace{1cm} +
\partial_u
\left[
\frac{Z_\spe\lambda_\spe\varphi_1^\lw}{u}
\left ( u \bun \cdot \nabla_\bR G_{\sigma 1}^\lw
- \mu \bun \cdot \nabla_\bR B \partial_u G_{\sigma 1}^\lw  \right ) \right ],
\end{eqnarray}
\begin{eqnarray}
\fl - u \bun \cdot \nabla_\bR \left [ \frac{I}{B}
\left ( \mu \partial_\psi B 
 + Z_\spe
 \partial_\psi\varphi_0  \right )
\right ] \partial_u G_{\sigma 1}^\lw=
\nonumber\\
\fl\hspace{1cm}
-
\left ( u \bun \cdot \nabla_\bR - \mu \bun \cdot \nabla_\bR B
\partial_u \right ) \left [
\frac{I}{B} \left ( \mu \partial_\psi B  + Z_\spe \partial_\psi\varphi_0  \right ) \partial_u
G_{\sigma 1}^\lw  \right ] \nonumber\\
\fl\hspace{1cm}
 -
\partial_u \left [ \frac{I}{B} \left
( \mu \partial_\psi B 
 +Z_\spe
\partial_\psi\varphi_0  \right ) \mu
\bun \cdot \nabla_\bR B \partial_u G_{\sigma 1}^\lw 
\right ] \nonumber\\
\fl\hspace{1cm}
+ \frac{I u}{B} \left ( \mu
\partial_\psi B 
+ Z_\spe
\partial_\psi\varphi_0  \right ) \bun \cdot
\nabla_\bR \left ( \partial_u G_{\sigma 1}^\lw  \right
),
\end{eqnarray}
\begin{eqnarray}
\fl \bun \cdot \nabla_\bR \Theta \partial_\psi 
\left ( \frac{Iu\mu}{\bB\cdot \nabla_\bR \Theta}
\bun \cdot \nabla_\bR B \right )
\partial_u G_{\sigma 1}^\lw  =
\nonumber\\
\fl\hspace{1cm}
 \bun \cdot \nabla_\bR
\Theta \partial_\psi  \left ( \frac{Iu\mu}{\bB
\cdot \nabla_\bR \Theta}\bun \cdot
\nabla_\bR B \partial_u G_{\sigma 1}^\lw  \right )
\nonumber\\
\fl\hspace{1cm}
- \frac{Iu\mu}{B} \bun \cdot \nabla_\bR B
\partial_\psi \partial_u G_{\sigma 1}^\lw,
\end{eqnarray}
and
\begin{eqnarray}
\fl \frac{Z_\spe\lambda_\spe I}{B}
\bun& \cdot
\nabla_\bR \varphi_1^\lw F_{\spe 0}
\Upsilon_\spe
\nonumber\\
\fl& -
\frac{Z_\spe\lambda_\spe}{T_\sigma B} F_{\spe 0}\left [
\mu \bun \times \nabla_\bR B +
u^2 \bun \times (\bun \cdot \nabla_\bR \bun)
\right ] \cdot \nabla_\bR \varphi_1^\lw\nonumber\\
\fl&= \frac{Z_\spe\lambda_\spe}{B} (\nabla_\bR \varphi_1^\lw \times \bun) \cdot
\nabla_\bR F_{\sigma 0}
\nonumber\\
\fl&
 + \frac{Z_\spe\lambda_\spe u}{B} [\bun \times (\bun \cdot
\nabla_\bR \bun)] \cdot \nabla_\bR \varphi_1^\lw \partial_u
F_{\sigma 0}.
\end{eqnarray}
Employing these results in \eq{eq:auxEqToBeSimplified} gives
\begin{eqnarray}\label{eq:EqSimplified}
\fl \frac{1}{B}\Bigg[ - Z_\spe
 \nabla_\bR \varphi_0 \times \bun +
\mu \bun \times \nabla_\bR B 
\nonumber\\
\fl\hspace{0.5cm} 
+
u^2 \bun \times (\bun \cdot \nabla_\bR \bun)
\Bigg] \cdot \nabla_\bR F_{ \sigma 1}^\lw \nonumber\\
\fl\hspace{0.5cm} 
- \Bigg\{
Z_\spe\lambda_\spe \bun \cdot \nabla_\bR \varphi_1^\lw +
\frac{u}{B} [\bun \times (\bun \cdot \nabla_\bR \bun)]
\nonumber\\
\fl\hspace{0.5cm} 
\cdot \left ( \mu \nabla_\bR B + Z_\spe
\nabla_\bR \varphi_0 \right ) \Bigg\} \partial_u
F_{\sigma 1}^\lw \nonumber\\
\fl\hspace{0.5cm} 
= -\bun\cdot\nabla_\bR\Theta
\partial_\psi
\Bigg[
\frac{Iu}{\bB\cdot\nabla_\bR\Theta}
\left ( u \bun
\cdot \nabla_\bR - \mu \bun \cdot \nabla_\bR B
\partial_u \right )G_{\spe 1}^\lw
\Bigg]
\nonumber\\
\fl\hspace{0.5cm} 
+\partial_u
\Bigg\{
\Bigg[\frac{I}{B}
(Z_\spe\partial_\psi\varphi_0 + \mu\partial_\psi B)
+
\frac{Z_\spe\lambda_\spe}{u}\varphi_1^\lw\Bigg]
\nonumber\\
\fl\hspace{0.5cm} 
\times
\left ( u \bun
\cdot \nabla_\bR - \mu \bun \cdot \nabla_\bR B
\partial_u \right )G_{\spe 1}^\lw
\Bigg\}
\nonumber\\
\fl\hspace{0.5cm} 
-
\frac{u}{B}
\left(
\bun\cdot\nabla_\bR\times\bun
 \right )
\left( u \bun
\cdot \nabla_\bR - \mu \bun \cdot \nabla_\bR B
\partial_u \right )
\Bigg[
G_{\spe 1}^\lw
\nonumber\\
\fl\hspace{0.5cm} 
-
\frac{Iu}{B}
\left(\frac{Z_\spe}{T_\spe}
\partial_\psi\varphi_0 + \Upsilon_\spe\right)
F_{\spe 0}
\Bigg]
\nonumber\\
\fl\hspace{0.5cm} 
+
\left( u \bun
\cdot \nabla_\bR - \mu \bun \cdot \nabla_\bR B
\partial_u \right )
\Bigg\{
\frac{Iu}{B}
\partial_\psi G_{\spe 1}^\lw
-
\frac{Z_\spe\lambda_\spe}{u}\varphi_1^\lw
\partial_u
G_{\spe 1}^\lw
\nonumber\\
\fl\hspace{0.5cm} 
-\frac{I}{B}
(Z_\spe\partial_\psi\varphi_0 + \mu\partial_\psi B)
\partial_u
G_{\spe 1}^\lw
-\frac{1}{2}
\left(
\frac{Z_\spe\lambda_\spe \varphi_1^\lw}{T_\spe}
\right)^2F_{\spe 0}
\nonumber\\
\fl\hspace{0.5cm} 
-
\frac{Z_\spe\lambda_\spe Iu}{T_\spe B}\varphi_1^\lw
F_{\spe 0}
\Bigg(
\frac{Z_\spe}{T_\spe}
\partial_\psi\varphi_0 +\Upsilon_\spe
 -\partial_\psi\ln T_\spe
\Bigg)
\nonumber\\
\fl\hspace{0.5cm} 
-\frac{1}{2}
\left(\frac{Iu}{B}\right)^2
\Bigg[
\partial_\psi\Upsilon_\spe
-\frac{\mu\partial_\psi B}{T_\spe}\partial_\psi\ln T_\spe
+\partial_\psi
\left(
\frac{Z_\spe}{T_\spe}\partial_\psi\varphi_0
\right)
\nonumber\\
\fl\hspace{0.5cm}
-
\frac{Z_\spe}{T_\spe^2}\partial_\psi\varphi_0
\partial_\psi T_\spe
+
\Bigg(
\frac{Z_\spe}{T_\spe}
\partial_\psi\varphi_0 +\Upsilon_\spe
\Bigg)^2
\Bigg]
\Bigg\}
\nonumber\\
\fl\hspace{0.5cm}
+
\frac{Z_\spe\lambda_\spe}{B}
\left(\nabla_\bR\varphi_1^\lw\times\bun\right)\cdot
\nabla_\bR F_{\spe 0}
\nonumber\\
\fl\hspace{0.5cm}
+
\frac{Z_\spe\lambda_\spe u}{B}
\left[\bun\times\left(\bun\cdot\nabla_\bR\bun\right)\right]
\cdot
\nabla_\bR\varphi_1^\lw
\partial_u F_{\spe 0}.
\end{eqnarray}
We also manipulate the terms containing $\bK$ (defined by
equation \eq{eq:defvectorK}) in \eq{eq:FPsecondorder}. First, note that
\begin{eqnarray}
\fl - \frac{u \mu}{B}&
(\nabla_\bR \times \bK)_\bot \cdot \nabla_\bR F_{\sigma 0}
\nonumber\\
\fl&
+
\frac{\mu}{B} (\nabla_\bR \times
\bK)_\bot \cdot \left ( \mu \nabla_\bR B
 + Z_\spe
\nabla_\bR \varphi_0 \right ) \partial_u
F_{\sigma 0}=
\nonumber\\
\fl&
-\frac{u\mu}{B}\left(
\frac{Z_\spe}{T_\sigma} \partial_\psi\varphi_0 
+\Upsilon_\spe
 \right)(\nabla_\bR \times \bK)\cdot\nabla_\bR\psi
F_{\spe 0}.
\end{eqnarray}
Using 
\ref{sec:proofofidentityvectorK} one finds
\begin{eqnarray}\label{eq:identityvectorK}
\fl (\nabla_\bR \times \bK) \cdot \nabla_\bR \psi = \bB \cdot
\nabla_\bR \Bigg( \frac{I}{2B} \bun \cdot \nabla_\bR \times \bun
\nonumber\\
\fl\hspace{1cm}
+ \frac{R}{|\nabla_\bR \psi|^2} \zun \cdot \nabla_\bR \nabla_\bR \psi
\cdot (\bun \times \nabla_\bR \psi) \Bigg),
\end{eqnarray}
so
\begin{eqnarray}\label{eq:AnotherAuxEq}
\fl - \frac{u \mu}{B}
(\nabla_\bR \times \bK)_\bot \cdot \nabla_\bR F_{\sigma 0}
\nonumber\\
\fl\hspace{1cm}+
\frac{\mu}{B} (\nabla_\bR \times
\bK)_\bot \cdot \left ( \mu \nabla_\bR B
 + Z_\spe
\nabla_\bR \varphi_0 \right ) \partial_u
F_{\sigma 0}=
\nonumber\\
\fl\hspace{1cm} - \left ( u \bun \cdot \nabla_\bR - \mu \bun \cdot
\nabla_\bR B \partial_u  \right )
\Bigg \{
 {\mu F_{\sigma 0}}
\nonumber\\
\fl\hspace{1cm} 
\times\Bigg(
\frac{Z_\spe}{T_\spe}\partial_\psi\varphi_0
+
\Upsilon_\spe
\Bigg)
\Bigg( \frac{I}{2B} \bun \cdot \nabla_\bR \times \bun
\nonumber\\
\fl\hspace{1cm}+ \frac{R}{|\nabla_\bR \psi|^2} \zun \cdot \nabla_\bR \nabla_\bR \psi
\cdot (\bun \times \nabla_\bR \psi) \Bigg)\Bigg \}
.
\end{eqnarray}

Hence, employing \eq{eq:EqSimplified} and \eq{eq:AnotherAuxEq}, and
reorganizing, the second-order Fokker-Planck equation,
\eq{eq:FPsecondorder}, becomes
\begin{eqnarray} \label{eq:secondorder2}
\fl -B\partial_\theta F_{\spe 3}^\lw
+ \frac{\lambda_\spe^2}{\tau_\spe}\partial_{\epsilon_s^2 t} F_{\spe 0} +
\Bigg ( u \bun \cdot \nabla_\bR  - \mu \bun \cdot \nabla_\bR
B \partial_u  \Bigg )
\Bigg\{
F_{\sigma 2}^\lw\nonumber\\
\fl\hspace{0.5cm}
+
\left[
\frac{Z_\spe\lambda_\spe^2}{T_\sigma}\varphi_2^\lw
+ \frac{Z_\spe}{T_\sigma}
\frac{\mu}{2B}(\matI - \bun \bun) : \nabla_\bR
\nabla_\bR \varphi_0
\right] F_{\spe 0}
 \nonumber\\
\fl\hspace{0.5cm}
+\frac{1}{T_\spe}
\left(
\Psi_{B,\spe} + Z_\spe\lambda_\spe \Psi_{\phi B,\spe}^\lw
 + Z_\spe^2\lambda_\spe^2\Psi_{\phi,\spe}^\lw
\right)
F_{\spe 0}
\nonumber\\
\fl\hspace{0.5cm}
+
\frac{Iu}{B}
\partial_\psi G_{\sigma 1}^\lw   -
\frac{Z_\spe\lambda_\spe \varphi_1^\lw}{u} \partial_u
G_{\sigma 1}^\lw 
\nonumber\\
\fl\hspace{0.5cm} -
\frac{I}{B} \left ( \mu \partial_\psi B  
+ Z_\spe \partial_\psi\varphi_0  \right )
\partial_u G_{\sigma 1}^\lw 
- \frac{1}{2}
\left(
\frac{Z_\spe\lambda_\spe\varphi_1^\lw}{T_\sigma}
\right)^2
F_{\sigma 0}
 \nonumber\\\fl\hspace{0.5cm}
 - \frac{Z_\spe\lambda_\spe I u}{T_\spe B}
\varphi_1^\lw F_{\spe 0} \Bigg [
\frac{Z_\spe}{T_\sigma}
\partial_\psi\varphi_0 +
\Upsilon_\spe
-
\frac{1}{T_\sigma} \partial_\psi
T_\sigma
\Bigg ] \nonumber\\\fl\hspace{0.5cm} - \frac{1}{2}
\left ( \frac{Iu}{B} \right )^2 F_{\spe 0} \Bigg [
\partial_\psi\Upsilon_\spe -\frac{\mu\partial_\psi B}{T_\spe}
\partial_\psi\ln T_\spe
\nonumber\\\fl\hspace{0.5cm} 
  +
\partial_\psi  \left (
\frac{Z_\spe}{T_\sigma}
\partial_\psi\varphi_0 
 \right ) - \frac{Z_\spe}{ T_\sigma^2}
 \partial_\psi\varphi_0 
\partial_\psi T_\sigma  
 \nonumber\\ \fl\hspace{0.5cm} + \Bigg (
\frac{Z_\spe}{T_\sigma}
\partial_\psi\varphi_0  + \Upsilon_\spe
  \Bigg )^2 \Bigg ]-  {\mu F_{\spe 0}}
\Bigg( \frac{I}{2B} \bun \cdot \nabla_\bR \times \bun
\nonumber\\\fl\hspace{0.5cm}
+ \frac{R}{|\nabla_\bR \psi|^2} \zun \cdot \nabla_\bR \nabla_\bR \psi
\cdot (\bun \times \nabla_\bR \psi) \Bigg)
 \left( 
\frac{Z_\spe}{T_\sigma}
\partial_\psi\varphi_0  + \Upsilon_\spe \right)
\nonumber\\
\fl\hspace{0.5cm} + \left[ \frac{Z_\spe\lambda_\spe }{u \bB \cdot
\nabla_\bR \Theta} F_{\sigma 1}^\sw (\bun\times
\nabla_{\bR_\perp/\epsilon_\spe}
\langle \phi_{\spe 1}^\sw \rangle) \cdot \nabla_\bR
\Theta \right]^\lw \Bigg \} \nonumber\\
\fl\hspace{0.5cm}
-\bun \cdot \nabla_\bR \Theta \partial_\psi  \left [
\frac{Iu}{\bB \cdot \nabla_\bR \Theta}
\left(
u \bun \cdot \nabla_\bR  - \mu \bun \cdot
\nabla_\bR B \partial_u
\right )G_{\spe 1}^\lw
\right ] \nonumber\\
\fl\hspace{0.5cm} + \partial_u  \left [
\frac{I}{B} \left ( \mu \partial_\psi B  
+ Z_\spe \partial_\psi\varphi_0  \right )
\left(
u \bun \cdot \nabla_\bR  - \mu \bun \cdot
\nabla_\bR B \partial_u
\right )G_{\spe 1}^\lw
\right ] \nonumber\\
\fl\hspace{0.5cm} +
\partial_u 
\left[
\frac{Z_\spe\lambda_\spe\varphi_1^\lw}{u}
\left(
u \bun \cdot \nabla_\bR  - \mu \bun \cdot
\nabla_\bR B \partial_u
\right )G_{\spe 1}^\lw
\right ] \nonumber \\
\fl\hspace{0.5cm} - \frac{u}{B}
\bun \cdot \nabla_\bR \times \bun
\left(
u \bun \cdot \nabla_\bR  - \mu \bun \cdot
\nabla_\bR B \partial_u
\right )G_{\spe 1}^\lw
\nonumber\\
\fl\hspace{0.5cm}
+\frac{Z_\spe\lambda_\spe}{B}
\bB \cdot \nabla_\bR \Theta
\partial_\psi 
\left[ \frac{F_{\spe 1}^\sw}{\bB \cdot
\nabla_\bR \Theta}
\left(\bun\times\nabla_{\bR_\perp/\epsilon_\spe}
\langle \phi_{\spe 1}^\sw \rangle \right)
\cdot \nabla_\bR\psi
\right]^\lw
\nonumber\\
\fl\hspace{0.5cm} - Z_\spe\lambda_\spe
\partial_u 
\Bigg[
\Bigg( 
\frac{\mu}{ u B}
(\bun \times \partial_\Theta B\nabla_\bR \Theta)
\cdot\nabla_{\bR_\perp/\epsilon_\spe}
\langle\phi_{\spe 1}\rangle^\sw \nonumber\\
\fl\hspace{0.5cm}
 + \frac{u}{B}
\left[\bun \times (\bun \cdot \nabla_\bR \bun)\right]
\cdot\nabla_{\bR_\perp/\epsilon_\spe}
\langle\phi_{\spe 1}\rangle^\sw
\nonumber\\
\fl\hspace{0.5cm}
+\bun\cdot\nabla_\bR\langle\phi_{\spe 1}\rangle^\sw
\Bigg)
F_{\spe 1}^\sw
\Bigg]^\lw
\nonumber\\
\fl\hspace{0.5cm}
=
\sum_{\spe'}\cT_{\spe,0}^{*}C_{\spe\spe'}^{(2)\lw}
+\sum_{\spe'}
\left[\cT_{\spe,1}^*C_{\spe\spe'}^{(1)}\right]^\lw.
\end{eqnarray}

Finally, defining
\begin{eqnarray}\label{eq:defG2}
\fl G_{\spe 2}^\lw & = \langle F_{\sigma 2}^\lw \rangle
\nonumber\\
\fl&
+
\left[
\frac{Z_\spe\lambda_\spe^2}{T_\sigma}\varphi_2^\lw
+ \frac{Z_\spe}{T_\sigma}
\frac{\mu}{2B}(\matI - \bun \bun) : \nabla_\bR
\nabla_\bR \varphi_0
\right] F_{\spe 0}
 \nonumber\\
\fl&
+\frac{1}{T_\spe}
\left(
\Psi_B + Z_\spe\lambda_\spe \Psi_{\phi B, \spe}^\lw
 + Z_\spe^2\lambda_\spe^2\Psi_{\phi,\spe}^\lw
\right)
F_{\spe 0}
\nonumber\\
\fl &
+
\frac{Iu}{B}
\partial_\psi G_{\sigma 1}^\lw   -
\frac{Z_\spe\lambda_\spe \varphi_1^\lw}{u} \partial_u
G_{\sigma 1}^\lw 
\nonumber\\
\fl & -
\frac{I}{B} \left ( \mu \partial_\psi B  + Z_\spe \partial_\psi\varphi_0  \right )
\partial_u G_{\sigma 1}^\lw 
- \frac{1}{2}
\left(
\frac{Z_\spe\lambda_\spe\varphi_1^\lw}{T_\sigma}
\right)^2
F_{\sigma 0}
 \nonumber\\\fl &
 - \frac{Z_\spe\lambda_\spe I u}{T_\spe B}
\varphi_1^\lw F_{\spe 0} \Bigg(
\frac{Z_\spe}{T_\sigma}
\partial_\psi\varphi_0 +\Upsilon_\spe - \frac{1}{T_\sigma} \partial_\psi
T_\sigma 
 \Bigg) \nonumber\\\fl & - \frac{1}{2}
\left ( \frac{Iu}{B} \right )^2 F_{\spe 0} \Bigg [
\partial_\psi\Upsilon_\spe
-
\frac{\mu\partial_\psi B}{T_\spe}
\partial_\psi\ln T_\spe \nonumber\\\fl & 
 +
\partial_\psi  \left (
\frac{Z_\spe}{T_\sigma}
\partial_\psi\varphi_0 
 \right ) - \frac{Z_\spe}{T_\sigma^2}
 \partial_\psi\varphi_0 
 \partial_\psi T_\sigma 
 + \Bigg (
 \frac{Z_\spe}{T_\sigma}
\partial_\psi\varphi_0 +\Upsilon_\spe\Bigg )^2 \Bigg ]
 \nonumber\\\fl &
-  {\mu F_{\spe 0}}
\Bigg( \frac{I}{2B} \bun \cdot \nabla_\bR \times \bun
 \nonumber\\\fl &
+ \frac{R}{|\nabla_\bR \psi|^2} \zun \cdot \nabla_\bR \nabla_\bR \psi
\cdot (\bun \times \nabla_\bR \psi) \Bigg)
 \Bigg(
\frac{Z_\spe}{T_\sigma}
\partial_\psi\varphi_0  +\Upsilon_\spe  \Bigg)
\nonumber\\
\fl& + \left[ \frac{Z_\spe\lambda_\spe}{
u \bB \cdot
\nabla_\bR \Theta} F_{\sigma 1}^\sw (\bun\times
\nabla_{\bR_\perp/\epsilon_\spe}
\langle \phi_{\spe 1}^\sw \rangle) \cdot \nabla_\bR
\Theta \right]^\lw
\nonumber\\
\fl& -
 \left \langle \frac{1}{ u \bun \cdot
  \nabla_\bR \Theta} \rhobf \cdot \nabla_\bR \Theta
\sum_{\sigma^\prime}\cT_{\spe,0}^*C_{\sigma
    \sigma^\prime}^{(1)\lw} \right \rangle;
\end{eqnarray}
using the results in \ref{sec:pullback} and
\begin{eqnarray}
B^{-1}\nabla_\bR\cdot(B\rhobf)
+\partial_u\hat{u}_1
+\partial_\mu\hat{\mu}_1
+\partial_\theta\hat{\theta}_1
=
\frac{u}{B}\bun\cdot\nabla_\bR\times\bun
\end{eqnarray}
to write
\begin{eqnarray}
\fl
\left\langle
\left[
\cT_{\sigma,1}^\ast C_{\sigma
\sigma^\prime}^{(1)}
\right]^\lw
\right\rangle = \nonumber\\
\fl\hspace{0.5cm}\partial_u 
\left(
\left \langle u \bun \cdot \nabla_\bR \bun \cdot \rhobf\,
\cT_{\spe,0}^*C_{\sigma \sigma^\prime}^{(1)\lw}
\right \rangle
-\mu
 \bun \cdot \nabla_\bR \times \bun \left \langle
\cT_{\spe,0}^*C_{\sigma \sigma^\prime}^{(1)\lw}
\right \rangle
\right)
\nonumber\\
\fl\hspace{0.5cm}
+ \partial_\mu
\Bigg\{
\frac{u \mu}{B} \bun \cdot \nabla_\bR \times \bun
\left \langle
\cT_{\spe,0}^*C_{\sigma \sigma^\prime}^{(1)\lw}
\right \rangle
\nonumber\\
\fl\hspace{0.5cm}
-
\left \langle \left (
\frac{Z_\spe}{B} \rhobf \cdot \nabla_\bR
\varphi_0 + \frac{\mu}{B} \rhobf \cdot \nabla_\bR B +
\frac{u^2}{B} \bun \cdot \nabla \bun \cdot \rhobf \right )
\cT_{\spe,0}^*C_{\sigma \sigma^\prime}^{(1)\lw}
\right \rangle
\Bigg\}
 \nonumber\\
\fl\hspace{0.5cm}
- \frac{u}{B}
\bun \cdot \nabla_\bR \times \bun
\left \langle
\cT_{\spe,0}^*C_{\sigma \sigma^\prime}^{(1)\lw}
\right \rangle
+ \frac{1}{B} \nabla_\bR \cdot \left \langle B
\rhobf \,
\cT_{\spe,0}^*C_{\sigma \sigma^\prime}^{(1)\lw}
 \right \rangle  \nonumber\\
\fl\hspace{0.5cm} +
\left[
\cT_{\spe,1}^\ast C_{\sigma
\sigma^\prime}^{(1)\sw}
\right]^\lw;
\end{eqnarray}
and employing \eq{eq:Vlasovorder1gyroav4} and \eq{eq:C1nc}, we can
write the gyroaveraged, long-wavelength second-order Fokker-Planck
equation as in \eq{eq:eqH2sigma}. However, for some purposes, mainly
in connection with the long-wavelength gyrokinetic quasineutrality
equation, it is useful to recast equation \eq{eq:defG2} in a
different fashion. After some straightforward algebra one gets
\begin{eqnarray}\label{eq:defG2again2}
\fl G_{\spe 2}^\lw & = \langle F_{\sigma 2}^\lw \rangle
+
\frac{Z_\spe\lambda_\spe^2}{T_\sigma}\varphi_2^\lw
F_{\spe 0}
+
\frac{Iu}{B}
\partial_\psi G_{\sigma 1}^\lw   -
\frac{Z_\spe\lambda_\spe \varphi_1^\lw}{u} \partial_u
G_{\sigma 1}^\lw 
\nonumber\\
\fl & -
\frac{I}{B} \left ( \mu \partial_\psi B  + Z_\spe \partial_\psi\varphi_0  \right )
\partial_u G_{\sigma 1}^\lw 
- \frac{1}{2}
\left(
\frac{Z_\spe\lambda_\spe\varphi_1^\lw}{T_\sigma}
\right)^2
F_{\sigma 0}
 \nonumber\\\fl &
 - \frac{Z_\spe\lambda_\spe I u}{T_\spe B}
\varphi_1^\lw F_{\spe 0} \Bigg [
\frac{Z_\spe}{T_\sigma}
\partial_\psi\varphi_0+\Upsilon_\spe
- \frac{1}{T_\sigma} \partial_\psi
T_\sigma  \Bigg ]\nonumber\\
\fl &
-\frac{1}{2B^2}\left(
\left ({Iu} \right )^2+\mu B|\nabla_\bR\psi|^2
\right)
\Bigg[
- \frac{2Z_\spe}{T_\sigma^2}
 \partial_\psi\varphi_0 
\partial_\psi T_\sigma
\nonumber\\
\fl &
- \frac{u^2/2 + \mu B}{T_\sigma}
\left( \partial_\psi \ln
T_\sigma \right )^2
+ \Bigg (
\frac{Z_\spe}{T_\sigma}
\partial_\psi\varphi_0 +
\Upsilon_\spe \Bigg )^2
\Bigg]F_{\spe 0}
 \nonumber\\
\fl & 
- \frac{1}{2}
\left ( \frac{Iu}{B} \right )^2 F_{\spe 0} \Bigg [
\partial^2_\psi \ln n_\sigma  +
\frac{Z_\spe}{T_\sigma}
\partial_\psi^2 \varphi_0
\nonumber\\
\fl &
+ \left (
\frac{u^2/2 + \mu B}{T_\sigma} - \frac{3}{2} \right )
\partial^2_\psi \ln T_\sigma 
\Bigg ] \nonumber\\
\fl &+
\mu\Bigg(\frac{1}{2 B^2}\nabla_\bR B\cdot\nabla_\bR\psi
-
\frac{I}{2B} \bun \cdot \nabla_\bR \times \bun
\nonumber\\
\fl &
- \frac{R}{|\nabla_\bR \psi|^2} \zun \cdot \nabla_\bR \nabla_\bR \psi
\cdot (\bun \times \nabla_\bR \psi)
 \Bigg)
\left(\frac{Z_\spe}{T_\sigma}
\partial_\psi\varphi_0+\Upsilon_\spe\right)F_{\spe 0}
\nonumber\\
\fl& + \left[ \frac{Z_\spe\lambda_\spe }{
u \bB \cdot
\nabla_\bR \Theta} F_{\sigma 1}^\sw (\bun\times
\nabla_{\bR_\perp/\epsilon_\spe}
\langle \phi_{\spe 1}^\sw \rangle) \cdot \nabla_\bR
\Theta \right]^\lw
\nonumber\\
\fl& -
 \left \langle \frac{1}{ u \bun \cdot
  \nabla_\bR \Theta} \rhobf \cdot \nabla_\bR \Theta
\sum_{\sigma^\prime} \cT_{\spe, 0}^{*} C_{\sigma
    \sigma^\prime}^{(1)\lw} \right \rangle
\nonumber\\
\fl&
-\frac{\mu}{2B} (\matI - \bun \bun) : \Bigg [
\nabla_\bR \nabla_\bR \ln n_\spe
\nonumber\\
\fl&
 + \left ( \frac{u^2/2+\mu B}{T_\spe} -
\frac{3}{2} \right ) \nabla_\bR \nabla_\bR \ln T_\spe 
\nonumber\\
\fl&
+\frac{Z_\spe}{T_\spe}\nabla_\bR\nabla_\bR\varphi_0
\Bigg ] F_{\spe 0}
- \frac{Z_\spe^2\lambda_\spe^2}{2T_\spe^2}
\left[ \left \langle
(\phiwig_{\spe 1}^\sw)^2 \right \rangle \right]^\lw F_{\spe
0}
\nonumber\\
\fl&
+\mu\bun\cdot\nabla_\bR\times\bun
\left(
\frac{u}{B}\partial_\mu
-\partial_u 
\right)F_{\spe 1}^\lw
\nonumber\\
\fl&
+
\left\langle\cT_{\spe,0}^{*}
\left[\cT_{\spe,1}^{-1*}F_{\spe 1}^\lw\right]^\lw
\right\rangle
+
\left\langle\cT_{\spe,0}^{*}
\left[\cT_{\spe, 2}^{-1*}F_{\spe 0}\right]^\lw
\right\rangle,
\end{eqnarray}
where $\left\langle \cT_{\spe, 0}^{*} \left[\cT_{\spe,1}^{-1*}F_{\spe
      1}^\lw\right]^\lw \right\rangle$ is given in
\eq{eq:T1F1lwAveraged} and $\left \langle \cT_{\spe, 0}^{*}
  \left[\cT_{\spe, 2}^{-1*}F_{\spe 0}\right]^\lw \right \rangle$ is
given in \eq{gyroaverageT2F0}. A less obvious calculation transforms
the previous equation into
\begin{eqnarray}\label{eq:defG2again3}
\fl G_{\spe 2}^\lw & = \langle F_{\sigma 2} \rangle^\lw
+
\frac{Z_\spe\lambda_\spe^2}{T_\sigma}\varphi_2^\lw
F_{\spe 0}
+
\frac{Iu}{B}
\partial_\psi G_{\sigma 1}^\lw   -
\frac{Z_\spe\lambda_\spe \varphi_1^\lw}{u} \partial_u
G_{\sigma 1}^\lw 
\nonumber\\
\fl & -
\frac{I}{B} \left ( \mu \partial_\psi B 
 + Z_\spe \partial_\psi\varphi_0  \right )
\partial_u G_{\sigma 1}^\lw 
- \frac{1}{2}
\left(
\frac{Z_\spe\lambda_\spe\varphi_1^\lw}{T_\sigma}
\right)^2
F_{\sigma 0}
 \nonumber\\\fl &
 - \frac{Z_\spe\lambda_\spe I u}{T_\spe B}
\varphi_1^\lw F_{\spe 0} \Bigg(
\frac{Z_\spe}{T_\sigma}
\partial_\psi\varphi_0 
+\Upsilon_\spe-
 \frac{1}{T_\sigma} \partial_\psi
T_\sigma  \Bigg)\nonumber\\
\fl &
-\frac{1}{2B^2}\left(
\left ({Iu} \right )^2+\mu B|\nabla_\bR\psi|^2
\right)
\Bigg[
- \frac{2Z_\spe}{T_\sigma^2}
 \partial_\psi\varphi_0 
\partial_\psi T_\sigma
\nonumber\\
\fl &
+ \Bigg(
\frac{Z_\spe}{T_\sigma}
\partial_\psi\varphi_0  +\Upsilon_\spe
 \Bigg )^2
+
\partial_\psi^2\ln n_\spe
\nonumber\\
\fl & 
+
\left(\frac{u^2/2 +\mu B}{T_\spe}-\frac{3}{2}\right)
\partial_\psi^2\ln T_\spe
\nonumber\\
\fl & 
-
\frac{u^2/2 +\mu B}{T_\spe}
(\partial_\psi \ln T_\spe)^2
 +
\frac{Z_\spe}{T_\sigma}
\partial_\psi^2 \varphi_0
\Bigg]F_{\spe 0}
 \nonumber\\
\fl& + \left[ \frac{Z_\spe\lambda_\spe}{
u \bB \cdot
\nabla_\bR \Theta} F_{\sigma 1}^\sw (\bun\times
\nabla_{\bR_\perp/\epsilon_\spe}
\langle \phi_{\spe 1}^\sw \rangle) \cdot \nabla_\bR
\Theta \right]^\lw\nonumber\\
\fl& -
\left \langle \frac{1}{ u \bun \cdot
  \nabla_\bR \Theta} \rhobf \cdot \nabla_\bR \Theta
\sum_{\sigma^\prime}\cT_{\spe,0}^*C_{\sigma
    \sigma^\prime}^{(1)\lw} \right \rangle
\nonumber\\
\fl
&- \frac{Z_\spe^2\lambda_\spe^2}{2T_\spe^2}
\left[
\left \langle 
(\phiwig_{\spe 1}^\sw)^2 \right \rangle \right]^\lw F_{\spe
0}
+\mu\bun\cdot\nabla_\bR\times\bun
\left(
\frac{u}{B}\partial_\mu
-\partial_u 
\right)G_{\spe 1}^\lw
\nonumber\\
\fl&
+ 
\left\langle \cT_{\spe,0}^{*}
\left[\cT_{\spe,1}^{-1*}F_{\spe 1}^\lw\right]^\lw
\right\rangle
+
\left\langle \cT_{\spe,0}^{*}
\left[\cT_{\spe,2}^{-1*}F_{\spe 0}\right]^\lw
\right\rangle.
\end{eqnarray}
To obtain \eq{eq:defG2again3} from
\eq{eq:defG2again2}  we used equation \eq{eq:defGspe1} and
\begin{eqnarray} \label{Bcancellation}
\fl \frac{1}{B} \nabla_\bR B \cdot \nabla_\bR \psi + I \bun \cdot
\nabla_\bR \times \bun
 \nonumber\\
\fl\hspace{1cm}
- \frac{2RB}{|\nabla_\bR \psi|^2} \zun \cdot
\nabla_\bR \nabla_\bR \psi \cdot (\bun \times \nabla_\bR \psi)
\nonumber\\
\fl\hspace{1cm}
- \nabla_\bR \nabla_\bR \psi : (\matI - \bun \bun) = 0.
\end{eqnarray}
Let us prove this. First, we have that
\begin{eqnarray}\label{eq:HorribleAppendixAuxEq1}
\fl \nabla_\bR B \cdot \nabla_\bR \psi = \frac{I}{R^2 B}
\nabla_\bR I \cdot \nabla_\bR \psi
\nonumber\\[5pt]
\fl\hspace{1cm}
+ \frac{1}{R^2 B} \nabla_\bR
\psi \cdot \nabla_\bR \nabla_\bR \psi \cdot \nabla_\bR \psi -
\frac{B}{R} \nabla_\bR R \cdot \nabla_\bR \psi,
\end{eqnarray}
where we have employed that $B^2 = (I^2 + |\nabla_\bR
\psi|^2)/R^2$. Noting that $\partial_\zeta (\nabla_\bR \psi) =
(\nabla_\bR \psi \cdot \nabla_\bR R) \zun$ we derive the following
identities
\begin{eqnarray}\label{eq:HorribleAppendixAuxEq2}
\fl \bun \cdot \nabla_\bR \times \bun = \frac{1}{B^2} \bB \cdot
(\nabla_\bR \times \bB) \nonumber\\
\fl\hspace{1cm}
= \frac{1}{B^2} \bB \cdot \left
[ \nabla_\bR I \times \nabla_\bR \zeta + \nabla_\bR \cdot \left (
\nabla_\bR \psi \nabla_\bR \zeta \right ) - \nabla_\bR \cdot \left
(\nabla_\bR \zeta \nabla_\bR \psi \right ) \right ] \nonumber\\
\fl\hspace{1cm}
=
\frac{1}{B^2} \bB \cdot \left [ \nabla_\bR I \times \nabla_\bR
\zeta + \left ( \nabla_\bR^2 \psi - \frac{2}{R} \nabla_\bR \psi
\cdot \nabla_\bR R \right )  \nabla_\bR \zeta \right ]
\nonumber\\
\fl\hspace{1cm}
= - \frac{1}{R^2 B^2} \nabla_\bR I \cdot \nabla_\bR
\psi + \frac{I}{R^2 B^2} \nabla_\bR^2 \psi - \frac{2I}{R^3 B^2}
\nabla_\bR \psi \cdot \nabla_\bR R,
\end{eqnarray}
\begin{eqnarray}\label{eq:HorribleAppendixAuxEq3}
\fl \zun \cdot \nabla_\bR \nabla_\bR \psi \cdot (\bun \times
\nabla_\bR \psi) = - \frac{|\nabla_\bR \psi|^2}{R B} \zun \cdot
\nabla_\bR \nabla_\bR \psi \cdot \zun
\nonumber\\
\fl\hspace{1cm}
 + \frac{I}{B} \zun \cdot
\nabla_\bR \nabla_\bR \psi \cdot (\nabla_\bR \zeta \times
\nabla_\bR \psi) \nonumber\\
\fl\hspace{1cm}
= - \frac{|\nabla_\bR \psi|^2}{R^2 B}
\nabla_\bR R \cdot \nabla_\bR \psi,
\end{eqnarray}
\begin{eqnarray}\label{eq:HorribleAppendixAuxEq4}
\fl \bun \cdot \nabla_\bR \nabla_\bR \psi \cdot \bun =
\frac{I^2}{R^2 B^2} \zun \cdot \nabla_\bR \nabla_\bR \psi \cdot
\zun 
\nonumber\\
\fl\hspace{1cm}
+\frac{2I}{RB^2} \zun \cdot \nabla_\bR \nabla_\bR \psi \cdot
(\nabla_\bR \zeta \times \nabla_\bR \psi) \nonumber\\
\fl\hspace{1cm}
+
\frac{1}{B^2} (\nabla_\bR \zeta \times \nabla_\bR \psi) \cdot
\nabla_\bR \nabla_\bR \psi \cdot (\nabla_\bR \zeta \times
\nabla_\bR \psi)  \nonumber\\
\fl\hspace{1cm}
= \frac{I^2}{R^3 B^2} \nabla_\bR R \cdot
\nabla_\bR \psi
 \nonumber\\
\fl\hspace{1cm}
+ \frac{1}{B^2} (\nabla_\bR \zeta
\times \nabla_\bR \psi) \cdot \nabla_\bR \nabla_\bR \psi \cdot
(\nabla_\bR \zeta \times \nabla_\bR \psi),
\end{eqnarray}
and
\begin{eqnarray}\label{eq:HorribleAppendixAuxEq5}
\fl (\nabla_\bR \zeta \times \nabla_\bR \psi) \cdot \nabla_\bR
\nabla_\bR \psi \cdot (\nabla_\bR \zeta \times \nabla_\bR \psi)
\nonumber\\
\fl\hspace{1cm}= - (\nabla_\bR \zeta \times \nabla_\bR \psi) \cdot
\nabla_\bR (\nabla_\bR \zeta \times \nabla_\bR \psi) \cdot
\nabla_\bR \psi \nonumber\\
\fl\hspace{1cm}= - \nabla_\bR \psi \cdot \nabla_\bR
(\nabla_\bR \zeta \times \nabla_\bR \psi) \cdot (\nabla_\bR \zeta
\times \nabla_\bR \psi) \nonumber\\
\fl\hspace{1cm}+ \nabla_\bR \psi \cdot \{
(\nabla_\bR \zeta \times \nabla_\bR \psi) \times [ \nabla_\bR
\times (\nabla_\bR \zeta \times \nabla_\bR \psi) ] \} \nonumber\\
\fl\hspace{1cm}=
- \nabla_\bR \psi \cdot \nabla_\bR \left ( \frac{|\nabla_\bR
\psi|^2}{2 R^2} \right ) 
\nonumber\\
\fl\hspace{1cm}
+ \frac{|\nabla_\bR \psi|^2}{R^2} \left (
\nabla_\bR^2 \psi - \frac{2}{R} \nabla_\bR R \cdot \nabla_\bR \psi
\right ) \nonumber\\
\fl\hspace{1cm}= - \frac{1}{R^2} \nabla_\bR \psi \cdot
\nabla_\bR \nabla_\bR \psi \cdot \nabla_\bR \psi 
\nonumber\\
\fl\hspace{1cm}
+
\frac{|\nabla_\bR \psi|^2}{R^2} \nabla_\bR^2 \psi -
\frac{|\nabla_\bR \psi|^2}{R^3} \nabla_\bR R \cdot \nabla_\bR
\psi.
\end{eqnarray}
Using relations \eq{eq:HorribleAppendixAuxEq1},
\eq{eq:HorribleAppendixAuxEq2}, \eq{eq:HorribleAppendixAuxEq3},
\eq{eq:HorribleAppendixAuxEq4}, and \eq{eq:HorribleAppendixAuxEq5}, it
is trivial to check that \eq{Bcancellation} is satisfied.

\section{Proof of \eq{eq:identityvectorK}}
\label{sec:proofofidentityvectorK}

First,
\begin{eqnarray}
\fl (\nabla_\bR \times \bK) \cdot \nabla_\bR \psi =
 \nonumber\\
 \fl\hspace{1cm}\bB \cdot
\nabla_\bR \Theta \partial_\Theta \left [
\frac{1}{\bB \cdot \nabla_\bR \Theta} \bK \cdot (\nabla_\bR
\psi \times \nabla_\bR \Theta) \right ]
\nonumber\\
\fl\hspace{1cm}
+\bB \cdot
\nabla_\bR \Theta \partial_\zeta \left [
\frac{1}{\bB \cdot \nabla_\bR \Theta} \bK \cdot (\nabla_\bR
\psi \times \nabla_\bR \zeta) \right ].
\end{eqnarray}
Employing that $\partial_\zeta\bR = (\nabla_\bR \psi
\times \nabla_\bR \Theta)/(\bB \cdot \nabla_\bR \Theta) = R\zun$ and
$\partial_\Theta\bR = - (\nabla_\bR \psi \times
\nabla_\bR \zeta)/(\bB \cdot \nabla_\bR \Theta)$, we find
\begin{eqnarray}
\fl (\nabla_\bR \times \bK) \cdot \nabla_\bR \psi = 
\nonumber\\
\fl\hspace{1cm}
\bB \cdot
\nabla_\bR \Theta \partial_\Theta \left (
\partial_\zeta\bR \cdot \bK \right ) - \bB \cdot
\nabla_\bR \Theta \partial_\zeta \left (
\partial_\Theta\bR \cdot \bK \right )
\nonumber\\
\fl\hspace{1cm}
= \bB \cdot \nabla_\bR \left ( \frac{I}{2B} \bun \cdot
\nabla_\bR \times \bun \right ) - \bB \cdot \nabla_\bR \Theta
\partial_\Theta \left ( \partial_\zeta
\eun_2 \cdot \eun_1 \right ) \nonumber\\
\fl\hspace{1cm}
+ \bB
\cdot \nabla_\bR \Theta \partial_\zeta \left (
\partial_\Theta \eun_2  \cdot \eun_1 \right ) =
\bB \cdot \nabla_\bR \left ( \frac{I}{2B} \bun \cdot \nabla_\bR
\times \bun \right ) \nonumber\\
\fl\hspace{1cm}
- \bB \cdot \nabla_\bR \Theta
\left (\partial_\zeta \eun_2 \cdot
\partial_\Theta \eun_1 - \partial_\Theta
\eun_2 \cdot \partial_\zeta \eun_1 \right ).
\end{eqnarray}

Now, with the help of the relations $\nabla_\bR \eun_1 = \nabla_\bR \eun_1
\cdot \bun \bun + \nabla_\bR \eun_1 \cdot \eun_2 \eun_2$ and
$\nabla_\bR \eun_2 = \nabla_\bR \eun_2 \cdot \bun \bun + \nabla_\bR
\eun_2 \cdot \eun_1 \eun_1$, one gets
\begin{eqnarray}
\fl 
\partial_\zeta \eun_2 \cdot
\partial_\Theta \eun_1 - \partial_\Theta
\eun_2 \cdot \partial_\zeta \eun_1
 = 
\nonumber\\
\fl\hspace{1cm}
\left
(\partial_\zeta \eun_2 \cdot \bun \right ) \left
(\partial_\Theta \eun_1 \cdot \bun \right ) -
\left ( \partial_\Theta \eun_2  \cdot \bun
\right ) \left ( \partial_\zeta \eun_1 \cdot \bun
\right ) \nonumber\\
\fl\hspace{1cm}
= \left (\partial_\zeta \bun
\cdot \eun_2 \right ) \left (\partial_\Theta \bun
 \cdot \eun_1 \right ) - \left ( \partial_\Theta
\bun \cdot \eun_2 \right ) \left (
\partial_\zeta \bun \cdot \eun_1 \right )
\nonumber\\
\fl\hspace{1cm}
= \left ( \partial_\Theta \bun
\times \partial_\zeta \bun \right ) \cdot \bun.
\end{eqnarray}
Since this quantity is independent of the choice of
$\eun_1$ and $\eun_2$, we can use $\eun_1 = \nabla_\bR \psi/|\nabla_\bR
\psi|$ and $\eun_2 = (\bun \times \nabla_\bR \psi)/|\nabla_\bR \psi|$
without loss of generality, giving
\begin{eqnarray}
\fl
\partial_\zeta \eun_2 \cdot
\partial_\Theta \eun_1 - \partial_\Theta
\eun_2 \cdot \partial_\zeta \eun_1=
\partial_\zeta \left ( \frac{\bun \times \nabla_\bR
\psi}{|\nabla_\bR \psi|} \right ) \cdot \partial_\Theta 
\left ( \frac{\nabla_\bR \psi}{|\nabla_\bR \psi|} \right )
\nonumber\\
\fl\hspace{1cm}- \partial_\Theta \left (
\frac{\bun \times \nabla_\bR \psi}{|\nabla_\bR \psi|} \right ) \cdot
\partial_\zeta \left ( \frac{\nabla_\bR \psi}{|\nabla_\bR
\psi|} \right )
\nonumber\\
\fl\hspace{1cm}
 = - \partial_\Theta \left (
\frac{1}{|\nabla_\bR \psi|^2} \frac{\partial \nabla_\bR \psi}{\partial
\zeta} \cdot (\bun \times \nabla_\bR \psi) \right ).
\end{eqnarray}
Thus,
\begin{eqnarray}
\fl (\nabla_\bR \times \bK) \cdot \nabla_\bR \psi = \bB \cdot
\nabla_\bR \Bigg( \frac{I}{2B} \bun \cdot \nabla_\bR \times \bun
\nonumber\\
\fl\hspace{1cm}
+ \frac{R}{|\nabla_\bR \psi|^2} \zun \cdot \nabla_\bR \nabla_\bR \psi
\cdot (\bun \times \nabla_\bR \psi) \Bigg).
\end{eqnarray}

\section{Some computations related to the long-wavelength 
quasineutrality equation}
\label{sec:computations_quasineutralitylw}

Firstly, let us show that \eq{eq:gyroPoissonlw2} can be rewritten as
in \eq{eq:gyroPoissonlw4}. Employ the relation (recall
\eq{mu1} and \eq{theta1})
\begin{equation}
\partial_\mu\mu_{\spe,1}+\partial_\theta\theta_{\spe,1} =
\frac{1}{B}\rhobf\cdot\nabla_\bR B,
\end{equation}
the identity
\begin{equation}
\langle\rhobf\rhobf\rangle = \frac{\mu}{B}(\matI-\bun\bun),
\end{equation}
and the long-wavelength limit of \eq{R2} and \eq{mu1},
\begin{eqnarray}\label{eq:R2<}
\fl\bR_{\spe,2}^\lw &=& - \frac{2u }{B} \bun \bun \cdot \nabla_\bR \bun \cdot
(\rhobf \times \bun) - \frac{1}{8} \bun \left [ \rhobf \rhobf -
(\rhobf \times \bun) (\rhobf \times \bun) \right ]: \nabla_\bR
\bun\nonumber\\
\fl&-&\frac{u }{B} \bun \times \nabla_\bR \bun \cdot \rhobf
- \frac{1}{2B} \rhobf \rhobf \cdot \nabla_\bR B +
O(\epsilon_\spe),
\end{eqnarray}
\begin{eqnarray}\label{eq:mu1<}
\fl\mu_{\spe,1}^\lw &=& - \frac{u^2}{B} \bun \cdot \nabla_{\bR} \bun
\cdot \rhobf + \frac{u}{4} \left [ \rhobf (\rhobf \times \bun) +
  (\rhobf \times \bun) \rhobf \right ]: \nabla_{\bR} \bun\\[5pt]
\fl&-&
\frac{Z_\spe}{B}
\rhobf\cdot\nabla_\bR\varphi_0 +
O(\epsilon_\spe),
\end{eqnarray}
to recast \eq{eq:gyroPoissonlw2} into
\begin{eqnarray}\label{eq:gyroPoissonlw3}
\fl\sum_\spe
\frac{Z_\spe}{\lambda_\spe^2}
\int
(BF_{\spe 2}^\lw -\mu\bun\cdot\nabla_\boldr\times\bK \, F_{\spe 0}
+u\bun\cdot\nabla_\boldr\times\bun \, F_{\spe 1}^\lw)
\dd u\dd\mu\dd\theta\nonumber\\[5pt]
\fl\hspace{1cm}
+2\pi\sum_\spe\frac{Z_\spe}{\lambda_\spe^2}
\Bigg[
\nabla_\boldr\cdot
\left(\frac{3}{2B} \nabla_{\boldr_\bot} B\int\mu F_{\spe 0}
\dd u\dd\mu\right)\nonumber\\[5pt]
\fl\hspace{1cm}
+\frac{1}{2}\nabla_\boldr\nabla_\boldr:
\left(
(\matI-\bun\bun)\int\mu F_{\spe 0}
\dd u\dd\mu
\right)\nonumber\\[5pt]
\fl\hspace{1cm}
-\nabla_\boldr\cdot
\left(
\frac{1}{B}\bun\cdot\nabla_\boldr\bun
\int u^2\mu \partial_\mu F_{\spe 0}
\dd u\dd\mu
\right)\nonumber\\[5pt]
\fl\hspace{1cm}
-
\nabla_\boldr\cdot
\left(
\frac{Z_\spe}{B}(\matI-\bun\bun)\cdot\nabla_\boldr
\varphi_0
\int \mu \partial_\mu F_{\spe 0}
\dd u\dd\mu
\right)
\Bigg] =0.
\end{eqnarray}
We can get  more explicit expressions by noting that the integrals
containing $F_{\spe 0}$ can be worked out
analytically. Namely, if
\begin{equation}\label{eq:maxwellian}
F_{\spe 0}=\frac{n_{\spe}}{(2\pi T_{\spe})^{3/2}}
\exp\left(-\frac{\mu B + u^2/2}{T_{\spe}}\right),
\end{equation}
then
\begin{equation}
\partial_{\mu}F_{\spe 0}=-\frac{B}{T_{\spe}}F_{\spe 0},
\end{equation}
and
\begin{equation}
\int \mu F_{\spe 0}\dd u\dd\mu=\frac{n_{\spe}T_{\spe}}{2\pi B^2},
\quad  \int u^2\mu F_{\spe 0}\dd u\dd\mu=\frac{n_{\spe}T_{\spe}^2}{2\pi B^2}.
\end{equation}
With these results, equation~\eq{eq:gyroPoissonlw3} finally becomes
\eq{eq:gyroPoissonlw4}.

Now, we proceed to recast \eq{eq:gyroPoissonlw4} into
\eq{eq:gyroPoissonlw4aux3} by using the function $G_{\spe 2}^\lw$
defined in \eq{eq:defG2again3}. A simple rewriting of
\eq{eq:gyroPoissonlw4} in terms of $G_{\spe 2}^\lw$ gives
\begin{eqnarray}\label{eq:gyroPoissonlw4aux2}
\fl\sum_\spe
\frac{Z_\spe}{\lambda_\spe^2}
\int B
\Bigg\{
G_{\spe 2}^\lw
-
\frac{Z_\spe\lambda_\spe^2}{T_\sigma}\varphi_2^\lw
F_{\spe 0}
-
\frac{Iu}{B}
\partial_\psi G_{\sigma 1}^\lw  
\nonumber\\[5pt]
\fl\hspace{1cm}
+
\frac{Z_\spe\lambda_\spe \varphi_1^\lw}{u} \partial_u
G_{\sigma 1}^\lw 
+
 \frac{1}{2}
\left(
\frac{Z_\spe\lambda_\spe\varphi_1^\lw}{T_\sigma}
\right)^2
F_{\sigma 0}
\nonumber\\[5pt]
\fl\hspace{1cm}
+\frac{1}{2B^2}\left(
\left ({Iu} \right )^2+\mu B|\nabla_\boldr\psi|^2
\right)
\Bigg[
- \frac{2Z_\spe}{T_\sigma^2}
 \partial_\psi\varphi_0 
\partial_\psi T_\sigma
\nonumber\\[5pt]
\fl\hspace{1cm}
+ \Bigg (
\frac{Z_\spe}{T_\sigma}
\partial_\psi\varphi_0  +\Upsilon_\spe
\Bigg )^2
+\partial_\psi^2\ln n_\spe
\nonumber\\[5pt]
\fl\hspace{1cm}
+
\left(
\frac{u^2/2+\mu B}{T_\spe}-\frac{3}{2}
\right)
\partial_\psi^2\ln T_\spe
\nonumber\\[5pt]
\fl\hspace{1cm}
-
\frac{u^2/2+\mu B}{T_\spe}
(\partial_\psi\ln T_\spe)^2
+
\frac{Z_\spe}{T_\sigma}
\partial_\psi^2 \varphi_0
\Bigg ]F_{\spe 0}
\nonumber\\[5pt]
\fl\hspace{1cm} 
- \left[ \frac{Z_\spe }{u \bB \cdot
    \nabla_\boldr \Theta} F_{\sigma 1}^\sw (\bun\times
  \nabla_{\boldr_\perp/\epsilon_\spe}
  \langle \phi_{\spe 1}^\sw \rangle) \cdot \nabla_\boldr
  \Theta \right]^\lw
\nonumber\\[5pt]
\fl\hspace{1cm}
+
\left \langle \frac{1}{ u \bun \cdot
    \nabla_\boldr\Theta} \rhobf
 \cdot \nabla_\boldr \Theta
  \sum_{\sigma^\prime}\cT_{\spe,0}^*C_{\sigma
    \sigma^\prime}^{(1)\lw} \right \rangle
\nonumber\\[5pt]
\fl\hspace{1cm}
+ \frac{Z_\spe^2\lambda_\spe^2}{2T_\spe^2}
\left[ \left\langle
    (\phiwig_{\spe 1}^\sw)^2 \right \rangle \right]^\lw F_{\spe
  0}
\Bigg\}
\dd u\dd\mu\dd\theta
\nonumber\\[5pt]
\fl\hspace{1cm}
+\sum_\spe
\frac{Z_\spe}{\lambda_\spe^2}\int u\bun\cdot(
\nabla_\boldr\times\bun)
G_{\spe 1}^\lw
\dd u\dd\mu\dd\theta
=0.
\end{eqnarray}
Here we have used \eq{eq:T1F1lwAveraged} to write
\begin{eqnarray}
\int B \cT_{\spe,0}^*
\left\langle
\left[
\cT_{\spe,1}^{-1 *}F_{\spe1}^\lw
\right]^\lw
\right\rangle\dd u\dd\mu\dd\theta
\nonumber\\[5pt]
\hspace{1cm}
=
\bun\cdot\nabla_\boldr\times\bun
\int u F_{\spe1}^\lw\,
\,\dd u\dd\mu\dd\theta,
\end{eqnarray}
and we have employed the result in \ref{sec:integralT2F0}:
\begin{eqnarray} \label{eq:intT2maxwellian} \fl \int B
  \left \langle \cT_{\spe,0}^*\left [
        \mathcal{T}^{-1\ast}_{\spe, 2} F_{\spe 0} \right ]^\lw \right
    \rangle  \,\dd u\dd\mu\dd\theta =
  \nonumber\\[5pt]
  \fl\hspace{1cm} \nabla_\boldr \nabla_\boldr : \left [ \frac{n_\spe
      T_\spe}{2B^2} (\matI - \bun \bun) \right ] + \nabla_\boldr \cdot
  \left ( \frac{Z_\spe n_\spe}{B^2} \nabla_{\boldr_\perp} \varphi_0
  \right )
  \nonumber\\[5pt]
  \fl\hspace{1cm} + \nabla_\boldr \cdot \left ( \frac{3n_\spe
      T_\spe}{2B^3} \nabla_{\boldr_\perp} B \right ) + \nabla_\boldr
  \cdot \left ( \frac{n_\spe T_\spe}{B^2} \bun \cdot \nabla_\boldr
    \bun \right )
  \nonumber\\[5pt]
  \fl\hspace{1cm} - \frac{n_\spe T_\spe}{B^2} \bun \cdot \nabla_\boldr
  \times \bK.
\end{eqnarray}

Equation \eq{eq:gyroPoissonlw4aux2} can be simplified even more employing
\begin{eqnarray}
\fl\int B \left((Iu)^2+\mu B|\nabla_\boldr\psi|^2
\right) F_{\spe 0}\dd u\dd\mu\dd\theta 
\nonumber\\[5pt]
= (RB)^2 n_\spe T_\spe\nonumber\\[5pt]
\fl\int B \left((Iu)^2+\mu B|\nabla_\boldr\psi|^2
\right)
\frac{u^2/2 + \mu B}{T_\spe}
 F_{\spe 0}\dd u\dd\mu\dd\theta 
\nonumber\\[5pt]
= \frac{5}{2}
(RB)^2 n_\spe T_\spe\nonumber\\[5pt]
\fl\int B \left((Iu)^2+\mu B|\nabla_\boldr\psi|^2
\right)
\left(
\frac{u^2/2 + \mu B}{T_\spe}-\frac{3}{2}
\right)
 F_{\spe 0}\dd u\dd\mu\dd\theta
\nonumber\\[5pt]
 =
(RB)^2 n_\spe T_\spe\nonumber\\[5pt]
\fl\int B \left((Iu)^2+\mu B|\nabla_\boldr\psi|^2
\right)
\left(
\frac{u^2/2 + \mu B}{T_\spe}-\frac{3}{2}
\right)^2
 F_{\spe 0}\dd u\dd\mu\dd\theta
\nonumber\\[5pt]
 =
\frac{7}{2}(RB)^2 n_\spe T_\spe,
\end{eqnarray}
finally giving \eq{eq:gyroPoissonlw4aux3}.

\section{Integral of the second-order piece of the
transformation of the Maxwellian}
\label{sec:integralT2F0}

In this Appendix we calculate the integral in
velocity space \eq{eq:intT2maxwellian}. The integrand $\left \langle \cT_{\spe, 0}^{*} \left [ \mathcal{T}^{-1\ast}_{\spe, 2} F_{\spe
0} \right ]^\lw \right \rangle$ is given in \eq{gyroaverageT2F0}.
Using
\begin{eqnarray}
\fl \frac{1}{T_\spe} \int B
\Psi_{B,\spe} F_{\spe 0}\, \dd u \dd \mu \dd \theta  =
\nonumber\\
\fl\hspace{0.5cm} 
 - \frac{3 n_\spe T_\spe}{2B^3}
\bun \cdot \nabla_\boldr \bun \cdot \nabla_\boldr B + \frac{n_\spe T_\spe}{2B^3}
(\matI - \bun \bun) : \nabla_\boldr \nabla_\boldr \bB \cdot \bun \nonumber\\
\fl\hspace{0.5cm} 
-
\frac{3 n_\spe T_\spe}{2B^4} |\nabla_{\boldr_\bot} B|^2 + \frac{n_\spe
T_\spe}{2B^2} \nabla_\boldr \bun : \nabla_\boldr \bun 
\nonumber\\
\fl\hspace{0.5cm} 
- \frac{n_\spe
T_\spe}{2B^2} (\nabla_\boldr \cdot \bun)^2
\end{eqnarray}
and
\begin{eqnarray}
\fl \int B\Bigg( \frac{Z_\spe^2\lambda_\spe^2}{2T_\spe^2} \left[ \left \langle
(\phiwig_{\spe 1}^\sw)^2 \right \rangle \right]^\lw
\nonumber\\[5pt]
\fl\hspace{0.5cm} 
 - 
\frac{Z_\spe^2\lambda_\spe^2}{2B T_\spe}
\partial_\mu \left[ \left \langle (\phiwig_{\spe 1}^\sw)^2 \right
\rangle \right]^\lw \Bigg) F_{\spe 0}\, \dd u \dd \mu \dd \theta =
\nonumber\\[5pt]
\fl\hspace{0.5cm}
 - \frac{Z_\spe^2\lambda_\spe^2}{2T_\spe} \int 
\partial_{\mu} \left (
\left[ \left \langle (\phiwig_{\spe 1}^\sw)^2 \right \rangle
\right]^\lw F_{\spe 0} \right )\, \dd u \dd
\mu \dd \theta = 0,
\end{eqnarray}
we obtain
\begin{eqnarray}\label{eq:auxEqAppendixT2F0}
\fl \int B \left \langle \cT_{\spe, 0}^{*}\left [ \mathcal{T}^{-1\ast}_{\spe, 2} F_{\spe 0} \right
]^\lw \right \rangle\, \dd u \dd \mu \dd \theta \nonumber\\
\fl\hspace{1cm}
= \frac{1}{2B^2} (\matI - \bun \bun) : \nabla_\boldr
\nabla_\boldr ( n_\spe T_\spe ) + \nabla_\boldr \cdot \left ( 
\frac{Z_\spe
n_\spe}{B^2} \nabla_{\boldr_\bot} \varphi_0 \right )
\nonumber\\
\fl\hspace{1cm}
-
\frac{1}{2B^3} \nabla_{\boldr_\bot} B \cdot \nabla_{\boldr_\bot}( n_\spe T_\spe ) -
\frac{3 n_\spe T_\spe}{2B^3} \bun \cdot \nabla_\boldr \bun \cdot \nabla_\boldr B
\nonumber\\
\fl\hspace{1cm}
+ \frac{n_\spe T_\spe}{2B^3} (\matI - \bun \bun) : \nabla_\boldr \nabla_\boldr
\bB \cdot \bun
- \frac{3 n_\spe T_\spe}{2B^4}
|\nabla_{\boldr_\bot} B|^2
\nonumber\\
\fl\hspace{1cm}
 + \frac{n_\spe T_\spe}{2B^2} \nabla_\boldr \bun :
\nabla_\boldr \bun - \frac{n_\spe T_\spe}{2B^2} (\nabla_\boldr \cdot \bun)^2.
\end{eqnarray}
Using
\begin{eqnarray}
\fl (\matI - \bun \bun) : \nabla_\boldr \nabla_\boldr \bB \cdot \bun =
\nonumber\\
\fl\hspace{1cm}
 (\matI -
\bun \bun): \nabla_\boldr \nabla_\boldr B - B \nabla_\boldr \bun : (\nabla_\boldr
\bun)^\mathrm{T} + B | \bun \cdot \nabla_\boldr \bun |^2
\end{eqnarray}
and
\begin{eqnarray}
\fl \frac{1}{2B^2} (\matI - \bun \bun) : \nabla_\boldr \nabla_\boldr (n_\spe
T_\spe) =
\nonumber\\
\fl\hspace{0.5cm}
 \nabla_\boldr \nabla_\boldr : \left [ \frac{n_\spe T_\spe}{2B^2}
(\matI - \bun \bun) \right ] + \frac{2}{B^3} \nabla_{\boldr_\bot} B \cdot
\nabla_{\boldr_\bot} (n_\spe T_\spe)
\nonumber\\
\fl\hspace{0.5cm}
+ \frac{1}{B^2} \bun \cdot
\nabla_\boldr \bun \cdot \nabla_\boldr (n_\spe T_\spe) + \frac{n_\spe
T_\spe}{B^3} (\matI - \bun \bun): \nabla_\boldr \nabla_\boldr B 
\nonumber\\
\fl\hspace{0.5cm}
+ \frac{5 n_\spe
T_\spe}{2 B^2} (\nabla_\boldr \cdot \bun)^2 - \frac{2 n_\spe T_\spe}{B^3}
\bun \cdot \nabla_\boldr \bun \cdot \nabla_\boldr B 
\nonumber\\
\fl\hspace{0.5cm}
+ \frac{n_\spe T_\spe}{B^2}
\bun \cdot \nabla_\boldr ( \nabla_\boldr \cdot \bun)
+ \frac{n_\spe
T_\spe}{2B^2} \nabla_\boldr \bun : \nabla_\boldr \bun
\nonumber\\
\fl\hspace{0.5cm}
- \frac{3 n_\spe
T_\spe}{B^4} |\nabla_{\boldr_\bot} B|^2
,
\end{eqnarray}
equation \eq{eq:auxEqAppendixT2F0} can be rewritten as
\begin{eqnarray}
\fl \int 
B \left \langle \cT_{\spe, 0}^{*}
 \left [ \mathcal{T}^{-1\ast}_{\spe, 2} F_{\spe 0} \right
]^\lw \right \rangle\, \dd u \dd\mu \dd\theta = 
\nonumber\\
\fl\hspace{0.5cm}
\nabla_\boldr \nabla_\boldr : \left [ \frac{n_\spe
T_\spe}{2B^2} (\matI - \bun \bun) \right ] + \nabla_\boldr \cdot \left (
\frac{Z_\spe n_\spe}{B^2} \nabla_{\boldr_\bot} \varphi_0 \right )
\nonumber\\
\fl\hspace{0.5cm}
+ \frac{3}{2B^3} \nabla_{\boldr_\bot} B \cdot \nabla_{\boldr_\bot}(
n_\spe T_\spe ) + \frac{1}{B^2} \bun \cdot \nabla_\boldr \bun \cdot
\nabla_\boldr ( n_\spe T_\spe )
\nonumber\\
\fl\hspace{0.5cm}
 - \frac{7 n_\spe T_\spe}{2B^3} \bun \cdot
\nabla_\boldr \bun \cdot \nabla_\boldr B 
+ \frac{3n_\spe
T_\spe}{2B^3} (\matI - \bun \bun) : \nabla_\boldr \nabla_\boldr B
\nonumber\\
\fl\hspace{0.5cm}
 - \frac{n_\spe
T_\spe}{2B^2} \nabla_\boldr \bun : (\nabla_\boldr \bun)^\mathrm{T} +
\frac{n_\spe T_\spe}{2B^2} | \bun \cdot \nabla_\boldr \bun |^2
\nonumber\\
\fl\hspace{0.5cm}
- \frac{9 n_\spe T_\spe}{2B^4} |\nabla_{\boldr_\bot} B|^2 +
\frac{n_\spe T_\spe}{B^2} \nabla_\boldr \bun : \nabla_\boldr \bun +
\frac{2n_\spe T_\spe}{B^2} (\nabla_\boldr \cdot \bun)^2
\nonumber\\
\fl\hspace{0.5cm}
+
\frac{n_\spe T_\spe}{B^2} \bun \cdot \nabla_\boldr ( \nabla_\boldr \cdot \bun).
\end{eqnarray}
With further manipulations, we find
\begin{eqnarray}
\fl \int 
B \left \langle \cT_{\spe, 0}^{*}
 \left [ \mathcal{T}^{-1\ast}_{\spe, 2} F_{\spe 0} \right
]^\lw \right \rangle \, \dd u \dd\mu \dd \theta = 
\nonumber\\
\fl\hspace{0.5cm}
\nabla_\boldr \nabla_\boldr : \left [ \frac{n_\spe
T_\spe}{2B^2} (\matI - \bun \bun) \right ] + \nabla_\boldr \cdot \left (
\frac{Z_\spe n_\spe}{B^2} \nabla_{\boldr_\bot} \varphi_0 \right )
\nonumber\\
\fl\hspace{0.5cm}
+ \nabla_\boldr \cdot \left ( \frac{3n_\spe T_\spe}{2B^3}
\nabla_{\boldr_\bot} B \right ) + \nabla_\boldr \cdot \left ( \frac{n_\spe
T_\spe}{B^2} \bun \cdot \nabla_\boldr \bun \right )
\nonumber\\
\fl\hspace{0.5cm}
-
\frac{n_\spe T_\spe}{2B^2} \nabla_\boldr \bun : (\nabla_\boldr \bun)^\mathrm{T}
+ \frac{n_\spe T_\spe}{2B^2} | \bun \cdot \nabla_\boldr \bun |^2 
\nonumber\\
\fl\hspace{0.5cm}
+
\frac{n_\spe T_\spe}{2B^2} (\nabla_\boldr \cdot \bun)^2.
\end{eqnarray}
Finally, we show that we can combine the last three terms of the
previous equation to give a more recognizable term. Employing
\begin{equation}
\fl \bun \cdot \nabla_\boldr \times \bK = \frac{1}{2} (\bun \cdot \nabla_\boldr
\times \bun)^2 - (\bun \times \nabla_\boldr \eun_1) : (\nabla_\boldr
\eun_2)^\mathrm{T},
\end{equation}
\begin{eqnarray}
\fl \nabla_\boldr \eun_1 \cdot (\nabla_\boldr \eun_2)^\mathrm{T} = (\nabla_\boldr
\eun_1 \cdot \bun) (\nabla_\boldr \eun_2 \cdot \bun) =\nonumber\\
 (\nabla_\boldr \bun \cdot
\eun_1) (\nabla_\boldr \bun \cdot \eun_2) = \frac{1}{2} \nabla_\boldr \bun \cdot
(\nabla_\boldr \bun \times \bun)^\mathrm{T},
\end{eqnarray}
and
\begin{equation}
\fl \bun \times \nabla_\boldr \bun \times \bun = (\nabla_\boldr \bun)^\mathrm{T}
- (\bun \cdot \nabla_\boldr \bun) \bun - (\nabla_\boldr \cdot \bun) (\matI -
\bun \bun),
\end{equation}
one finds the identity
\begin{equation}
\fl \bun \cdot \nabla_\boldr \times \bK = \frac{1}{2} (\bun \cdot
\nabla_\boldr \times \bun)^2 + \frac{1}{2} \nabla_\boldr \bun : \nabla_\boldr \bun -
\frac{1}{2} (\nabla_\boldr \cdot \bun)^2.
\end{equation}
Since $\nabla_\boldr \bun : (\nabla_\boldr \bun)^\mathrm{T} - \nabla_\boldr \bun :
\nabla_\boldr \bun = |\nabla_\boldr \times \bun|^2$ and $\nabla_\boldr \times \bun =
\bun \bun \cdot \nabla_\boldr \times \bun + \bun \times (\bun \cdot
\nabla_\boldr \bun)$, we obtain
\begin{equation}
\fl \bun \cdot \nabla_\boldr \times \bK = \frac{1}{2} \nabla_\boldr \bun
: (\nabla_\boldr \bun)^\mathrm{T} - \frac{1}{2} |\bun \cdot \nabla_\boldr
\bun|^2 - \frac{1}{2} (\nabla_\boldr \cdot \bun)^2,
\end{equation}
giving equation \eq{eq:intT2maxwellian}. We point out that
$\nabla_\boldr\times\bK$ was computed for the first time by Littlejohn in
reference \cite{Littlejohn81}.

\section{Proof of  \eq{eq:cancellationenergy}}
\label{app:cancellationenergy}

In this Appendix we prove
  \eq{eq:cancellationenergy}. To do so, we take the short-wavelength
  quasineutrality equation to first order, given in
  \eq{eq:quasinautralitySWorder1appendix}, apply the operator
  $\partial_t + B^{-1} (\bun \times \nabla_\boldr
  \varphi_0(\boldr, t)) \cdot \nabla_{\boldr_\bot/\epsilon_s}$, and
  multiply it by $\varphi_1^\sw (\boldr, t)$ to find
\begin{eqnarray}\label{eq:dtquasineutrality}
\fl\sum_\spe \frac{Z_\spe}{\lambda_\spe}
 \varphi_1^\sw (\boldr, t) \left ( \partial_t + \frac{Z_\spe \tau_\spe}{B} (\bun \times \nabla_\boldr \varphi_0(\boldr, t) ) \cdot \nabla_{\boldr_\bot/\epsilon_\spe} \right ) \int B
\Bigg[ \nonumber\\[5pt]
\fl\hspace{0.5cm} -Z_\spe\lambda_\spe \phiwig_{\spe
      1}^\sw \left(\boldr
-\epsilon_\spe\rhobf(\boldr,\mu,\theta),\mu,\theta,t\right)
\frac{F_{\spe 0}(\boldr,u,\mu,t)}{T_{\spe}(\boldr,t)}\nonumber\\[5pt]
\fl\hspace{0.5cm}
 + F_{\spe 1}^\sw
\left(\boldr
-\epsilon_\spe\rhobf(\boldr,\mu,\theta),u,\mu,t
\right)
\Bigg]
  \dd u \dd \mu \dd \theta = 0.
\end{eqnarray}
Since
\begin{equation} \label{eq:varphiphiapprox}
  \varphi_1^\sw(\boldr,t) =
  \phi_{\spe 1}^\sw(\boldr-\epsilon_\spe\rhobf(\boldr,\mu,\theta),\mu,\theta,t)
 + O(\epsilon_\spe),
\end{equation}
\begin{equation}
\nabla_\boldr \varphi_0 (\boldr, t) = \nabla_\bR \varphi_0(\bR, t) + O(\epsilon_\spe)
\end{equation}
and
\begin{equation}
\nabla_{\boldr_\bot/\epsilon_\sigma} = \nabla_{\bR_\bot/\epsilon_\spe} + O(\epsilon_\spe),
\end{equation}
we find that \eq{eq:dtquasineutrality} becomes
\begin{eqnarray}\label{eq:dtquasineutrality2}
\fl
\sum_\spe \frac{Z_\spe}{\lambda_\spe} \int B
\Bigg[
\phi_{\spe 1}^\sw
\Bigg ( \partial_t + \frac{Z_\spe \tau_\spe}{B} (\bun \times \nabla_\bR \varphi_0 ) \cdot \nabla_{\bR_\bot/\epsilon_\spe} \Bigg )  \nonumber\\[5pt]
\fl\hspace{0.5cm}
\Bigg(-\frac{Z_\spe\lambda_\spe \phiwig_{\spe
      1}^\sw}{T_{\spe}} F_{\spe 0}
+ F_{\spe 1}^\sw\Bigg )
\Bigg]^\lw
  \dd u \dd \mu \dd \theta = O(\epsilon_s).
\end{eqnarray}
In this expression, the functions $\phi_{\spe 1}^\sw$ and $F_{\spe
  1}^\sw$ are evaluated at $\bR = \boldr - \epsilon_\spe \rhobf
(\boldr, \mu, \theta)$, but after the coarse-grain average we can
Taylor expand and, to lowest order, they can be evaluated at
$\bR = \boldr$. Thus, we find
\begin{eqnarray}\label{eq:dtquasineutrality3}
\fl -\sum_\spe Z_\spe^2  \int
\frac{B  F_{\spe 0} }{2 T_{\spe}} 
\Bigg ( \partial_t \left[ (\phiwig_{\spe 1}^\sw)^2\right]^\lw
\nonumber\\[5pt] \fl
\hspace{0.5cm}
+ \frac{Z_\spe \tau_\spe}{B} (\bun \times \nabla_\bR \varphi_0 ) \cdot \nabla_{\bR_\bot/\epsilon_\spe}
\left[ (\phiwig_{\spe 1}^\sw)^2\right]^\lw
 \Bigg )
   \dd u \dd \mu \dd \theta
\nonumber\\[5pt] \fl
\hspace{0.5cm}
+ \sum_\spe
\frac{Z_\spe}{\lambda_\spe} \int B \Bigg[\langle\phi_{\spe 1}^\sw\rangle
\Bigg ( \partial_t F_{\spe 1}^\sw 
\nonumber\\[5pt] \fl
\hspace{0.5cm}
+ \frac{Z_\spe \tau_\spe}{B} (\bun \times \nabla_\bR \varphi_0 ) \cdot \nabla_{\bR_\bot/\epsilon_\spe} 
F_{\spe 1}^\sw 
\Bigg )
\Bigg]^\lw
  \dd u \dd \mu \dd \theta = O(\epsilon_s).
\end{eqnarray}
Employing that the time derivative of a long-wavelength contribution
is small by $\epsilon_s^2$ and that $\nabla_{\bR_\bot/\epsilon_\spe}
g^\lw \sim \epsilon_\spe$, we obtain that
\begin{eqnarray}\label{eq:dtquasineutrality4}
\fl \sum_\spe \frac{Z_\spe}{\lambda_\spe}
 \int B \Bigg[\langle\phi_{\spe 1}^\sw\rangle
\Bigg ( \partial_t
F_{\spe 1}^\sw  
\nonumber\\[5pt] \fl
\hspace{0.5cm}
+ \frac{Z_\spe \tau_\spe}{B} (\bun \times \nabla_\bR \varphi_0 ) \cdot \nabla_{\bR_\bot/\epsilon_\spe}
F_{\spe 1}^\sw 
 \Bigg ) 
\Bigg]^\lw
  \dd u \dd \mu \dd \theta = O(\epsilon_s).
\end{eqnarray}
Now, we use \eq{eq:sworder1distfunction} in
\eq{eq:dtquasineutrality4}, getting
\begin{eqnarray}\label{eq:dtquasineutrality5}
\fl - \sum_\spe  \int B \Bigg [\langle\phi_{\spe 1}^\sw\rangle \Bigg ( u\bun\cdot\nabla_\bR F_{\spe 1}^\sw
\nonumber\\[5pt] \fl\hspace{0.5cm}
+
\frac{u^2}{B} (\bun \times (\bun\cdot\nabla_\bR\bun))\cdot\nabla_{\bR_\perp/\epsilon_\spe}F_{\spe 1}^\sw
\nonumber\\[5pt] \fl\hspace{0.5cm} +\frac{\mu}{B}(\bun\times\nabla_\bR B)
\cdot\nabla_{\bR_\perp/\epsilon_\spe}F_{\spe 1}^\sw \Bigg ) \Bigg ]^\lw \dd u \dd \mu \dd \theta
\nonumber\\[5pt]
\fl\hspace{0.5cm}
+ \sum_{\spe,\spe'}  \int B 
 \Bigg [\langle\phi_{\spe 1}^\sw\rangle \Bigg ( \cT_{NP, \spe}^* C_{\sigma \sigma^\prime}
\Bigg[\modTinv F_{\sigma 1}^\sw
\nonumber\\[5pt] \fl\hspace{0.5cm}
 -
\frac{Z_\spe\lambda_\spe}{T_\spe}
\modTinv\tilde\phi_{\spe 1}^\sw \cT_{\spe,0}^{-1
*} F_{\spe 0}, \cT_{\spe',0}^{-1
*}F_{\spe' 0}
 \Bigg]\nonumber\\[5pt]
\fl\hspace{0.5cm}
 + \
\frac{\lambda_\spe}{\lambda_{\spe'}}
\cT_{NP, \spe}^*
C_{\sigma
\sigma^\prime} \Bigg[ \cT_{\spe,0}^{-1 *}F_{\spe 0} ,
\modTinvprime F_{\sigma' 1}^\sw
\nonumber\\[5pt] \fl\hspace{0.5cm}
 -
\frac{Z_{\spe'}\lambda_{\spe'}}{T_{\spe'}}
\modTinvprime\tilde\phi_{\spe'
1}^\sw \cT_{\spe',0}^{-1 *}F_{\spe' 0}
 \Bigg] \Bigg ) \Bigg ]^\lw \dd u \dd \mu \dd \theta
=
O(\epsilon_s)
.
\end{eqnarray}
Here, we have used the fact that $\langle \phi_{\spe 1}^\sw \rangle$
does not depend on $u$, and the relations
\begin{equation}
\left [ \langle \phi_{\spe 1}^\sw \rangle \nabla_{\bR_\bot/\epsilon_\spe} \langle \phi_{\spe 1}^\sw \rangle \right ]^\lw = \frac{1}{2}  \nabla_{\bR_\bot/\epsilon_\spe} \left [ \langle \phi_{\spe 1}^\sw \rangle^2 \right ]^\lw = O(\epsilon_\spe)
\end{equation}
and
\begin{eqnarray}
\fl \left [ \langle \phi_{\spe 1}^\sw \rangle (\bun \times \nabla_{\bR_\bot/\epsilon_\spe} \langle \phi_{\spe 1}^\sw \rangle) \cdot \nabla_{\bR_\bot/\epsilon_\spe} F_{\spe 1}^\sw \right ]^\lw= \nonumber\\
\fl\hspace{1cm}
 - \frac{1}{2} \bun \cdot \nabla_{\bR_\bot/\epsilon_\spe} \times \left [ F_{\spe 1}^\sw \nabla_{\bR_\bot/\epsilon_\spe} \langle \phi_{\spe 1}^\sw \rangle^2 \right ]^\lw 
=O( \epsilon_\spe).
\end{eqnarray}

Finally, to relate \eq{eq:dtquasineutrality5} to
\eq{eq:cancellationenergy}, we employ that, up to terms of order
$\epsilon_\spe$,
\begin{eqnarray} \label{eq:phicollisions}
\fl \int B \left [ \langle
    \phi_{\spe 1}^\sw \rangle \cT_{NP, \spe} C_{\spe \spe^\prime}^\sw
  \right ]^\lw \dd u \dd \mu \dd \theta =
\nonumber\\[5pt]
\fl\hspace{1.5cm}
- \int B \left [
    \tilde\phi_{\spe 1}^\sw \cT_{NP, \spe} C_{\spe \spe^\prime}^\sw
  \right ]^\lw \dd u \dd \mu \dd \theta,
\end{eqnarray}
where $C_{\spe \spe^\prime}^\sw (\boldr, \bv, t)$ is the collision
operator applied on a function with wavelengths on the order of the
sound gyroradius. To prove this, we begin with the particle
conservation property of the collision operator, that gives
\begin{eqnarray}
  \left [ \varphi_1^\sw (\boldr, t) \int C_{\spe \spe^\prime}^\sw (\boldr, \bv, t) \dd^3 v \right ]^\lw = 0.
\end{eqnarray}
Using \eq{eq:varphiphiapprox} this equation becomes
\begin{eqnarray}
  \fl \int B \Bigg[ \phi_{\spe 1}^\sw (\boldr - \epsilon_\spe \rhobf (\boldr, \mu, \theta), \mu, \theta, t)\times
  \nonumber\\[5pt]
  \fl\hspace{0.5cm}
  C_{\spe \spe^\prime}^\sw (\boldr, u \bun + \rhobf (\boldr, \mu, \theta) \times \bB, t)
  \Bigg]^\lw \dd u \dd\mu\dd\theta = O(\epsilon_s).
\end{eqnarray}
Since we are only considering the long wavelength component, we can
Taylor expand around $\boldr$, leaving
\begin{eqnarray}
  \fl \int B \Bigg[ \phi_{\spe 1}^\sw (\boldr, \mu, \theta, t)
  \times
  \nonumber\\[5pt]
  \fl\hspace{0.5cm}
  C_{\spe \spe^\prime}^\sw (\boldr + \epsilon_\spe \rhobf (\boldr, \mu, \theta), u \bun + \rhobf (\boldr, \mu, \theta) \times \bB, t) \Bigg]^\lw \dd u \dd\mu\dd\theta
  \nonumber\\[5pt]
  \fl\hspace{0.5cm}
  = 
  O(\epsilon_s)
  ,
\end{eqnarray}
which is equivalent to \eq{eq:phicollisions}.

Substituting \eq{eq:phicollisions} into \eq{eq:dtquasineutrality5},
flux surface averaging and integrating by parts finally yields
\begin{eqnarray}\label{eq:dtquasineutrality6}
\fl \sum_\spe  \Bigg \langle \int B \Bigg [F_{\spe 1}^\sw \Bigg ( u\bun\cdot\nabla_\bR \langle\phi_{\spe 1}^\sw\rangle
\nonumber\\[5pt] \fl\hspace{0.5cm}
+
\frac{u^2}{B} (\bun \times (\bun\cdot\nabla_\bR\bun))\cdot\nabla_{\bR_\perp/\epsilon_\spe}\langle\phi_{\spe 1}^\sw\rangle
\nonumber\\[5pt] \fl\hspace{0.5cm} +\frac{\mu}{B}(\bun\times\nabla_\bR B)
\cdot\nabla_{\bR_\perp/\epsilon_\spe}\langle\phi_{\spe 1}^\sw\rangle \Bigg ) \Bigg ]^\lw \dd u \dd \mu \dd \theta \Bigg \rangle_\psi
\nonumber\\[5pt]
\fl\hspace{0.5cm}
- \sum_{\spe,\spe'}  \Bigg \langle \int B 
 \Bigg [\tilde\phi_{\spe 1}^\sw \Bigg ( \cT_{NP, \spe}^* C_{\sigma \sigma^\prime}
\Bigg[\modTinv F_{\sigma 1}^\sw 
\nonumber\\[5pt] \fl\hspace{0.5cm}
-
\frac{Z_\spe\lambda_\spe}{T_\spe}
\modTinv\tilde\phi_{\spe 1}^\sw \cT_{\spe,0}^{-1
*} F_{\spe 0}, \cT_{\spe',0}^{-1
*}F_{\spe' 0}
 \Bigg]\nonumber\\[5pt]
\fl\hspace{0.5cm}
 + \
\frac{\lambda_\spe}{\lambda_{\spe'}}
\cT_{NP, \spe}^*
C_{\sigma
\sigma^\prime} \Bigg[ \cT_{\spe,0}^{-1 *}F_{\spe 0} ,
\modTinvprime F_{\sigma' 1}^\sw 
\nonumber\\[5pt] \fl\hspace{0.5cm}
-
\frac{Z_{\spe'}\lambda_{\spe'}}{T_{\spe'}}
\modTinvprime\tilde\phi_{\spe'
1}^\sw \cT_{\spe',0}^{-1 *}F_{\spe' 0}
 \Bigg] \Bigg ) \Bigg ]^\lw \dd u \dd \mu \dd \theta \Bigg \rangle_\psi.
\end{eqnarray}
From this expression and \eq{eq:turbpiececollgyroave} we obtain
\eq{eq:cancellationenergy} by integrating by parts in $\mu$.

\section{Solvability conditions of the gyrokinetic 
Fokker-Planck equation of any order}
\label{app:solvabilityconditions}

We want to prove that the solvability conditions in subsection
\ref{sec:transportEqs} are the only ones to second order. We do this
by showing that, to general order, only the flux-surface averaged
zeroth and second moments of the Fokker-Planck equation can give
solvability conditions. To order $\epsilon_s^j$ the gyrokinetic
Fokker-Planck equation for species $\spe$ can be written as (we drop
the superindex lw in this appendix)
\begin{eqnarray}\label{eq:FPequationsSolvCond}
\fl\tau_\spe\lambda_\spe^{-j}&\left(
u\bun\cdot\nabla_\bR - 
\mu\bun\cdot\nabla_\bR B\partial_u
\right) G_{\spe j}
\nonumber\\[5pt]
\fl &-
\tau_\spe\sum_{\spe'}
\Bigg(
\cT_{\spe,0}^*
C_{\spe\spe'}
\Bigg[
\lambda_\spe^{-j}
\cT_{\spe,0}^{-1*}G_{\spe j},
\cT_{\spe',0}^{-1*}F_{\spe' 0}
\Bigg]
\nonumber\\[5pt]
\fl &
+
\cT_{\spe,0}^*
 C_{\spe\spe'}
\Bigg[
\cT_{\spe,0}^{-1*}F_{\spe 0},
\lambda_{\spe'}^{-j}
\cT_{\spe',0}^{-1*}G_{\spe' j}
\Bigg]
\Bigg)
=
\tau_\spe\lambda_\spe^{-j} R_{\spe j},
\end{eqnarray}
where $R_{\spe j}$ collects terms that do not contain $\langle F_{\spe
  j}\rangle$ for any $\spe$. To any order, $\langle G_{\spe
  j}\rangle = G_{\spe j}$ differs from $\langle F_{\spe j}\rangle$, at
most, in terms that have been determined by lowest order equations. We
recall that the gyrophase-dependent piece of the distribution function
to order $O(\epsilon_s^j)$, $F_{\spe j} - \langle F_{\spe j}\rangle$,
has been determined by the Fokker-Planck equation of order
$O(\epsilon_s^{j-1})$.  Also, we point out that we have introduced the
factor $\tau_\spe\lambda_\spe^{-j}$ in \eq{eq:FPequationsSolvCond}
because it is convenient for the proof that follows.

We must study the solvability conditions for the set of equations
\eq{eq:FPequationsSolvCond} when $\spe$ runs from $1$ to $N$, with $N$
the number of different species. To this end, it is appropriate to
work in the vector space
\begin{equation}
{\cal F}^N :={\cal
  F}({\cal P}_1)\times\dots\times{\cal F}({\cal P}_\spe)\times\dots\times
{\cal
  F}({\cal P}_N) ,
\end{equation}
which is the cartesian product of the sets of functions on the phase
spaces of the different species. Define $G_j = [ G_{1 j},\dots,  G_{\spe
  j},\dots, G_{N j}]$ and $S_j = [\tau_1 \lambda_1^{-j}
R_1,\dots,\tau_\spe \lambda_\spe^{-j} R_\spe,\dots, \tau_N
\lambda_N^{-j} R_N] \in {\cal F}^N$. On ${\cal F}^N$, the
set of equations \eq{eq:FPequationsSolvCond} can be rewritten as
\begin{equation}\label{eq:FPequationVeryCompact}
{\cal L}_j G_j = S_j,
\end{equation}
where
\begin{eqnarray}\label{eq:FPequationsCompactNotation}
\fl
({\cal L}_j G_j)_\spe
&:=
\tau_\spe\lambda_\spe^{-j}\left(
u\bun\cdot\nabla_\bR - 
\mu\bun\cdot\nabla_\bR B\partial_u
\right) G_{\spe j}
\nonumber\\[5pt]
\fl &-
\tau_\spe
\sum_{\spe'}
\Bigg(
\cT_{\spe,0}^*
C_{\spe\spe'}
\Bigg[
\lambda_\spe^{-j}
\cT_{\spe,0}^{-1*}G_{\spe j},
\cT_{\spe',0}^{-1*}F_{\spe' 0}
\Bigg]
\nonumber\\[5pt]
\fl &
+
\cT_{\spe,0}^*
C_{\spe\spe'}
\Bigg[
\cT_{\spe,0}^{-1*}F_{\spe 0},
\lambda_{\spe'}^{-j}
\cT_{\spe',0}^{-1*}G_{\spe' j}
\Bigg]
\Bigg).
\end{eqnarray}
The solvability conditions are defined by functions $K\in{\cal F}^N$ satisfying
\begin{eqnarray}
\sum_\spe
\int B K_\spe({\cal L}_j G_j)_\spe\dd^3 R\dd u\dd\mu\dd\theta = 0
\end{eqnarray}
for every $G_j\in{\cal F}^N$. Then, equation
\eq{eq:FPequationVeryCompact} implies that
\begin{eqnarray}
\sum_\spe\tau_\spe\lambda_\spe^{-j}
\int B K_\spe R_{\spe j}\dd^3 R\dd u\dd\mu\dd\theta = 0.
\end{eqnarray}

Let us denote by $K =
[K_1,\dots,K_\spe,\dots, K_N]$ and $G =
[G_1,\dots,G_\spe,\dots, G_N]$ two arbitrary elements of ${\cal
  F}^N$. Then, a natural scalar product is defined by
\begin{equation}
\fl(K \vert G)
=\sum_\spe \int B(\bR)K_\spe(\bR,u,\mu,\theta)G_\spe(\bR,u,\mu,\theta)
\dd^3 R
\dd u\dd\mu\dd\theta.
\end{equation}
The question about solvability conditions can be expressed in
terms of the scalar product. Our aim is to find those $K\in{\cal F}^N$
such that
\begin{equation}
(K\vert {\cal L}_j G_j) = 0
\end{equation}
for every $G_j\in {\cal F}^N$. Since the scalar product is
non-degenerate, the condition for $K$ is equivalent to $({\cal
  L}_j^\dagger K \vert G_j) = 0$, where ${\cal L}_j^\dagger$ is the
adjoint of ${\cal L}_j$. Therefore, the solvability conditions derived
to $j$th order are given by the equations
\begin{equation}\label{eq:solvcondintermsofKernel}
(K\vert S_j)
= 0, \quad K\in \mbox{Ker}({\cal L}_j^\dagger).
\end{equation}
Of course, it might happen that some of these equations be trivial
identities that do not add new conditions on lower-order
quantities. The important point is that every non-trivial
solvability condition is found by calculating all of the equations
\eq{eq:solvcondintermsofKernel}.

We turn to compute ${\cal L}_j^\dagger$. It is obvious that the piece
of ${\cal L}_j$ associated to parallel streaming, ${\cal L}_{j
  ||}$, defined by
\begin{equation}
\fl
({\cal L}_{j ||} G_j)_\spe
:=
\tau_\spe\lambda_\spe^{-j}\left(
u\bun\cdot\nabla_\bR - 
\mu\bun\cdot\nabla_\bR B\partial_u
\right) G_{\spe j},
\end{equation}
is antisymmetric. That is,
\begin{equation}
(K\vert {\cal L}_{j ||} G_j) = - ({\cal L}_{j ||} K\vert G_j),
\end{equation}
for any $K\in{\cal F}^N$. In other words, ${\cal L}_{j ||}^\dagger =
-{\cal L}_{j ||}$. It will also be useful to note that
\begin{equation}
 ({\cal L}_{j ||} G_j)_\spe = F_{\spe 0}
 ({\cal L}_{j ||} {\hat G}_j)_\spe
,
\end{equation}
with ${\hat G}_{\spe j} :=  G_{\spe j}/F_{\spe 0}$, and
\begin{eqnarray}
 \fl F_{\spe 0}(\bR,u,\mu)
=
\frac{n_{\spe}(\psi)}{(2\pi T_{\spe}(\psi))^{3/2}}
\exp\left(-\frac{\mu B(\psi,\Theta) + u^2/2}{T_{\spe}(\psi)}\right).
\end{eqnarray}

In order to find the adjoint of the piece of ${\cal L}_j$
corresponding to collisions,
\begin{equation}
{\cal L}_{j \, coll} = {\cal L}_j -
{\cal L}_{j ||}, 
\end{equation}
we need to prove a preliminary property. Define
\begin{eqnarray}
\fl {\hat C}_{\spe\spe'}[{\hat G}_{\spe j},{\hat G}_{\spe' j}]
&=
C_{\spe\spe'}
\Bigg[
\lambda_\spe^{-j}
\cT_{\spe,0}^{-1*}G_{\spe j},
\cT_{\spe',0}^{-1*}F_{\spe' 0}
\Bigg]
\nonumber\\[5pt]
\fl &
+
C_{\spe\spe'}
\Bigg[
\cT_{\spe,0}^{-1*}F_{\spe 0},
\lambda_{\spe'}^{-j}
\cT_{\spe',0}^{-1*}G_{\spe' j}
\Bigg].
\end{eqnarray}
From definition
\eq{eq:collisionoperatornondim} one obtains
\begin{eqnarray}\label{eq:collisionsgspegspeprime}
\fl {\hat C}_{\spe\spe'}[{\hat G}_{\spe j},{\hat G}_{\spe' j}]=
\nonumber\\[5pt]
\fl\hspace{0.5cm}
\gamma_{\spe\spe'}
\nabla_{\bv}
\cdot
\int
\matW
(\tau_\spe\bv - \tau_{\spe'}\bv')\cdot
\nonumber\\[5pt]
\fl\hspace{0.5cm}
\Bigg(
\frac{\tau_\spe}{\lambda_\spe^{j}}\nabla_\bv \hat g_\spe -
\frac{\tau_{\spe'}}{\lambda_{\spe'}^{j}}\nabla_{\bv'} \hat g_{\spe'}
\Big)f_{\spe 0}(\bv)f_{\spe' 0}(\bv')\dd^3v'.
\end{eqnarray}
Here, to ease the notation, we understand $f_{\spe 0} \equiv
\cT_{\spe,0}^{-1*}F_{\spe 0}$ and $\hat g_\spe \equiv \cT_{\spe,0}^{-1*}
{\hat G}_{\spe j}$. To get \eq{eq:collisionsgspegspeprime} we have
used
\begin{eqnarray}
\nabla_\bv f_{\spe 0} = -\frac{1}{T_\spe}\bv f_{\spe 0},
\end{eqnarray}
\begin{eqnarray}
T_\spe = T_{\spe'}, \mbox{ for every pair $\spe$, $\spe'$,}
\end{eqnarray}
and
\begin{eqnarray}
  \matW
  (\tau_\spe\bv - \tau_{\spe'}\bv')\cdot 
  (\tau_\spe\bv - \tau_{\spe'}\bv') \equiv 0, 
\mbox{ for every pair $\spe$, $\spe'$.}
\end{eqnarray}
The operator ${\hat C}_{\spe\spe'}[{\hat G}_{\spe j},{\hat G}_{\spe'
  j}]$ does not have nice symmetry properties with respect to the
scalar product, but its symmetrization in $\spe$ and $\spe'$ does. A
simple integration by parts yields the following symmetric expression
for any pair of functions $k_\spe(\bv)$ and $k_{\spe'}(\bv)$:
\begin{eqnarray}
\fl
\tau_\spe \int k_\spe(\bv)
\cT_{\spe,0}^{*}{\hat C}_{\spe\spe'}[{\hat G}_{\spe j},{\hat G}_{\spe' j}]\dd^3 v
\nonumber\\[5pt]
\fl\hspace{0.5cm}
+
\tau_{\spe'} \int k_{\spe'}(\bv)
\cT_{\spe',0}^{*}{\hat C}_{\spe'\spe}[{\hat G}_{\spe' j},{\hat G}_{\spe j}]\dd^3 v=
\nonumber\\[5pt]
\fl\hspace{0.5cm}
-
\gamma_{\spe\spe'}
\int \left(\tau_\spe\nabla_\bv k_\spe - \tau_{\spe'}\nabla_{\bv'} k_{\spe'} \right)\cdot
\matW
(\tau_\spe\bv - \tau_{\spe'}\bv')\cdot
\nonumber\\[5pt]
\fl\hspace{0.5cm}
\left( \tau_\spe
\nabla_\bv\left(\lambda_\spe^{-j} \hat g_\spe\right) -
\tau_{\spe'}\nabla_{\bv'}\left(
\lambda_{\spe'}^{-j} \hat g_{\spe'}
\right)
\right)f_{\spe 0}(\bv)f_{\spe' 0}(\bv')\dd^3v\dd^3v'.
\end{eqnarray}
Hence, denoting $K_\spe\equiv \cT_{\spe,0}^{*}k_\spe $, we can write
\begin{eqnarray}
\fl\tau_\spe\int k_\spe(\bv)
\cT_{\spe,0}^{*}{\hat C}_{\spe\spe'}[{\hat G}_{\spe j},{\hat G}_{\spe' j}]\dd^3 v
\nonumber\\[5pt]
\fl\hspace{0.5cm}
+
\tau_{\spe'}\int
 k_{\spe'}(\bv)
\cT_{\spe',0}^{*}{\hat C}_{\spe'\spe}[{\hat G}_{\spe' j},{\hat G}_{\spe j}]\dd^3 v
=
\nonumber\\[5pt]
\fl\hspace{0.5cm}
\frac{\tau_\spe}{\lambda_\spe^{j}}\int
 \cT_{\spe,0}^{*}{\hat C}_{\spe\spe'}
\left[\lambda_\spe^{j} K_\spe,
\lambda_{\spe'}^{j} K_{\spe'}
\right]
 \hat g_\spe(\bv)\dd^3 v
\nonumber\\[5pt]
\fl\hspace{0.5cm}
+
\frac{\tau_{\spe'}}{\lambda_{\spe'}^{j}} \int
\cT_{\spe',0}^{*}{\hat C}_{\spe'\spe}
\left[\lambda_{\spe'}^{j} K_{\spe'},
\lambda_{\spe}^{j} K_{\spe}
\right]
\hat g_{\spe'}(\bv)\dd^3 v.
\end{eqnarray}

Thus, for every $K\in{\cal F}^N$,
\begin{eqnarray}
\fl  (K\vert {\cal L}_j G)=
\nonumber\\[5pt]
\fl\hspace{0.5cm}
-
\sum_\spe
\int
\Bigg[\frac{\tau_\spe B}{\lambda_\spe^j F_{\spe 0}}
\left(
u\bun\cdot\nabla_\bR - 
\mu\bun\cdot\nabla_\bR B\partial_u
\right)F_{\spe 0}K_\spe\Bigg]
G_{\spe j}\dd^3 R
\dd u\dd\mu\dd\theta
\nonumber\\[5pt]
\fl\hspace{0.5cm}
-
\sum_{\spe,\spe'}
\int
\frac{\tau_\spe B}{\lambda_{\spe}^{j}F_{\spe 0}} 
\cT_{\spe,0}^{*}{\hat C}_{\spe\spe'}
\left[\lambda_{\spe}^{j}K_{\spe},
\lambda_{\spe'}^{j}K_{\spe'}
\right]
G_{\spe j}\dd^3 R
\dd u\dd\mu\dd\theta.
\end{eqnarray}
This means that for any $K\in{\cal F}^N$, the action
of the adjoint of ${\cal L}_j$ is given by
\begin{eqnarray}
\fl
({\cal L}_j^\dagger K)_\spe=
-
\Bigg[\frac{\tau_\spe}{\lambda_\spe^j F_{\spe 0}}
\left(
u\bun\cdot\nabla_\bR - 
\mu\bun\cdot\nabla_\bR B\partial_u
\right)F_{\spe 0}K_\spe\Bigg]
\nonumber\\[5pt]
\fl\hspace{0.5cm}
-
\sum_{\spe'}
\frac{\tau_\spe}{\lambda_{\spe}^{j}F_{\spe 0}} 
\cT_{\spe,0}^{*}{\hat C}_{\spe\spe'}
\left[\lambda_{\spe}^{j}K_{\spe},
\lambda_{\spe'}^{j}K_{\spe'}
\right].
\end{eqnarray}
An entropy argument similar to the one employed in subsection
\ref{sec:FokkerPlancklong0} can be used here to obtain the solutions of
${\cal L}_j^\dagger K = 0$. Multiplying the equation ${\cal
  L}_j^\dagger K = 0$ by $\lambda_\spe^j B K_\spe F_{\spe 0}$,
integrating over $u,\mu$, and $\theta$, flux-surface averaging, and
summing over all of the species gives
\begin{eqnarray}
\fl
-
\left\langle\sum_{\spe,\spe'}\tau_{\spe}
\int B K_\spe
\cT_{\spe,0}^{*}{\hat C}_{\spe\spe'}
\left[\lambda_{\spe}^{j}K_{\spe},
\lambda_{\spe'}^{j}K_{\spe'}
\right]\dd u\dd\mu\dd\theta\right\rangle_\psi = 0,
\end{eqnarray}
that can be recasted into
\begin{eqnarray}
\fl\Bigg\langle\sum_{\spe,\spe'}
\frac{\gamma_{\spe\spe'}}{2}
\int
\left(\tau_\spe\nabla_\bv k_\spe-
\tau_{\spe'}\nabla_{\bv'} k_{\spe'}
 \right)\cdot\matW(\tau_\spe\bv - \tau_{\spe'}\bv')\cdot
\nonumber\\[5pt]
\fl\hspace{1cm}
\left(\tau_\spe\nabla_\bv k_\spe-
\tau_{\spe'}\nabla_{\bv'} k_{\spe'}
 \right)
f_{\spe 0}
f_{\spe' 0}
\dd^3v\dd^3v'
\Bigg\rangle_\psi = 0.
\end{eqnarray}
This equation has the following types of solutions: $k_\spe =
q_\spe(\boldr)$, $k_\spe = \tau_\spe^{-1} \bV(\boldr)\cdot\bv$, and
$k_\spe = Q(\boldr) \bv^2/2$. Again, in analogy with the calculation
of subsection \ref{sec:FokkerPlancklong0}, it is easy to show that
${\cal L}^\dagger_j K = 0$ implies that $\bV(\bR)\equiv 0$ and that
$\{q_\spe$, $\spe = 1,\dots,N\}$ and $Q$ are flux functions, but
otherwise arbitrary. In other words, every $K\in\mbox{Ker}({\cal
  L}_j^\dagger)$ can be written as a linear combination of elements of
the form
\begin{eqnarray}
\fl [q_1(\psi),0,\dots,0],\, \dots,\, [0,\dots, q_\spe(\psi),\dots,0],\,
 \dots, \, [0,\dots, q_N(\psi)],
\nonumber\\[5pt]
\fl
Q(\psi)\left(u^2/2 + \mu B\right)[1,
\, \dots, 
1,\, \dots, \, 1
],
\end{eqnarray}
where the functions $\{q_\spe$, $\spe = 1,\dots,N\}$ and $Q$ are
arbitrary. Then, the solvability conditions are given by
\eq{eq:solvcondintermsofKernel}. Equivalently, due to the
arbitrariness of the functions $\{q_\spe$, $\spe = 1,\dots,N\}$ and $Q$,
the solvability conditions can be expressed as
\begin{eqnarray}\label{eq:solvcondFSA}
  \sum_\spe\left\langle
\tau_\spe\lambda_\spe^{-j}
\int B K_\spe R_{\spe j}\dd u\dd\mu\dd\theta
\right\rangle_\psi = 0, \quad K\in\mbox{Ker}({\cal
    L}_j^\dagger).
\end{eqnarray}
More concretely, all the solvability conditions of the Fokker-Planck
equation to order $O(\epsilon_s^j)$ are obtained by working out
\begin{eqnarray}\label{eq:solvcondConcretely}
  \fl
  \left\langle
    \tau_\spe\lambda_\spe^{-j} \int B R_{\spe j}\dd u\dd\mu\dd\theta
  \right\rangle_\psi = 0, \mbox{ for each $\spe$, and}\nonumber\\[5pt]
  \fl
  \left\langle
    \sum_\spe
    \tau_\spe\lambda_\spe^{-j}\int B \left(u^2/2 + \mu B\right)
 R_{\spe j}\dd u\dd\mu\dd\theta
  \right\rangle_\psi = 0.
\end{eqnarray}

The proof in this appendix guarantees that transport equations for
particle and total energy density are the only solvability conditions
for the long-wavelength second-order Fokker-Planck equation,
\eq{eq:eqH2sigma}. Finally, the reader can immediately check that when
\eq{eq:solvcondConcretely} is applied to the first-order equations,
\eq{eq:Vlasovorder1gyroav4}, no condition is obtained.

\end{document}